\documentclass[8.5pt,twoside,twocolumn]{article}
\usepackage[super,sort&compress,comma]{natbib}
\usepackage{mhchem}
\usepackage{times,mathptmx}
\usepackage{sectsty}
\usepackage{balance}
\usepackage{extarrows}
\usepackage{graphicx} 
\usepackage{lastpage}
\usepackage[format=plain,justification=raggedright,singlelinecheck=false,font=small,labelfont=bf,labelsep=space]{caption}
\usepackage{fancyhdr}
\usepackage{color}

\graphicspath{{/media/ioan/Elements/gt/thermal_corr/data/}}
\DeclareGraphicsExtensions{.jpg, .eps, .ps, .png, .pdf}

\begin{document}

\thispagestyle{plain}
\fancypagestyle{plain}{
\renewcommand{\headrulewidth}{1pt}}
\renewcommand{\thefootnote}{\fnsymbol{footnote}}
\renewcommand\footnoterule{\vspace*{1pt}%
\hrule width 3.4in height 0.4pt \vspace*{5pt}}
\setcounter{secnumdepth}{5}

\makeatletter
\def\subsubsection{\@startsection{subsubsection}{3}{10pt}{-1.25ex plus -1ex minus -.1ex}{0ex plus 0ex}{\normalsize\bf}}
\def\paragraph{\@startsection{paragraph}{4}{10pt}{-1.25ex plus -1ex minus -.1ex}{0ex plus 0ex}{\normalsize\textit}}
\renewcommand\@biblabel[1]{#1}
\renewcommand\@makefntext[1]%
{\noindent\makebox[0pt][r]{\@thefnmark\,}#1}
\makeatother
\renewcommand{\figurename}{\small{Fig.}~}
\sectionfont{\large}
\subsectionfont{\normalsize}

\fancyfoot{}
\fancyfoot[RO]{\footnotesize{\sffamily{1--\pageref{LastPage} ~\textbar  \hspace{2pt}\thepage}}}
\fancyfoot[LE]{\footnotesize{\sffamily{\thepage~\textbar\hspace{3.45cm} 1--\pageref{LastPage}}}}
\fancyhead{}
\renewcommand{\headrulewidth}{1pt}
\renewcommand{\footrulewidth}{1pt}
\setlength{\arrayrulewidth}{1pt}
\setlength{\columnsep}{6.5mm}
\setlength\bibsep{1pt}
\newcommand{\paper}{paper}
\newcommand{\alt}{\raisebox{-0.3ex}{$\stackrel{<}{\sim}$}}
\newcommand{\agt}{\raisebox{-0.3ex}{$\stackrel{>}{\sim}$}}
\newcommand{\figname}{\small{Fig.}~}
\newcommand{\figsname}{\small{Figs.}~}
\newcommand{\secname}{Sec.~}
\newcommand{\esi}{ESI$^\dagger$}
\twocolumn[
  \begin{@twocolumnfalse}
    \noindent\LARGE{\textbf{Can room temperature data for tunneling molecular junctions be analyzed within a theoretical framework assuming zero temperature? 
    }}
\vspace{0.6cm}

\noindent\large{\textbf{Ioan B\^aldea 
\textit{$^{a \ast}$}
}}\vspace{0.5cm}

\noindent
\normalsize{Routinely, experiments on tunneling molecular junctions report values of conductances ($G_{RT}$) and currents ($I_{RT}$) measured at room temperature.
  On the other side, theoretical approaches based on simplified models
  provide analytic formulas for the conductance ($G_{0K}$) and current ($I_{0K}$) valid at zero temperature.
  Therefore, interrogating the applicability of the theoretical results
  deduced in the zero temperature limit to real experimental situations at room temperature
  (i.e., $G_{RT} \approx G_{0K}$ and  $I_{RT} \approx I_{0K}$) is a relevant aspect.
  Quantifying the pertaining 
  temperature impact on the transport properties computed within
  the ubiquitous single level model with Lorentzian transmission
  is the specific aim of the present work. Comprehensive results are presented for broad ranges of the relevant parameters
  (level's energy offset $\varepsilon_0$ and width $\Gamma_a $, and applied bias $V$)
  that safely cover values characterizing currently fabricated junctions. They demonstrate that the strongest
  thermal effects occur at biases below resonance
  ($2 \left\vert \varepsilon_0 \right\vert - \delta\varepsilon_0 \alt \vert e V\vert \alt 2 \left\vert \varepsilon_0 \right\vert$). At fixed $V$,
  they affect an $\varepsilon_0$-range whose largest width $\delta\varepsilon_0 $
  is about nine times larger than the thermal energy
  ($\delta\varepsilon_0 \approx 3 \pi k_B T$) at $\Gamma_a \to 0$. 
  The numerous figures included aim at conveying a quick overview on the applicability
  of the zero temperature limit to a specific real junction.
  In quantitative terms, the conditions of applicability are expressed as mathematical inequalities involving elementary functions.
  They constitute the basis of an interactive data fitting procedure proposed, which aims at guiding
  experimentalists interested in data processing in a specific case.
}
$ $ \\  

  {{\bf Keywords}:
molecular electronics, nanojunctions, single level model, thermal effects}
\vspace{0.5cm}
  \end{@twocolumnfalse}
]

\footnotetext{\textit{$^{a}$~Theoretical Chemistry, Heidelberg University, Im Neuenheimer Feld 229, D-69120 Heidelberg, Germany}}
\footnotetext{$^\ast$~E-mail: ioan.baldea@pci.uni-heidelberg.de
}
\section{Introduction}
\label{sec:intro}
Routinely, charge transport experiments on molecular junctions are carried out at room temperature (RT).
Nevertheless, most theoretical approaches--- especially those based on model simulations
\cite{Schmickler:86,Metzger:01b,Stafford:96,Buttiker:03,Baldea:2012a}---
were developed for zero temperature ($T = 0$).
In fact, a series of studies conducted at variable temperature ($T$)
\cite{Poot:06,Choi:08,Reed:09,Tao:11,Lambert:11,Zandvliet:12,McCreery:13a,Asadi:13,Lewis:13,Tao:16b,McCreery:16a,Nijhuis:16b,Nijhuis:16d,Guo:17,McCreery:17a,Baldea:2018a,Guo:21} revealed a significant $T$-dependence of transport properties,
which turned out to be fully compatible with a tunneling mechanism
\cite{Poot:06,Nijhuis:16b,Baldea:2017d,Baldea:2022c,Baldea:2022j}.

In this vein, interrogating the applicability of the theoretical results
deduced in the zero temperature limit to real experimental situations at room temperature
is a relevant aspect. Quantifying the pertaining thermal corrections to transport properties both at
low bias (ohmic regime) and higher bias (nonlinear regime) is the aim of the present paper.

Although extensive numerical results will be reported below, this theoretical study is not merely intended
to be a comprehensive numerical simulation experiment. 
Equally important, we aim at (i) clearly formulating simple conditions (mathematical inequalities)
legitimating the applicability of formulas deduced theoretically
for $T=0$ to process transport measurements performed at room temperature
and (ii) at proposing a practical receipt guiding experimentalists through an interactive data fitting procedure
able to extract reliable model parameters.
\section{Model and working equations}
\label{sec:model}
In order to make the paper self-contained, let us start with a short recap \cite{baldea:comment}.
Important insight into charge transport through tunneling molecular junctions
can be gained by assuming a single dominant molecular orbital MO
(usually, HOMO or LUMO) coupled via energy independent couplings $\Gamma_{s,t}$ to
wide, flat metallic electrodes (hence Lorentzian-shaped transmission) subject to an applied bias $V$
\cite{Meir:92,Stafford:96,Metzger:01b,HaugJauho,Zotti:10,Baldea:2012a,CuevasScheer:17}.

The exact expression of the tunneling current for this single level model can be
written as a particular case of the general formula
deduced for the charge transport by tunneling \cite{Caroli:71a}
\begin{eqnarray}
  I_{exact} & = & \frac{2 e}{h} \int_{-\infty}^{\infty} \mathcal{T}(\varepsilon)
  \left[f\left(\varepsilon - \frac{e V}{2}\right) - f\left(\varepsilon + \frac{e V}{2} \right)\right] d \varepsilon \nonumber \\  
  \label{eq-Iexact}
  & = & \frac{2 e}{h} \Gamma_{g}^2 \int_{-\infty}^{\infty}
  \frac{f\left(\varepsilon - e V/2\right) - f\left(\varepsilon + e V/2 \right)}{\left(\varepsilon - \varepsilon_0\right)^2 + \Gamma_{a}^2} d \varepsilon 
\end{eqnarray}
Above, $\varepsilon_0 \equiv E_{MO} - E_F$ is the MO energy offset relative to electrodes' Fermi energy,
$f(\varepsilon) = 1/\left[1 + \exp(\beta \varepsilon)\right]$ is the Fermi distribution
($1/\beta = k_B T$). $\Gamma_g$ and $\Gamma_a$ are the geometric and arithmetic MO-electrode couplings 
\begin{subequations}
\begin{equation}
  \label{eq-Gamma-g}
  \Gamma_{g} = \sqrt{\Gamma_s \Gamma_t}
  \end{equation}
\begin{equation}
  \label{eq-Gamma-a}
  \Gamma_{a} = \left(\Gamma_s + \Gamma_t\right)/2
\end{equation}
\end{subequations}

In general, the energy offset entering above is bias dependent, e.g., \cite{Datta:03,Baldea:2012a}
$\varepsilon_0 \to \varepsilon_0(V) = \left . \varepsilon_0(V)\right\vert_{V=0}  + \gamma e V $
A nonvanishing $\gamma$ yields an asymmetric current voltage curve ($I(-V) \neq -I(V)$).
Although the formulas for the current (eqn~(\ref{eq-Iexact}), (\ref{eq-I0K}), and (\ref{eq-I0Koff}))
hold for arbitrary bias dependent MO offsets,
given the fact that current rectification is not our main focus here, below we only present numerical
results for a bias independent $\varepsilon_0$ ($\gamma \equiv 0$).

For MO's symmetrically coupled to electrodes, all $\Gamma$'s are equal: $\Gamma_a = \Gamma_g = \Gamma_s =\Gamma_t$.
To avoid confusions (see ref.~\citenum{baldea:comment}),
we note that $\Gamma$'s used by us differ by a factor two from
quantities denoted by the same symbol by other authors (e.g., ref.~\citenum{CuevasScheer:17}).

The expression of the current at zero temperature $I_{0K}$ follows as an exact result from eqn~(\ref{eq-Iexact})
wherein  the Fermi distribution $f(\varepsilon)$ reduces to the Heaviside step function
\cite{Meir:92,Stafford:96,Metzger:01b,HaugJauho,Zotti:10,Baldea:2012a,CuevasScheer:17}
\begin{equation}
  \label{eq-I0K}
  I_{0K} = \frac{2 e}{h \Gamma_a} \Gamma_{g}^2
  \left(
  \tan^{-1}\frac{\varepsilon_0 + eV/2}{\Gamma_{a}} -
  \tan^{-1}\frac{\varepsilon_0 - eV/2}{\Gamma_{a}}
  \right)
\end{equation}

Because the zero bias (also referred to as ohmic or low bias) conductance defined by
\begin{equation}
  \label{eq-g}
  G = \lim_{V\to 0} I(V)/V = \lim_{V\to 0} \partial I(V)/\partial V
  \end{equation}
represents the focus of most experiments
done in molecular electronics, it is meaningful to consider the expression of $G$
pertaining to the currents expressed by eqn~(\ref{eq-Iexact}) and (\ref{eq-I0K}).
Eqn~(\ref{eq-I0K}) straightforwardly yields
\begin{eqnarray}
  \label{eq-g0K}
  \frac{G}{G_0}  & \simeq & \frac{G_{0K}}{G_0} = \frac{\Gamma_{g}^2}{\varepsilon_0^2 + \Gamma_a^2} 
 \end{eqnarray}
Sommerfeld expansions \cite{Sommerfeld:33,AshcroftMermin} were employed to
derive thermal corrections to $G$ in closed analytic form \cite{Baldea:2022c,Baldea:2022j}.
Although they may suffice for the present paper, wherein we aim at considering rather modest
deviations of transport properties at room temperature from those at $T = 0$,
we prefer to use the zero bias (ohmic) value of the exact conductance
pertaining to the exact current $I_{exact}$, because it can be 
expressed analytically for arbitrary values of the model parameters\cite{Baldea:2022c}.
In terms of the real part of Euler's trigamma function of complex argument
function $\psi^{\prime}(z)$ \cite{JahnkeEmde:45,AbramowitzStegun:64},
the exact conductance reads 
\begin{equation}
  \label{eq-g-exact}
  \frac{G_{exact}}{G_0} = 
\frac{\Gamma_{g}^2}{2 \pi \Gamma_{a} k_B T} \mbox{Re}\, \psi^{\prime} \left(\frac{1}{2} + \frac{\Gamma_{a}}{2\pi k_B T} + i\,\frac{\varepsilon_0}{2\pi k_B T}\right) 
\end{equation}
The trigamma function represents the derivative of the digamma function,
$\psi^{\prime}(z) \equiv \psi(1; z) \equiv \frac{d}{d\,z}\psi(z)$,
which, in turn, is the logarithmic derivative of Euler's gamma function
\cite{AbramowitzStegun:64}.

In view of the foregoing analysis, situations wherein $I_{0K}$ (eqn~(\ref{eq-I0K}))
and  $G_{0K}$ (eqn~(\ref{eq-g0K})) 
represent good approximation of $I_{exact}$ (eqn~(\ref{eq-Iexact})) and $G_{exact}$  (eqn~(\ref{eq-g-exact})), respectively
can be referred to as the low temperature limit
\begin{equation}
\label{eq-def-low-T}
  \mbox{Low temperature limit: } \left\{
  \begin{array}{ll}
    I_{0K} \approx I_{exact} \\
    G_{0K} \approx G_{exact} \\
  \end{array}
  \right .
\end{equation}
The low temperature limit applies in cases where 
the transmission function (whose shape is controlled by $\varepsilon_0$ and $\Gamma_{a}$) exhibits a negligible 
variation within energy ranges of widths of the order $\sim k_B T$ around the electrode Fermi levels ($\pm e V/2$).
Consequently, MO levels need be sufficiently far away from the Fermi levels 
for the low temperature limit to apply;
$\left\vert\left\vert \varepsilon_0\right\vert -  \vert e V\vert/2\right\vert $ should be sufficiently larger than $k_B T$.

Provided that the arguments of the inverse trigonometric functions entering
eqn~(\ref{eq-I0K}) are sufficiently large ($x > x_0 = 2.928$, see \figname\ref{fig:arctan})
\begin{subequations}
  \begin{equation}
    \label{eq-arctan}
    x \equiv \frac{\left\vert \varepsilon_0 \right\vert - \vert e V\vert/2}{\Gamma_a} > x_0 = 2.928
    \end{equation}
  \begin{equation}
    \tan^{-1}(x) \simeq 
    \frac{\pi}{2} - \frac{1}{x} 
  \end{equation}
\end{subequations}
eqn~(\ref{eq-I0K}) reduces to \cite{Baldea:2012a,baldea:comment}
\begin{equation}
  \label{eq-I0Koff}
  I_{0K,\mbox{\small off}} = G_0 \frac{\Gamma_{g}^2}{\varepsilon_0^2 - (e V/2)^2} V
\end{equation}
where $G_0 = 2 e^2/h = 77.48\,\mu$S is the universal conductance quantum.

As a rule of thumb, for applying eqn~(\ref{eq-I0Koff})
we suggested (see ref.~\citenum{baldea:comment} and citations therein)
an upper bias limit 
\begin{equation}
  \label{eq-1.4}
  \vert V \vert \alt V_{1.4} \equiv 1.4\,\left\vert\varepsilon_0\right\vert / e
\end{equation}
Eqn~(\ref{eq-1.4}) is justified by the the fact that most molecular
junctions have a conductance much smaller than $G_0$
\begin{equation}
  \label{eq-Gamma-vs-e0}
  G/G_0 \alt 0.01 \xlongrightarrow{\Gamma_g \approx \Gamma_a} \Gamma_a \alt \left\vert\varepsilon_0\right\vert / 10
\end{equation}
In cases where eqn~(\ref{eq-Gamma-vs-e0}) holds, eqn~(\ref{eq-1.4}) follows 
via eqn~(\ref{eq-arctan}).

Eqn~(\ref{eq-1.4}) and (\ref{eq-Gamma-vs-e0}) express the rationale
of using the term ``off-resonant single level model'' for the transport by tunneling
modeled using eqn~(\ref{eq-I0Koff}): at the biases envisaged (eqn~(\ref{eq-1.4})),
the energy mismatch between (MO) level and the closest electrodes' electrochemical potential
($\left \vert \varepsilon_0 \right\vert \agt 0.7 \vert e V\vert $ versus $\vert e V\vert /2$
is much larger than the level broadening $\Gamma_a$ due to the MO-coupling to electrodes
($\left \vert \varepsilon_0 \right\vert - \vert e V\vert /2 \gg \Gamma_a$).

Eqn~(\ref{eq-I0Koff}) straightforwardly yields
\begin{eqnarray}
   \label{eq-g0Koff}
  \frac{G}{G_0} & \simeq & \frac{G_{0K,off}}{G_0} = \frac{\Gamma_{g}^2}{\varepsilon_0^2}
\end{eqnarray}

For large values of the argument in the RHS of eqn~(\ref{eq-g-exact})  
\begin{equation*}
  \vert z \vert \equiv \sqrt{\left(\frac{1}{2} + \frac{\Gamma_a}{2 \pi k_B T}\right)^2 + \left(\frac{\varepsilon_0}{2 \pi k_B T}\right)^2}
    \gg 1 \ , \ \vert\arg z\vert < \pi
\end{equation*}
the trigamma function can be approximated by the first term of its asymptotic expansion \cite{AbramowitzStegun:64}
\begin{equation*}
  \psi^{\prime}(z) = \frac{1}{z} + \mathcal{O}\left(z^{-2}\right)
\end{equation*}
This shows that the exact 
eqn~(\ref{eq-g-exact}) of the zero bias conductance $G_{exact}$
recovers the expression of $G_{0K}$ at $T=0$ (eqn~(\ref{eq-g0K})) in the (low temperature) limit
\begin{equation*}
  \sqrt{\left(\pi k_B T + \Gamma_a\right)^2 + \varepsilon_0^2} \gg 2 \pi k_B T
\end{equation*}

The above equation 
has a precise physical content: thermal effects do not substantially affect the
charge transport in cases where the transmission function does not appreciably varies at energies accessed by electrons
thermally excited above electrode's Fermi energy.
It 
expresses the low temperature condition at $V = 0$ but the physical insight gained in this way is 
clear and this makes generalization at $V \neq 0$ straightforward.

At finite biases ($V \neq 0$), the low temperature limit is justified in situations where
energy ranges around $\varepsilon_0$ having widths $\sim \Gamma_a$ 
(wherein the transmission function rapidly varies) do not notably overlap with energy ranges around electrodes' electrochemical potential
smeared out by thermic excitations. An applied bias brings the MO closer in energy 
to the closest electrode's electrochemical potential.
Mathematically, this amounts to replace in the above equation $\varepsilon_0$ 
by $\left\vert \varepsilon_0 \right\vert - \vert e V \vert /2$
\begin{subequations}
\label{eq-low-T-V} 
  \begin{equation}
  \label{eq-low-T-V-gg}
  \sqrt{\left(\pi k_B T + \Gamma_a\right)^2 + \left(  \left\vert \varepsilon_0 \right\vert - \vert e V \vert /2\right)^2} \gg 2 \pi k_B T
\end{equation}
or, equivalently,
  \begin{equation}
  \label{eq-low-T-V-q}
  \sqrt{\left(\pi k_B T + \Gamma_a\right)^2 + \left(  \left\vert \varepsilon_0 \right\vert - \vert e V \vert /2\right)^2} > q \pi k_B T
\end{equation}
\end{subequations}
where $q$ is a dimensionless number sufficiently larger than two.

To sum up, while the foregoing analysis allows one to understand that the applicability of
the zero temperature limit can be expressed in terms of certain mathematical inequalities,
what does ``sufficiently small'' or ``sufficiently large'' precisely mean in the foregoing analysis
is a question that cannot merely be settled based on qualitative considerations like those delineated above.

To address this question and find out, e.g., what is a ``good'' numerical value of $q$
to be used in eqn~(\ref{eq-low-T-V-q}), we conducted extensive numerical simulations,
as detailed in the next section.
\section{Results of numerical simulations}
\label{sec:results}
The impact of a variable temperature on the charge transport by tunneling
is an interesting topic
(see, e.g., the Arrhenius-Sommerfeld transition \cite{Baldea:2022b} or the possibility of estimating
the number of molecules in large area molecular junctions \cite{Baldea:2022j}), but 
a full analysis of thermal effects at variable temperature will not be attempted here.
Rather, in view of unpleasant flaws in recent analysis of the thermal effects 
(see discussion in ref.~\citenum{baldea:comment}), 
we find it useful and aim at comprehensively characterizing physical situations wherein 
transport properties at room temperature can be reasonably estimated via more facile
computations assuming $T=0$.
Therefore, in all numerical results presented below, ``temperature'' means ``room temperature''
(RT, $T=T_{RT} = 298.15$\,K, $k_B T = k_B T_{RT} = 25.7$\,meV).
``Exact'' current and zero bias conductance values are values at room temperature:
$I_{exact} = I_{RT}$ and $G_{exact} = G_{RT}$.

The results reported below quantify the impact of the
three parameters which are relevant for the present study:
$\varepsilon_0$, $\Gamma_a$, and $V$.
Noteworthily, $\Gamma_g$ plays no role in discussing relative deviations between the cases
$T=0$ and RT; it merely enters all above formulas for $G$'s and $I$'s as a multiplicative factor.
The broad ranges of model parameters $\varepsilon_0$ and $\Gamma_a$ considered safely cover experimentally
estimated values for real molecular junctions
\cite{Zotti:10,Scheer:12,LukaGuth:16,Tao:13,Baldea:2018a,Baldea:2019d,Baldea:2019h,Frisbie:21a,Frisbie:21b,Gu:21,Chiechi:21}.

A nonvanishing temperature has an insignificant effect on the resonant current.
Results for this case ($e V = 2 \varepsilon_0$) depicted in \figname\ref{fig:shift}a reveal a slight current reduction
($I_{RT} < I_{0K}$) limited a very narrow range ($\varepsilon_0 = e V/2 \alt k_B T$).

A significant impact on the current occurs only slightly away from strict resonance but not strictly on resonance
($\varepsilon_0 = e V/2$). 
In general, the impact is qualitatively different, depending on whether the energy level lies outside or within
the Fermi window ($\left\vert \varepsilon_0\right\vert >  \vert e V\vert / 2$ or
$\left\vert \varepsilon_0\right\vert <  \vert e V\vert / 2$, respectively).
Indeed, as visualized in \figname\ref{fig:iv}, except for very small values of $\varepsilon_0$ 
(cf.~\figname\ref{fig:reversal-very-small-e0}),
thermal effects enhance the current ($I_{RT} > I_{0K}$) in the former case 
while diminishing it ($I_{RT} < I_{0K}$) in the latter case.
Because $\left\vert \varepsilon_0\right\vert =  \vert e V\vert / 2$ corresponds to resonant tunneling
(MO energy equal to the electrochemical potential of one electrode), biases for which
$\left\vert \varepsilon_0\right\vert >  \vert e V\vert / 2$ will be referred to as ``below resonance'',
while those for which $\left\vert \varepsilon_0\right\vert <  \vert e V\vert / 2$ will be referred to as ``above resonance''.

Before proceeding with specific results, we make two remarks to clarify why the various
figures presented below contain white (empty) regions.

First, the various figures that follow comprise one panel depicting positive values of the
relative deviations (always in percent) $ I_{RT}/I_{0K} - 1 (>0)$ for situations below resonance
($\left\vert \varepsilon_0\right\vert >  \vert e V\vert / 2$),
and another panel depicting values of $ I_{0K}/I_{RT} - 1 (>0)$ for situations above resonance
($\left\vert \varepsilon_0\right\vert <  \vert e V\vert / 2$). However, in order to convey an overall picture, 
both of these panels depict the full ranges of $V$ or $\varepsilon_0$ (i.e., both below resonance and above resonance).
For this reason, e.g., the left (right) part of the panels depicting
``above resonance'' (``below resonance'') situations at fixed $\varepsilon_0$
like those in \figname\ref{fig:errors-e0-1.0}a (\figname\ref{fig:errors-e0-1.0}b)
is empty. (In the empty area of \figname\ref{fig:errors-e0-1.0}a  $ I_{RT}/I_{0K} - 1$ is negative; likewise,
in the empty area of \figname\ref{fig:errors-e0-1.0}b $ I_{0K}/I_{RT} - 1 $ is also negative.)
We think that this presentation is more expressive than that (mathematically equivalent)
of \figsname\ref{fig:presentation}a and \ref{fig:presentation}c (or \figsname\ref{fig:presentation}b and \ref{fig:presentation}d)
of the {\esi}, where situations where $I_{RT} > I_{0K}$ and $I_{RT} < I_{0K}$
(or $I_{RT} < I_{0K}$ and $I_{RT} > I_{0K}$, respectively) are depicted in the same panel.

Second, emphasis in this paper is on specifying situations wherein the zero temperature limit applies.
Therefore, in the various figures shown below
we will only depict parameter regions corresponding to ``thermal corrections'' (i.e., ``reasonably weak'' thermal effects),
namely those wherein 
$\left\vert I_{0K}\right\vert /2 < \left\vert I_{RT} \right\vert < 2 \left\vert I_{0K}\right\vert $.
For this reason, situations wherein the relative deviations
$ I_{0K}/I_{RT} - 1 $ or $ I_{RT}/I_{0K} - 1 $ exceed 100\% appear as white (empty) regions
in the various diagrams presented.
\subsection{Thermal effects at fixed MO energy offset}
\label{sec:fixed-e0}
Because the model with $\gamma \equiv 0$ possesses charge conjugation symmetry
and all physical observables are invariant under an $\varepsilon_0 \to -\varepsilon_0$
transformation, in the presentation that follows we can and will restrict ourselves to positive
values of $\varepsilon_0$ and $ V $.
Whenever confusion can be excluded, we will write
$\varepsilon_0 $ and $V$ instead of $\left\vert \varepsilon_0 \right\vert $ and $ \vert V\vert $.

\begin{figure}[htb]
  \centerline{\includegraphics[width=0.22\textwidth,height=0.42\textwidth,angle=-90]{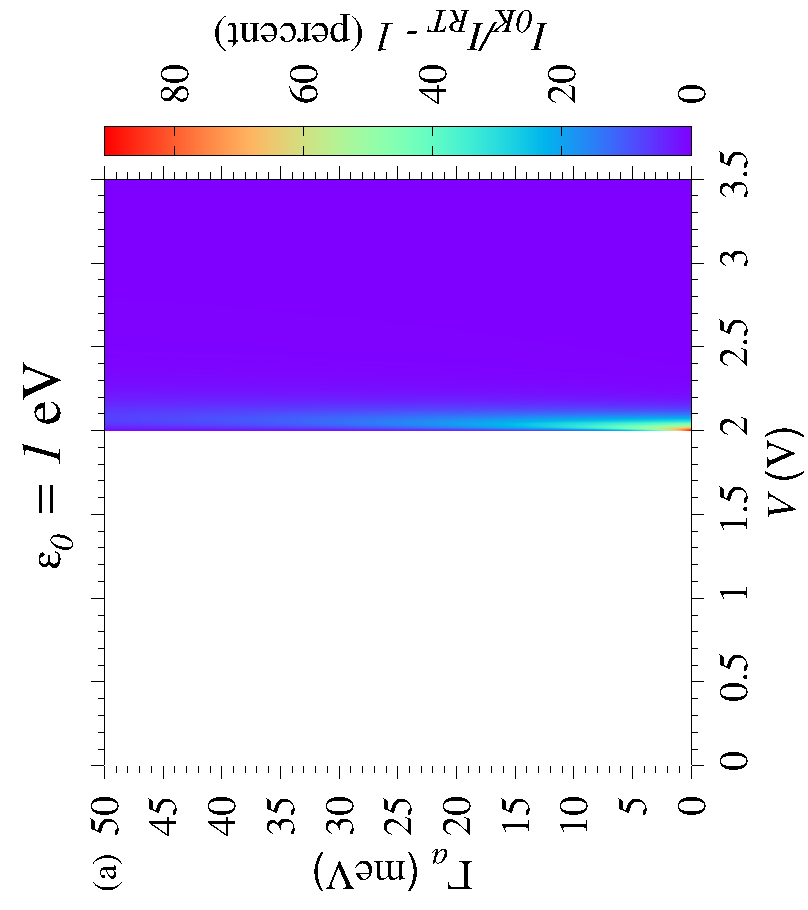}}
  \centerline{\includegraphics[width=0.22\textwidth,height=0.42\textwidth,angle=-90]{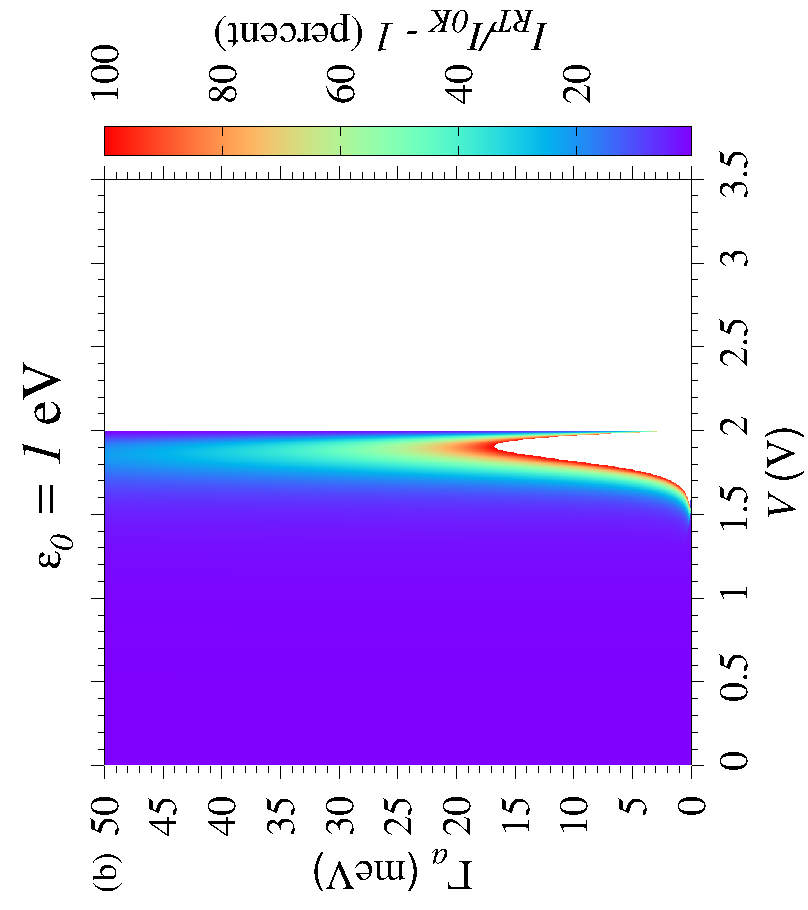}}
  \centerline{\includegraphics[width=0.22\textwidth,height=0.42\textwidth,angle=-90]{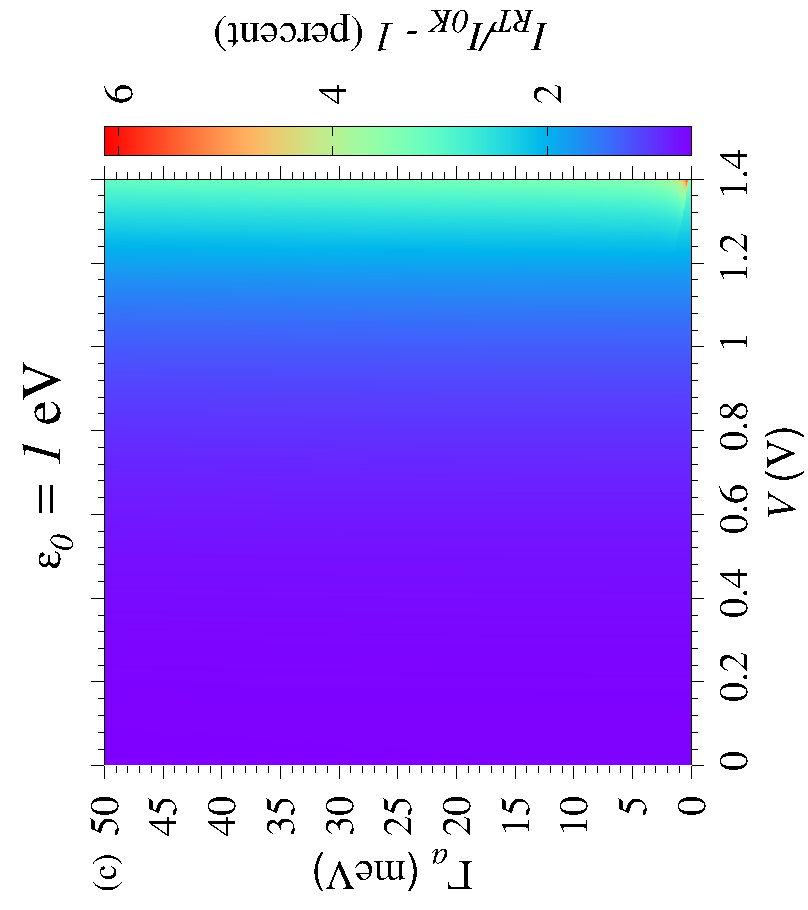}}
  \centerline{\includegraphics[width=0.22\textwidth,height=0.42\textwidth,angle=-90]{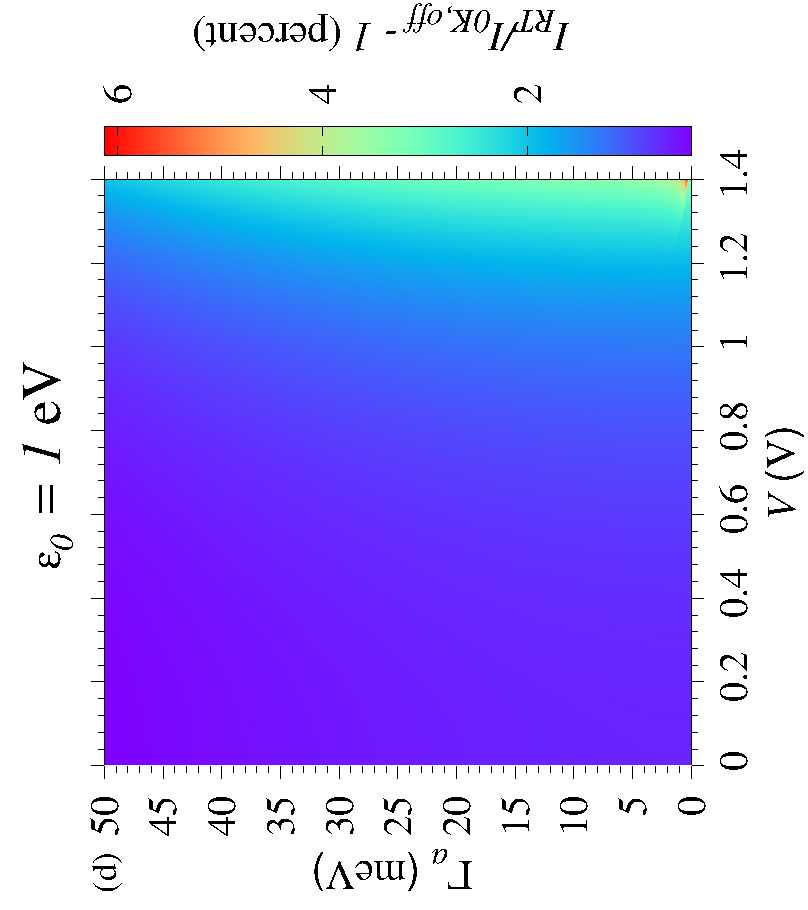}}
  \caption{The colored regions in the plane ($V, \Gamma_a$) depict situations where,
    at the fixed value of the MO energy offset indicated ($\varepsilon_0 = 1$\,eV), 
    the current $I_{0K}$ computed at $T=0$ using eqn~(\ref{eq-I0K})
    is larger ($\vert eV\vert > 2 \left\vert\varepsilon_0 \right\vert$, panel a) or smaller ($\vert eV\vert < 2 \left\vert\varepsilon_0 \right\vert$, panel b)
    than the exact current $I_{RT}$ computed from eqn~(\ref{eq-Iexact}) at room temperature ($T = 298.15$\,K).
    For parameter values compatible with eqn~(\ref{eq-1.4}), the current $I_{0K,off}$ computed using eqn~(\ref{eq-I0Koff}) is very accurate (panel d);
    it is as accurate as $I_{0K}$ (panel c). Relative deviations (shown only when not exceeding 100\%) are indicated in the color box.
    To facilitate comparison between $I_{0K,off}$ and $I_{0K}$, abscissas in panel c depicting $I_{0K}$ are restricted to those in panel d.
    Notice that the $z$-range in panels c and d is different from panel b.}
  \label{fig:errors-e0-1.0}
\end{figure}

\begin{figure}[htb]
  \centerline{\includegraphics[width=0.22\textwidth,height=0.42\textwidth,angle=-90]{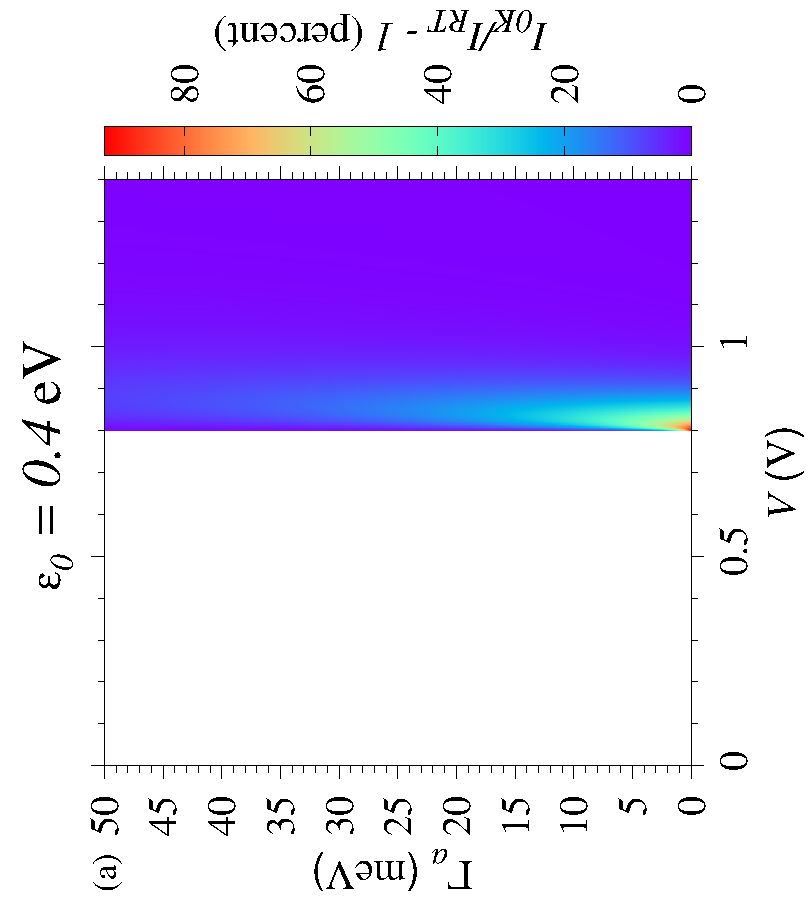}}
  \centerline{\includegraphics[width=0.22\textwidth,height=0.42\textwidth,angle=-90]{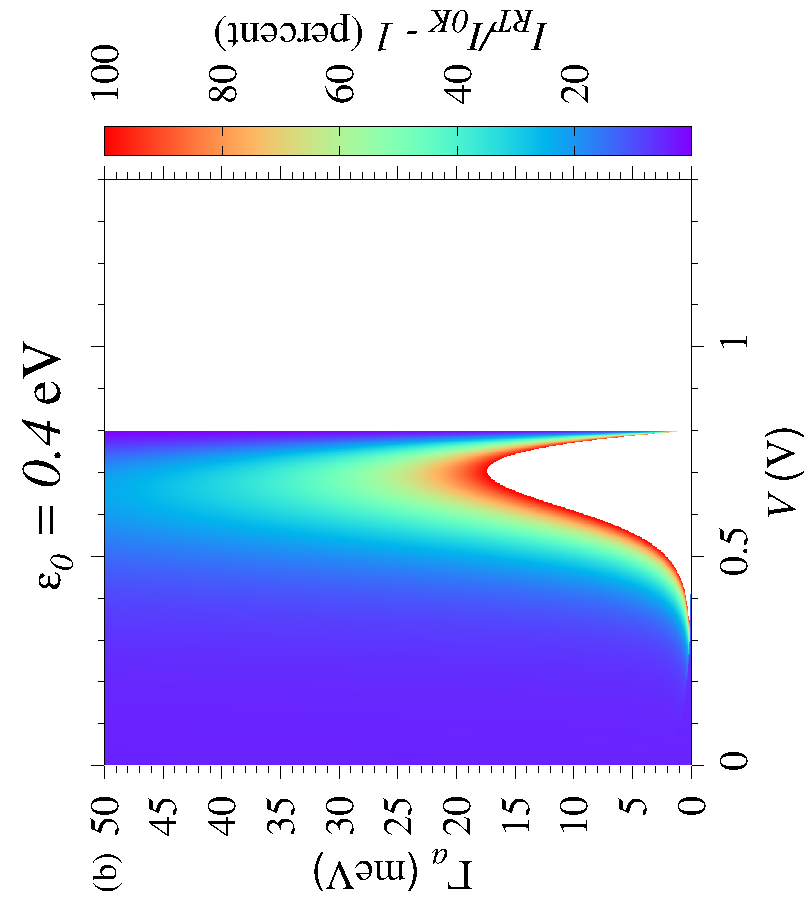}}
  \centerline{\includegraphics[width=0.22\textwidth,height=0.42\textwidth,angle=-90]{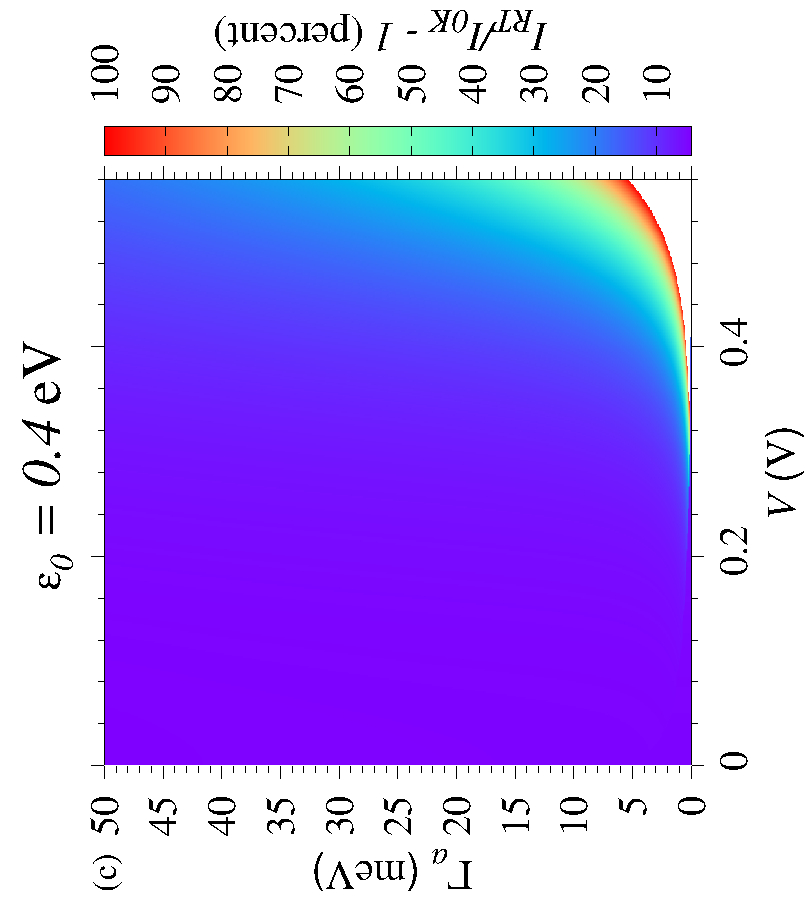}}
  \centerline{\includegraphics[width=0.22\textwidth,height=0.42\textwidth,angle=-90]{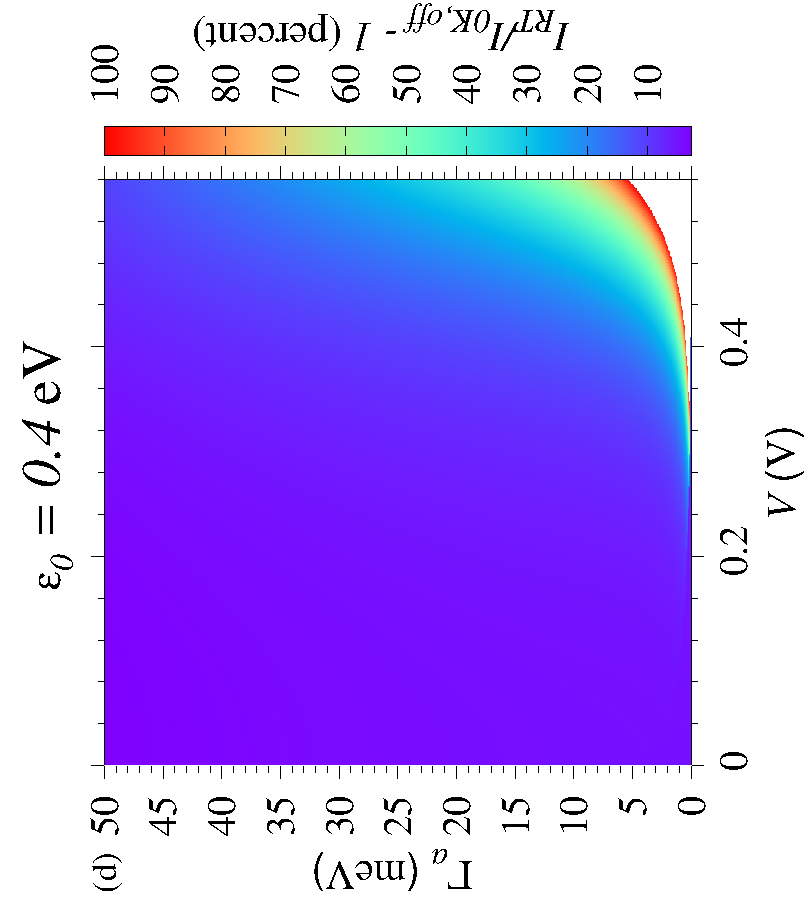}}
  \caption{The colored regions in the plane ($V, \Gamma_a$) depict situations where,
    at the fixed value of the MO energy offset indicated ($\varepsilon_0 = 0.4$\,eV), 
    the current $I_{0K}$ computed at $T=0$ using eqn~(\ref{eq-I0K})
    is larger ($\vert eV\vert > 2 \left\vert\varepsilon_0 \right\vert$, panel a) or smaller ($\vert eV\vert < 2 \left\vert\varepsilon_0 \right\vert$, panel b)
    than the exact current $I_{RT}$ computed from eqn~(\ref{eq-Iexact}) at room temperature ($T = 298.15$\,K).
    For parameter values compatible with eqn~(\ref{eq-1.4}) and (\ref{eq-Gamma-vs-e0}),
    the current $I_{0K,off}$ computed using eqn~(\ref{eq-I0Koff}) (panel d)
    is as accurate as $I_{0K}$ (panel c). Relative deviations (shown only when not exceeding 100\%) are indicated in the color box.
    To facilitate comparison between $I_{0K,off}$ and $I_{0K}$, abscissas in panel c depicting $I_{0K}$ are restricted to those in panel d.
    Notice that the $z$-range in panels c and d is different from panel b.}
  \label{fig:errors-e0-0.4}
\end{figure}

\begin{figure*}[htb]
  \centerline{\includegraphics[width=0.22\textwidth,height=0.42\textwidth,angle=-90]{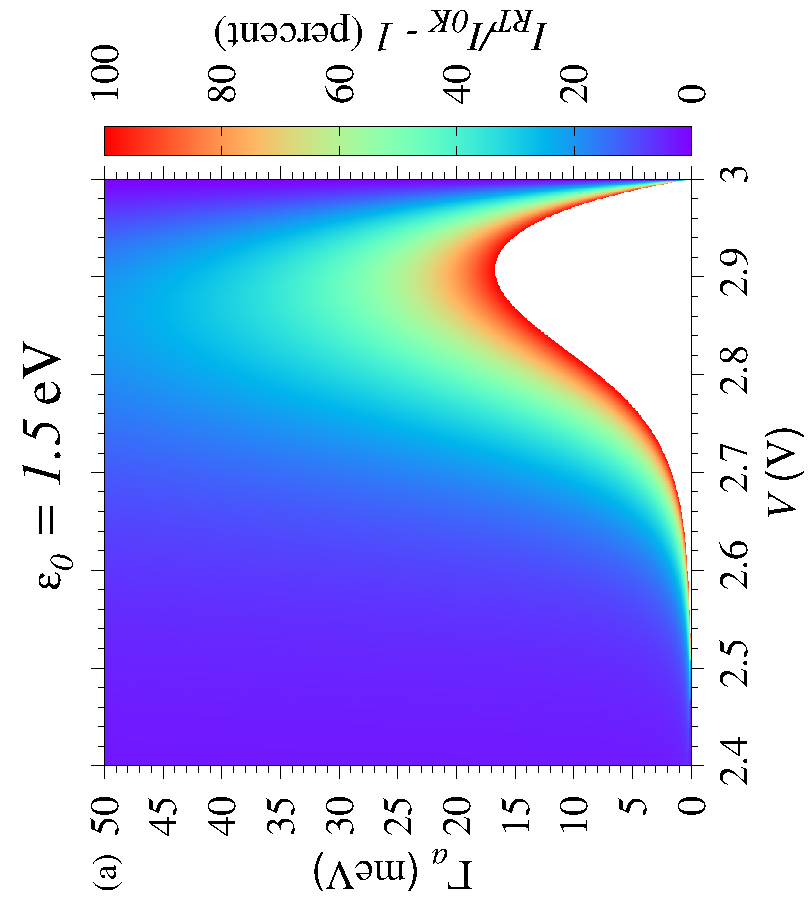}
    \includegraphics[width=0.22\textwidth,height=0.42\textwidth,angle=-90]{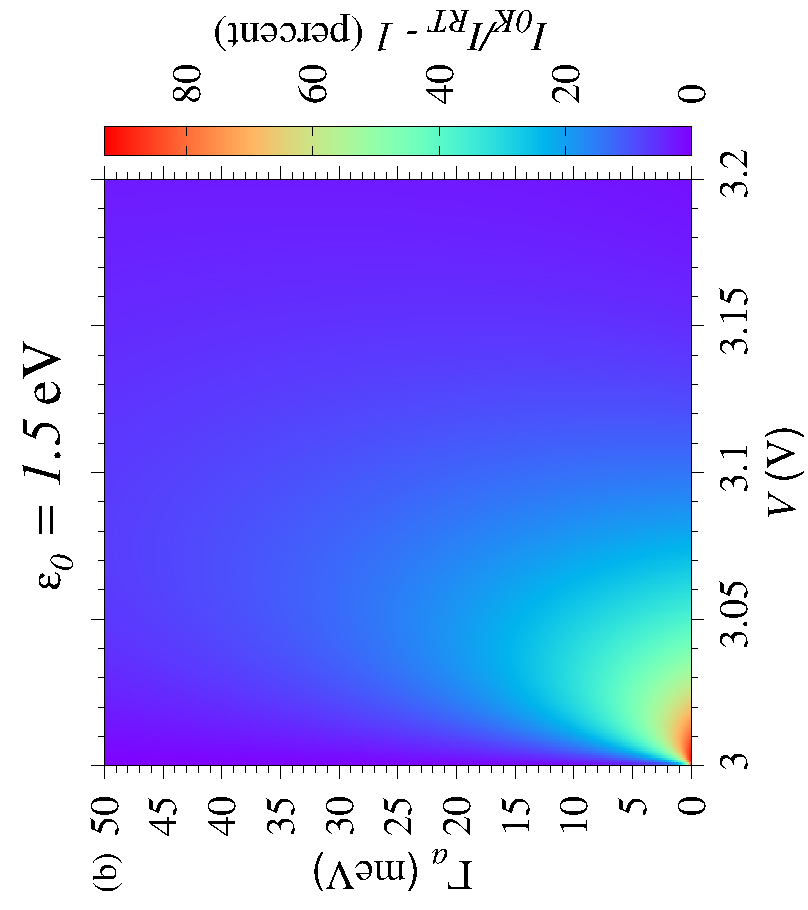}}
  \centerline{\includegraphics[width=0.22\textwidth,height=0.42\textwidth,angle=-90]{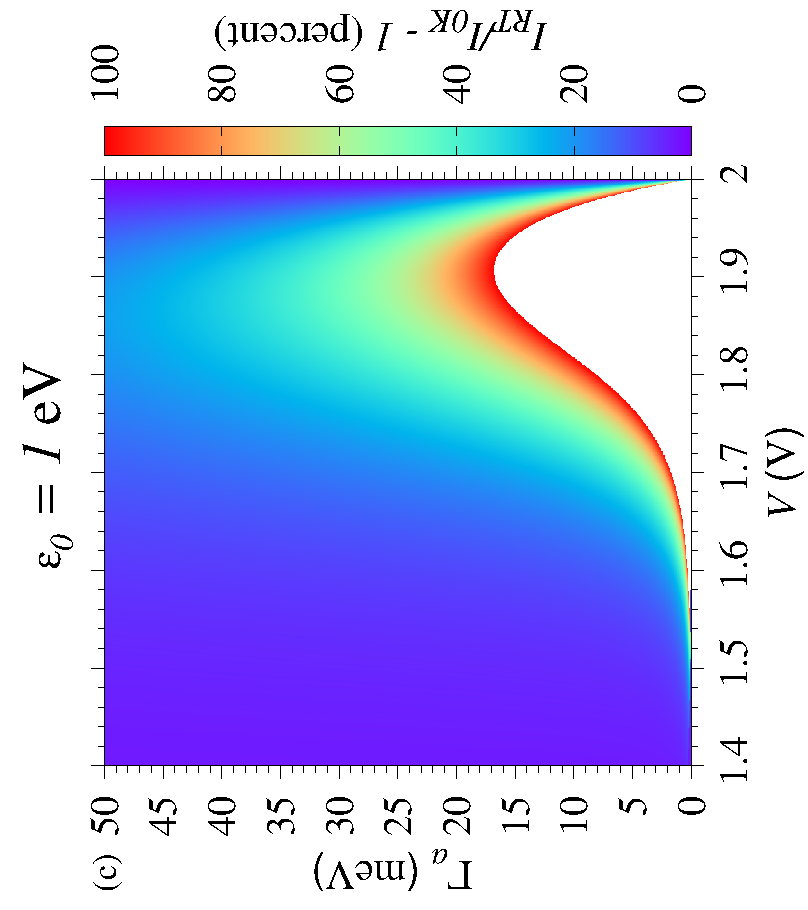}
    \includegraphics[width=0.22\textwidth,height=0.42\textwidth,angle=-90]{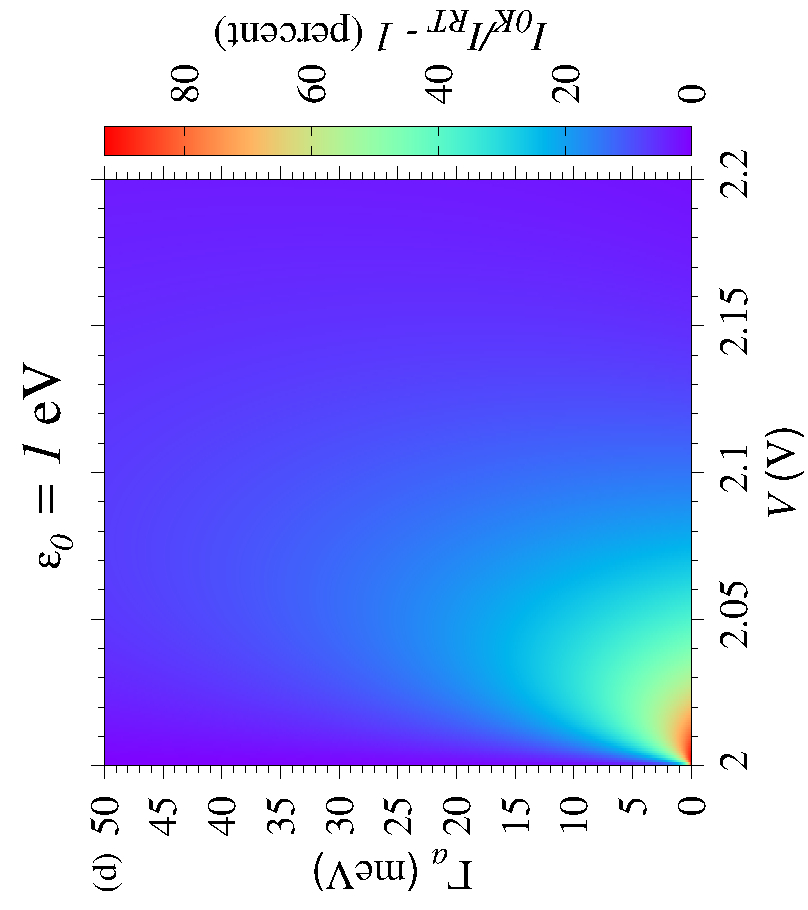}}  
  \centerline{\includegraphics[width=0.22\textwidth,height=0.42\textwidth,angle=-90]{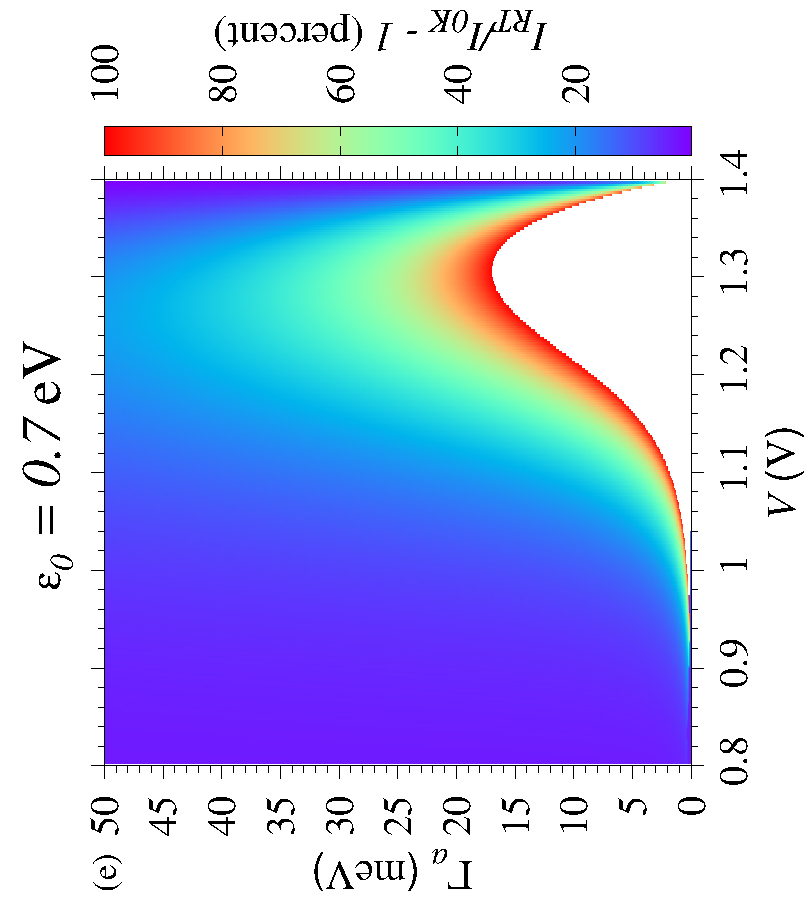}
    \includegraphics[width=0.22\textwidth,height=0.42\textwidth,angle=-90]{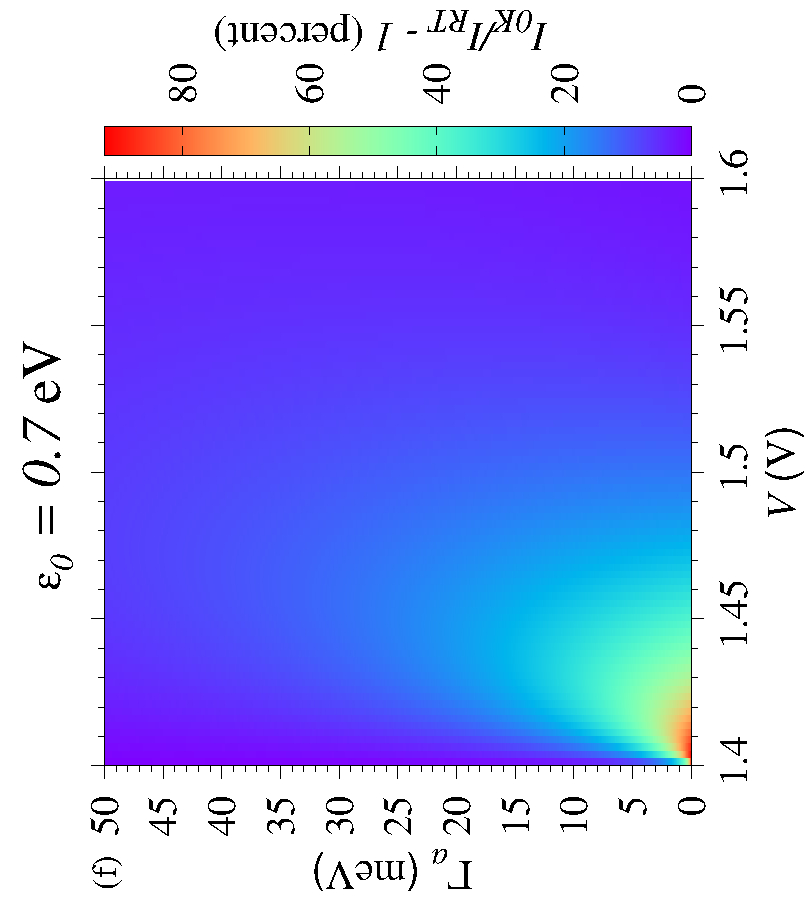}}
  \centerline{\includegraphics[width=0.22\textwidth,height=0.42\textwidth,angle=-90]{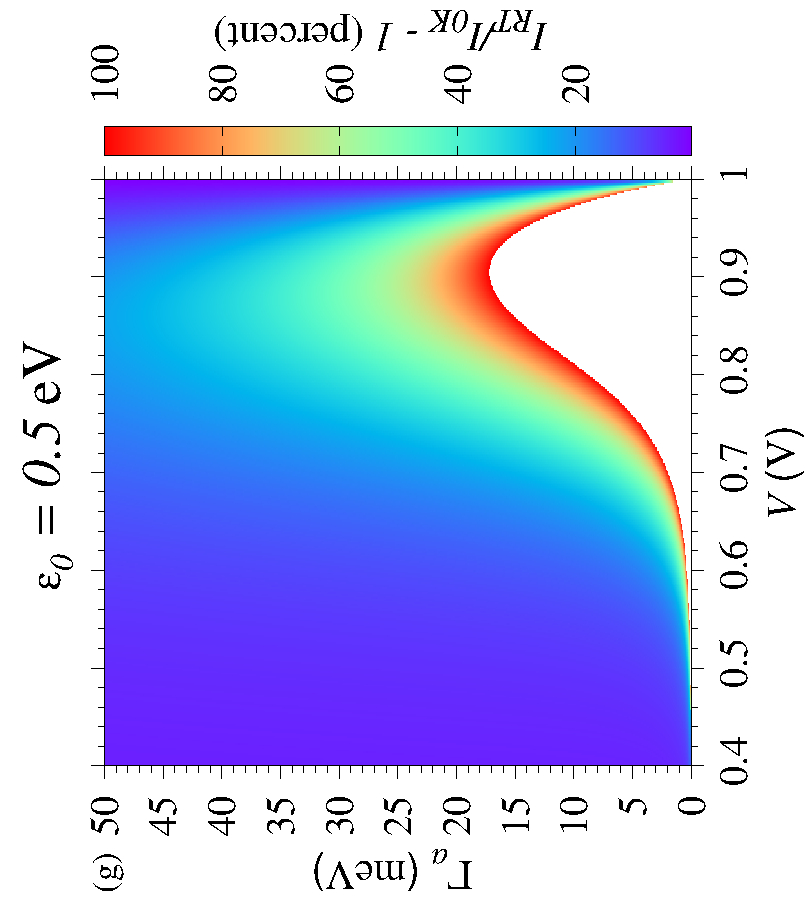}
    \includegraphics[width=0.22\textwidth,height=0.42\textwidth,angle=-90]{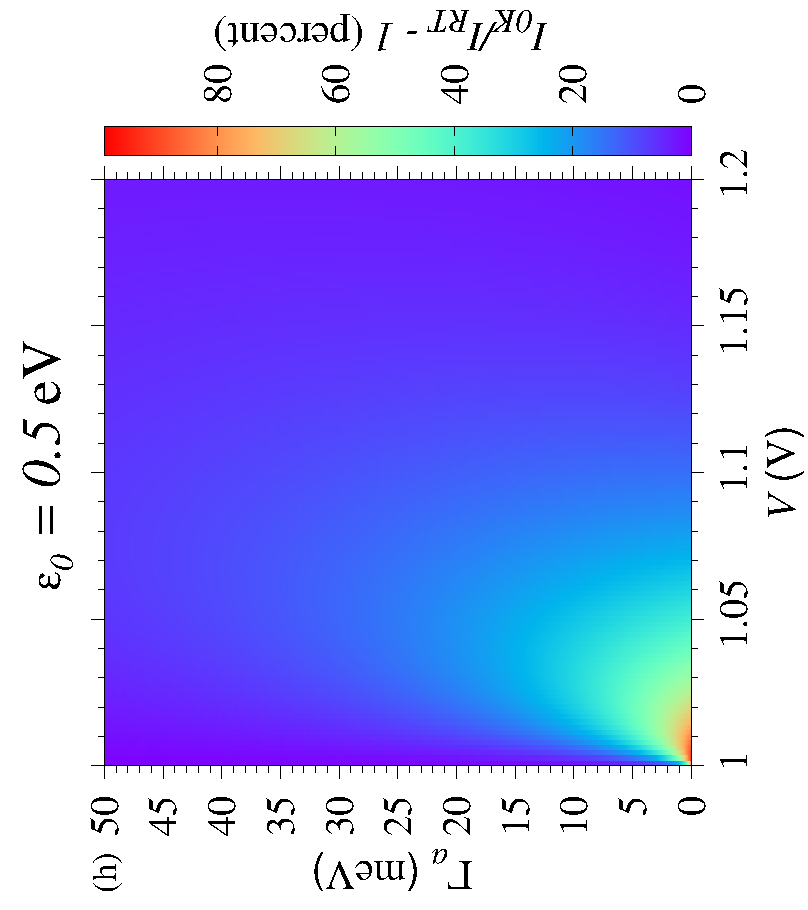}}
  \centerline{\includegraphics[width=0.22\textwidth,height=0.42\textwidth,angle=-90]{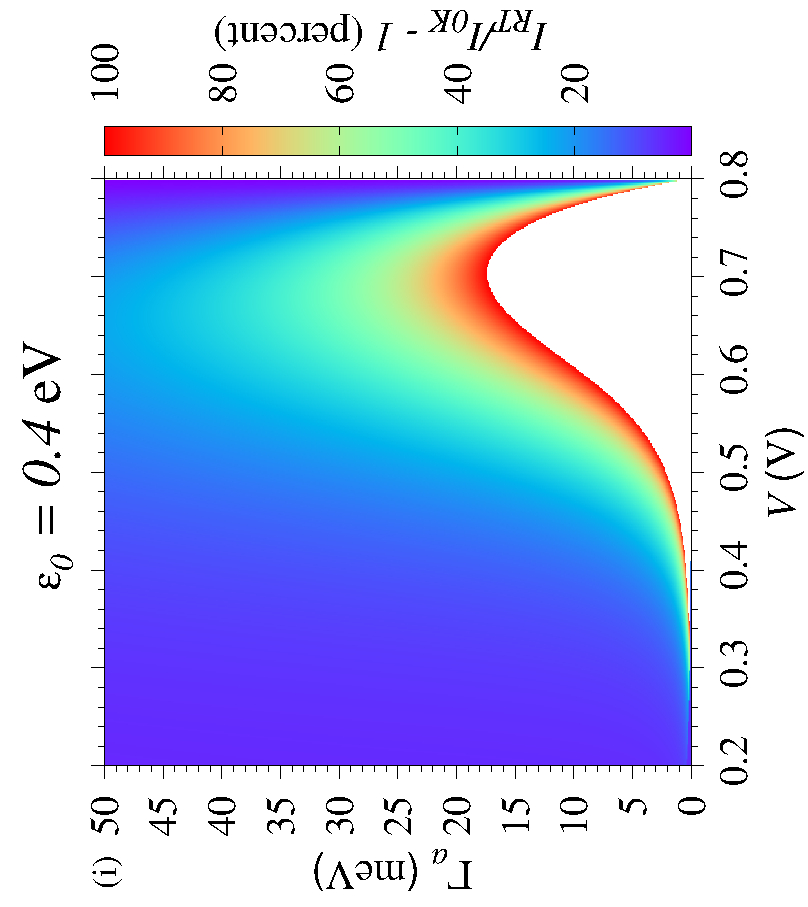}
    \includegraphics[width=0.22\textwidth,height=0.42\textwidth,angle=-90]{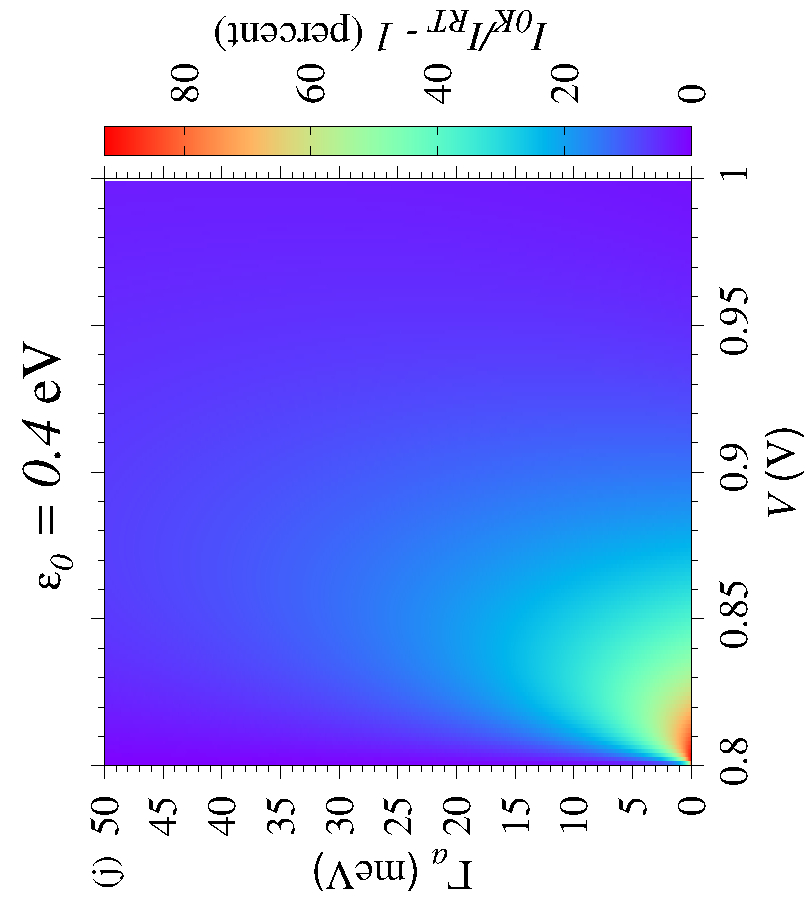}}
  \caption{For biases below resonance ($\vert e V\vert < 2 \left\vert\varepsilon_0 \right\vert$, left panels),
     thermal effects enhance the room temperature current $I_{RT}$
     (eqn~(\ref{eq-Iexact})) with respect to the zero-temperature current $I_{0K}$ (eqn~(\ref{eq-I0K})),
     while slightly reducing it ($I_{RT} < I_{0K}$) above resonance ($\vert e V\vert > 2 \left\vert\varepsilon_0 \right\vert$, right panels).
     For sufficiently large values of $\left\vert\varepsilon_0\right\vert$($\agt 0.4$\,eV), the location around resonance
     ($\vert e V\vert = 2 \left\vert\varepsilon_0 \right\vert$) of the
     regions of the ($V, \Gamma_a$)-plane affected is nearly independent of $\varepsilon_0$.
     Notice that all rightmost (leftmost) positions of the left (right) panels are aligned to resonance.
     In the white (empty) regions of the left panels the relative deviations exceed 100\%.}
  \label{fig:errors-e0-aligned}
\end{figure*}

\figsname\ref{fig:errors-e0-1.0} to \ref{fig:errors-e0-aligned} as well as
\figsname\ref{fig:errors-e0-0.7-si} to \ref{fig:errors-e0-0.1-si} of the {\esi} depict
thermal effects for variable bias $V$ at
fixed values of the MO energy offset $\varepsilon_0$.

In accord with the general considerations delineated in \secname\ref{sec:model}, 
these figures show that the current $I_{0K}$ computed at $T = 0$ deviates from the current $I_{RT}$ at room temperature for biases
around the resonance value $e V = 2 \varepsilon_0 $. 
Lowering of the MO offset from the value $\varepsilon_0 = 1$\,eV
(\figname\ref{fig:errors-e0-1.0}a)
to $\varepsilon_0 = 0.7; 0.5; 0.4; 0.3; 0.2; 0.1$\,eV
(\figsname\ref{fig:errors-e0-0.7-si}a, \ref{fig:errors-e0-0.5-si}a, \ref{fig:errors-e0-0.4}a,
\ref{fig:errors-e0-0.3-si}a, \ref{fig:errors-e0-0.2-si}a, and \ref{fig:errors-e0-0.1-si}a, respectively)
shifts the predominantly red region (corresponding to deviations up to 100\%)
from $V \alt 2$\,V to bias ranges around the smaller values $V (= 2\varepsilon_0/e) = 1.4; 1; 0.8; 0.6; 0.4; 0.2$\,V.

To make more evident the fact that thermal effects are intimately related to the resonance condition
($e V \approx 2 \varepsilon_0 $), \figname\ref{fig:errors-e0-aligned} and
\figname\ref{fig:errors-e0-aligned-si} of the {\esi} depict
relative deviations of $I_{0K}$ from $I_{RT}$ computed for various MO offsets
aligned to the same abscissa value (namely, $e V - 2 \varepsilon_0 $).
\figsname\ref{fig:errors-e0-aligned} and \figsname\ref{fig:shift}c and d
make it clear that thermal effects around resonance
($\vert e V\vert \approx 2 \left\vert \varepsilon_0\right\vert$) are insensitive to
$\varepsilon_0$ provided that the latter is reasonably large with respect to the thermic energy $k_B T = 25.7$\,meV.
Loosely speaking, this means $\varepsilon_0 \agt 0.4$\,eV. At smaller values of $\varepsilon_0$, slightly broader
parameter areas are affected, which extend towards larger values of $\Gamma_a$ (\figname\ref{fig:errors-e0-aligned-si}).

In all cases, temperature's impact above resonance ($V > 2 \varepsilon_0/e$)
is weaker than below resonance. Most significantly affected is the bias range below resonance
\begin{subequations}
  \label{eq-delta-V}
  \begin{equation}
    \max\left(2 \varepsilon_0 / e - \delta V\left(\Gamma_a\right), 0\right) \alt V < 2\varepsilon_0 / e,
    \ \delta V\left(\Gamma_a\right) < \left . \delta V\right\vert_{\Gamma_a \to 0}
  \end{equation}
  whose width $\delta V \equiv \delta V(\Gamma_a) $ is nearly independent of $\varepsilon_0$.
  At small $\Gamma_a$, it amounts to
  \begin{equation}
    \left . \delta V\right\vert_{\Gamma_a \to 0} \approx 0.5\,\mbox{V}
  \end{equation}
\end{subequations}
Moving upwards to larger values of $\Gamma_a$, 
the bias range $\delta V$ where thermal effects are significant becomes gradually narrower. 

Along with deviations with respect to the exact current $I_{RT}$ at room
temperature of the current $I_{0K}$ computed at $T=0$
(\figsname\ref{fig:errors-e0-1.0}a and \ref{fig:errors-e0-0.4}a as well as
\figsname\ref{fig:errors-e0-0.7-si}a to \ref{fig:errors-e0-0.1-si}a of the {\esi}),
in
\figsname\ref{fig:errors-e0-1.0}b and \ref{fig:errors-e0-0.4}b as well as in 
\figsname\ref{fig:errors-e0-0.7-si}a to \ref{fig:errors-e0-0.1-si}b of the {\esi}
we also show deviations from $I_{RT}$ of the current $I_{0K,off}$ computed via eqn~(\ref{eq-I0Koff}).
Inspection of these figures reveals that, for biases $V$ sufficiently below resonance and
sufficiently large MO energy offsets ($\varepsilon_0 \agt 0.4$\,eV;
noteworthy, the same numerical value as encountered above in the analysis based on eqn~(\ref{eq-I0K})),
eqn~(\ref{eq-I0Koff}) is very accurate irrespective of the value of $\Gamma_a$.

Always in the bias range compatible with eqn~(\ref{eq-1.4}),
deviations of $I_{0K,off}$ from $I_{RT}$ become progressively significant as $\varepsilon_0$ decreases below
$\varepsilon_0 \alt 0.4$\,eV (\figsname\ref{fig:errors-e0-0.4}b and \ref{fig:errors-e0-0.3-si}b to \ref{fig:errors-e0-0.1-si}b).
Still, even in such situations, the approximation $I_{0K,off} \approx I_{RT}$ is as good as the approximation
$I_{0K} \approx I_{RT}$ as long as $\Gamma_a$ is sufficiently small to comply with eqn~(\ref{eq-Gamma-vs-e0});
compare among themselves panels (b) and (c) in \figsname\ref{fig:errors-e0-0.4} and \ref{fig:errors-e0-0.1-si},
and in \figsname\ref{fig:errors-e0-0.3-si} to \ref{fig:errors-e0-0.1-si} of the {\esi}.
\subsection{Thermal effects at fixed bias voltage}
\label{sec:fixed-V}
\begin{figure}[htb]
  \centerline{\includegraphics[width=0.22\textwidth,height=0.42\textwidth,angle=-90]{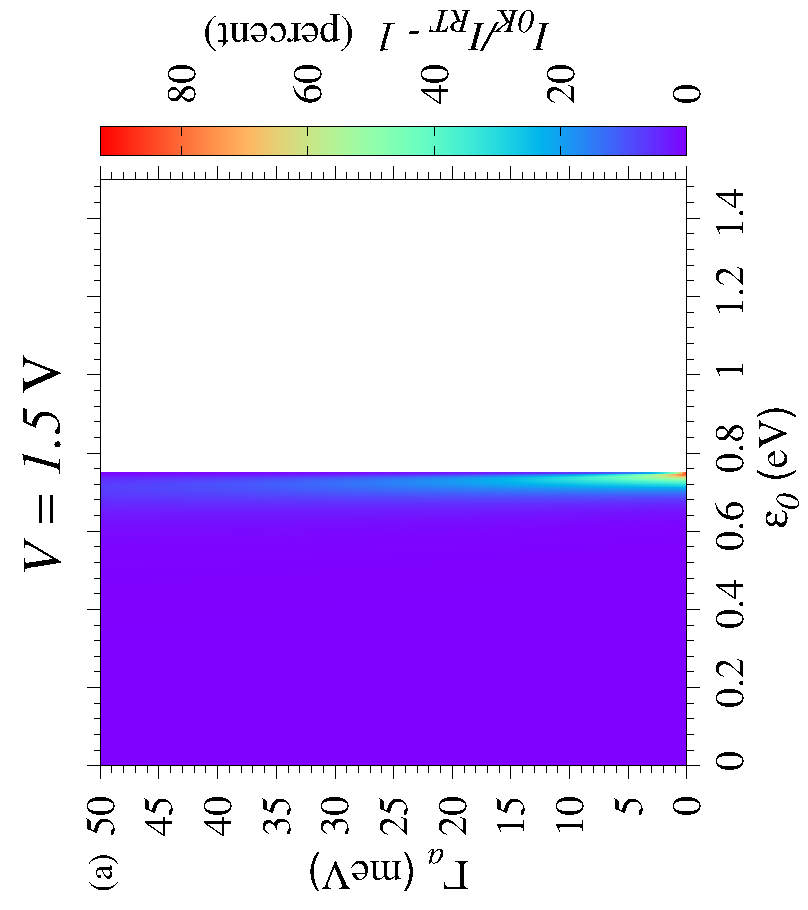}}
  \centerline{\includegraphics[width=0.22\textwidth,height=0.42\textwidth,angle=-90]{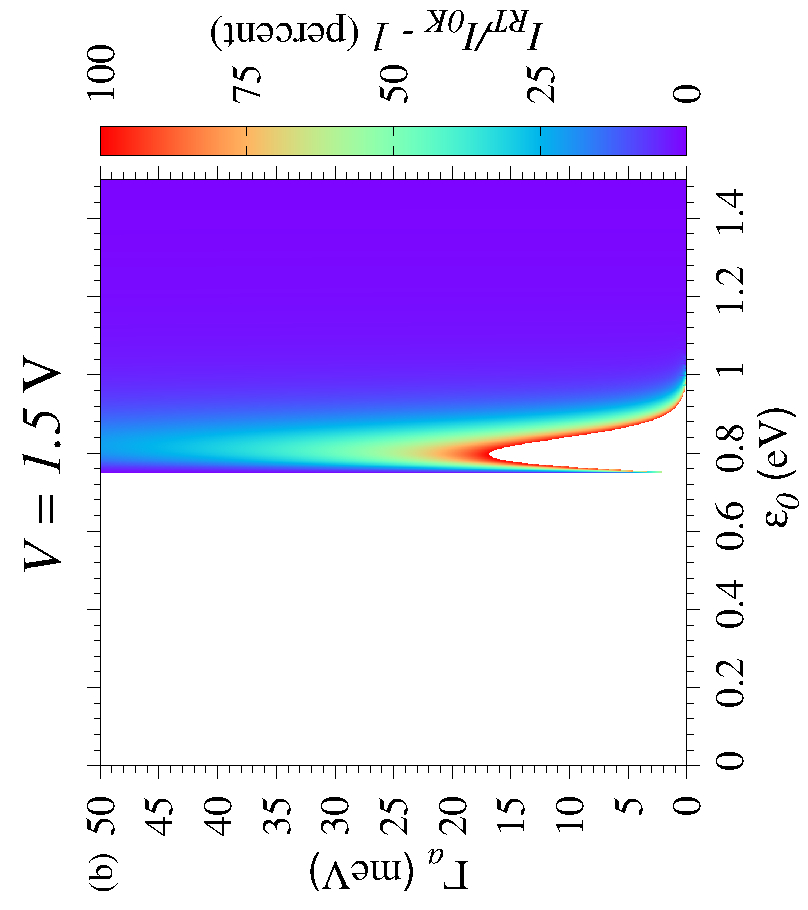}}
  \centerline{\includegraphics[width=0.22\textwidth,height=0.42\textwidth,angle=-90]{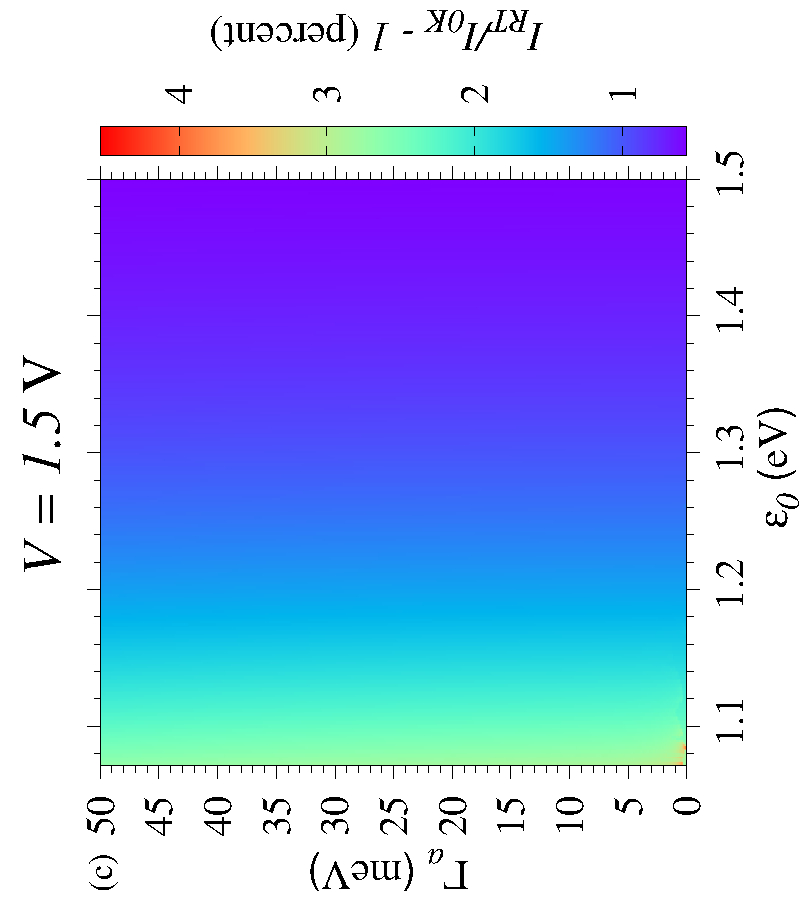}}
  \centerline{\includegraphics[width=0.22\textwidth,height=0.42\textwidth,angle=-90]{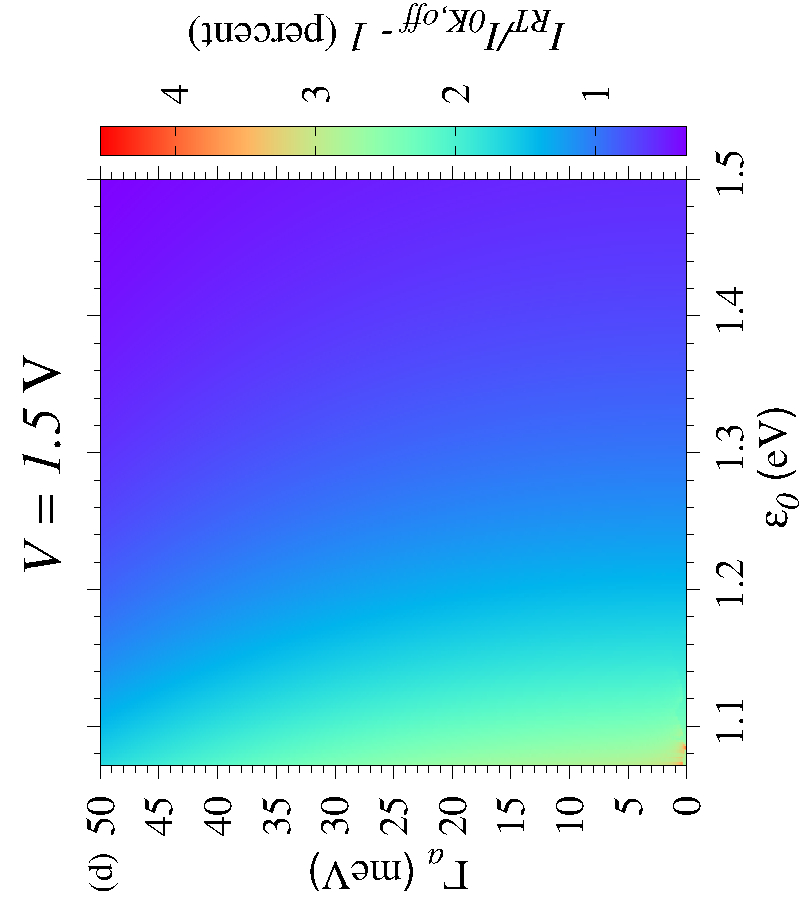}}
  \caption{The colored regions in the plane ($\varepsilon_0, \Gamma_a$) depict situations where,
    at the fixed bias indicated ($V = 1.5$\,V), 
    the current $I_{0K}$ computed at $T=0$ using eqn~(\ref{eq-I0K})
    is larger ($\vert eV\vert > 2 \left\vert\varepsilon_0 \right\vert$, panel a) or smaller ($\vert eV\vert < 2 \left\vert\varepsilon_0 \right\vert$, panel b)
    than the exact current $I_{RT}$ computed from eqn~(\ref{eq-Iexact}) at room temperature ($T = 298.15$\,K).
    For parameter values compatible with eqn~(\ref{eq-1.4}),
    the current $I_{0K,off}$ computed using eqn~(\ref{eq-I0Koff}) is very accurate (panel d);
    it is as accurate as $I_{0K}$ (panel c). Relative deviations (shown only when not exceeding 100\%) are indicated in the color box.
    To facilitate comparison between $I_{0K,off}$ and $I_{0K}$, abscissas in panel c depicting $I_{0K}$ are restricted to those in panel d.
    Notice that the $z$-range in panels c and d is different from panel b.}
  \label{fig:errors-1.5V}
\end{figure}

\begin{figure}[htb]
  \centerline{\includegraphics[width=0.22\textwidth,height=0.42\textwidth,angle=-90]{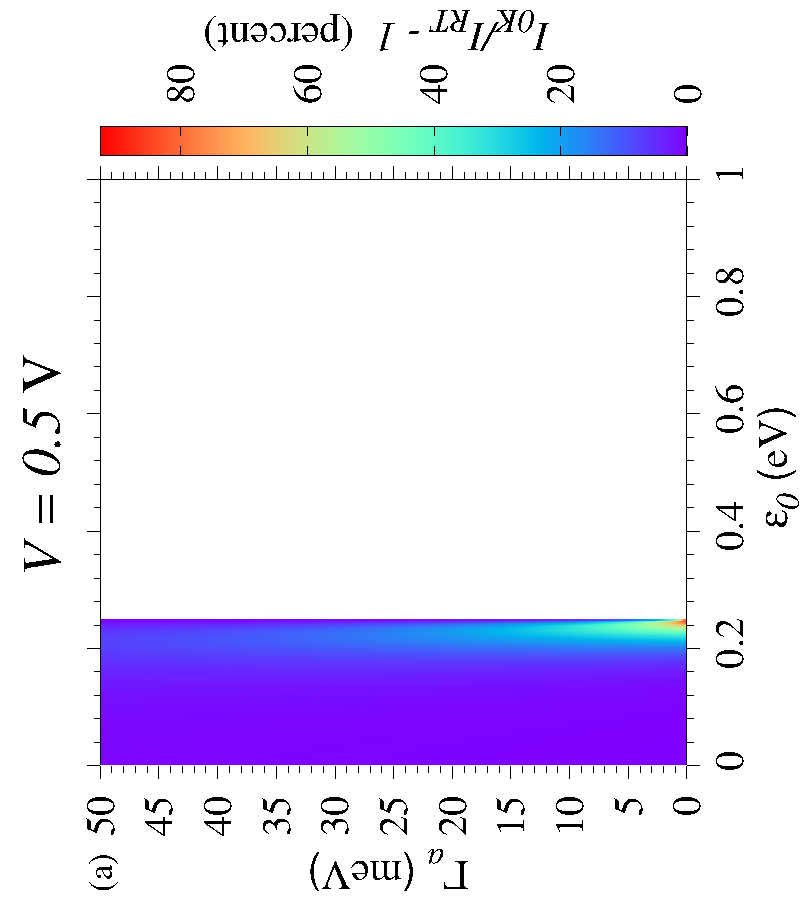}}
  \centerline{\includegraphics[width=0.22\textwidth,height=0.42\textwidth,angle=-90]{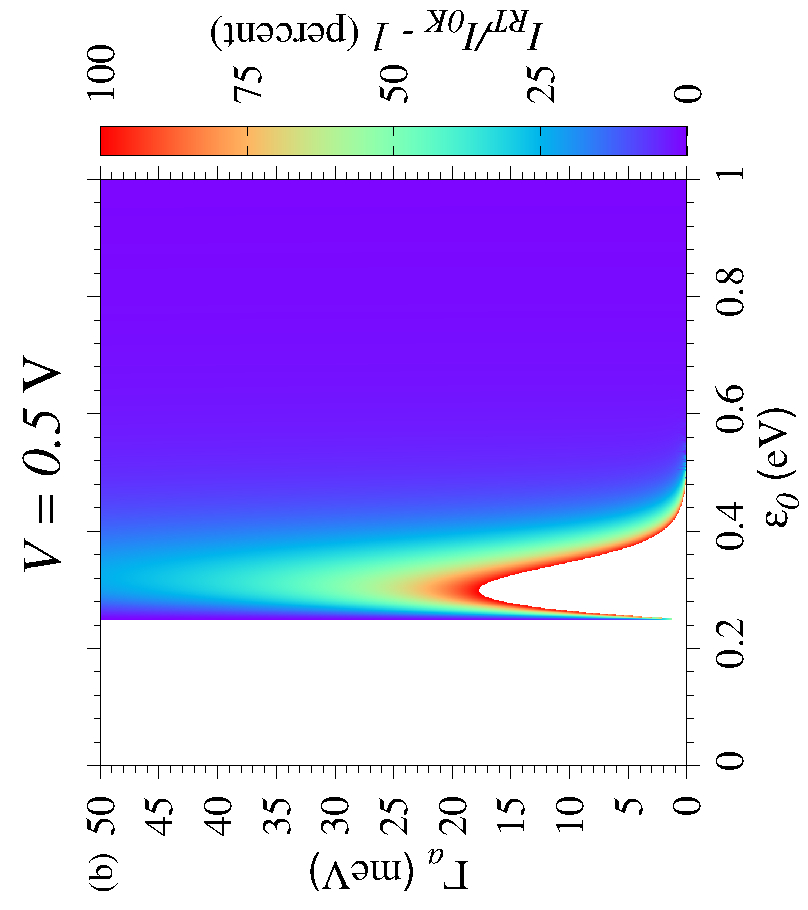}}
  \centerline{\includegraphics[width=0.22\textwidth,height=0.42\textwidth,angle=-90]{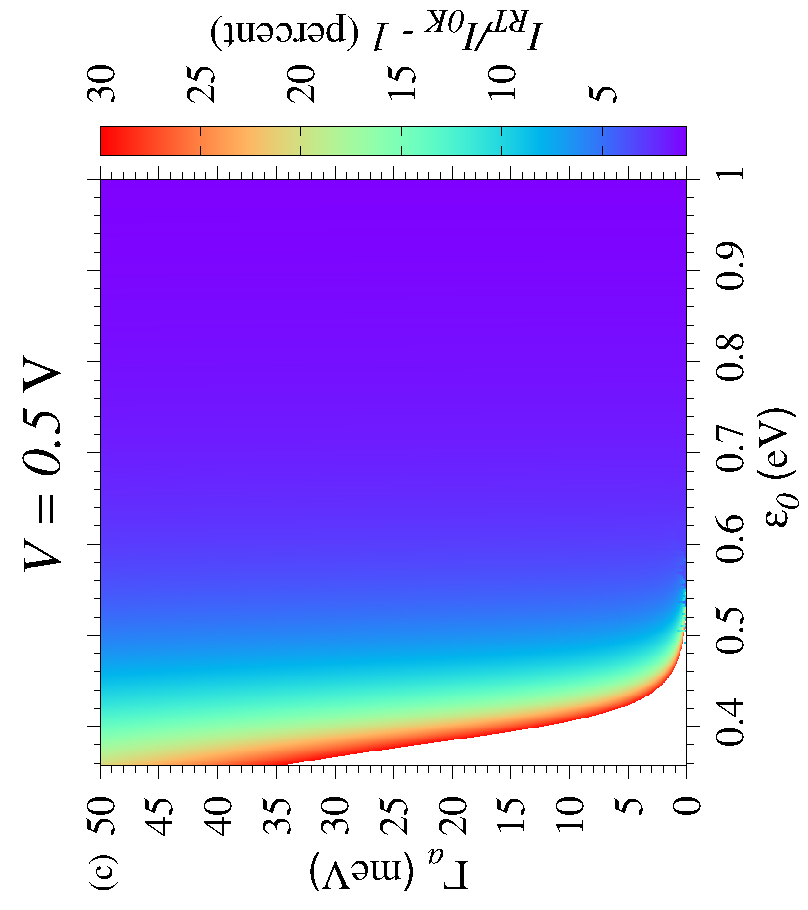}}
  \centerline{\includegraphics[width=0.22\textwidth,height=0.42\textwidth,angle=-90]{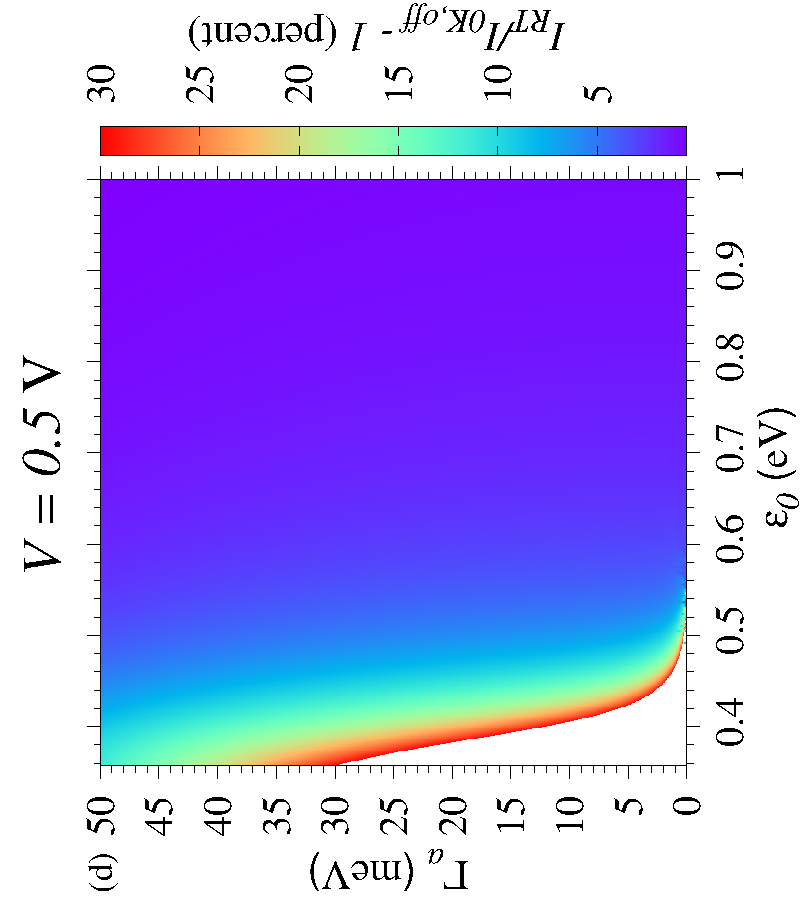}}
  \caption{The colored regions in the plane ($\varepsilon_0, \Gamma_a$) depict situations where,
    at the fixed bias indicated ($V = 0.5$\,V), 
    the current $I_{0K}$ computed at $T=0$ using eqn~(\ref{eq-I0K})
    is larger ($\vert eV\vert > 2 \left\vert\varepsilon_0 \right\vert$, panel a) or smaller ($\vert eV\vert < 2 \left\vert\varepsilon_0 \right\vert$, panel b)
    than the exact current $I_{RT}$ computed from eqn~(\ref{eq-Iexact}) at room temperature ($T = 298.15$\,K).
    For parameter values compatible with eqn~(\ref{eq-1.4}) and (\ref{eq-Gamma-vs-e0}),
    the current $I_{0K,off}$ computed using eqn~(\ref{eq-I0Koff}) (panel d)
    is as accurate as $I_{0K}$ (panel c). Relative deviations (shown only when not exceeding 100\%) are indicated in the color box.
    To facilitate comparison between $I_{0K,off}$ and $I_{0K}$, abscissas in panel c depicting $I_{0K}$ are restricted to those in panel d.
    Notice that the $z$-range in panels c and d is different from panel b.}
  \label{fig:errors-0.5V}
\end{figure}

\begin{figure*}[htb]
  \centerline{\includegraphics[width=0.22\textwidth,height=0.42\textwidth,angle=-90]{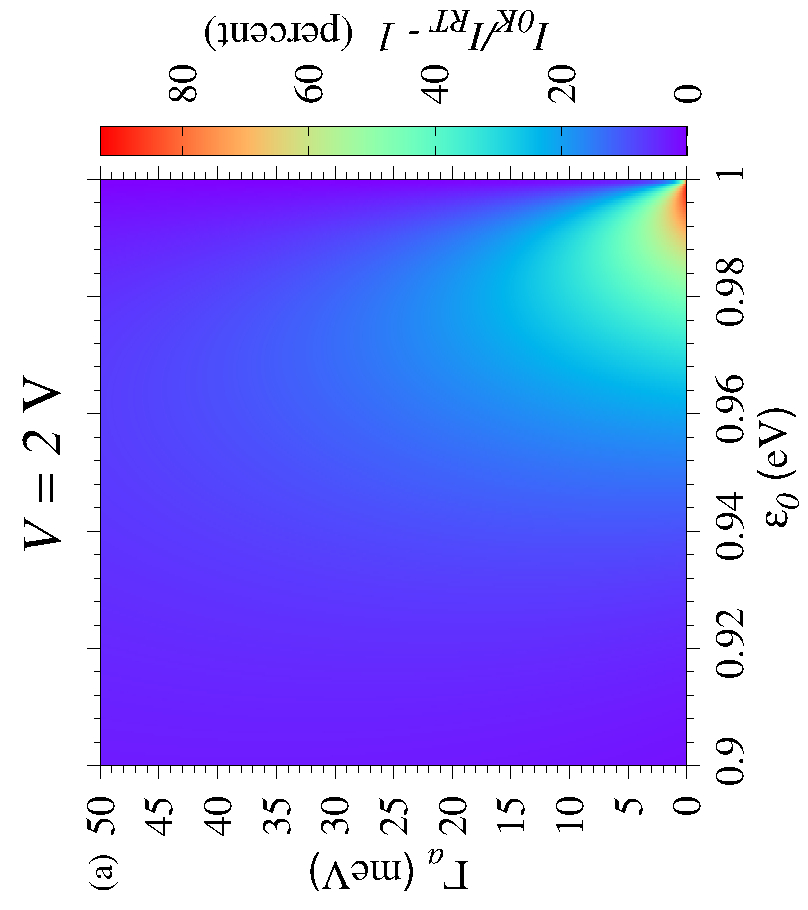}
    \includegraphics[width=0.22\textwidth,height=0.42\textwidth,angle=-90]{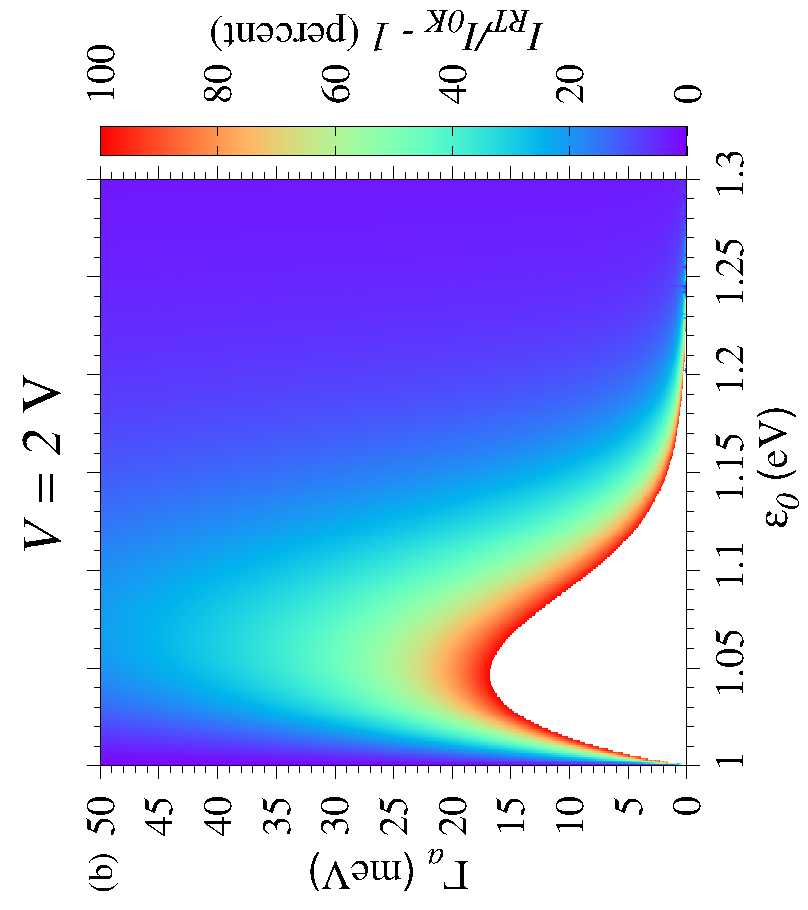}}
  \centerline{\includegraphics[width=0.22\textwidth,height=0.42\textwidth,angle=-90]{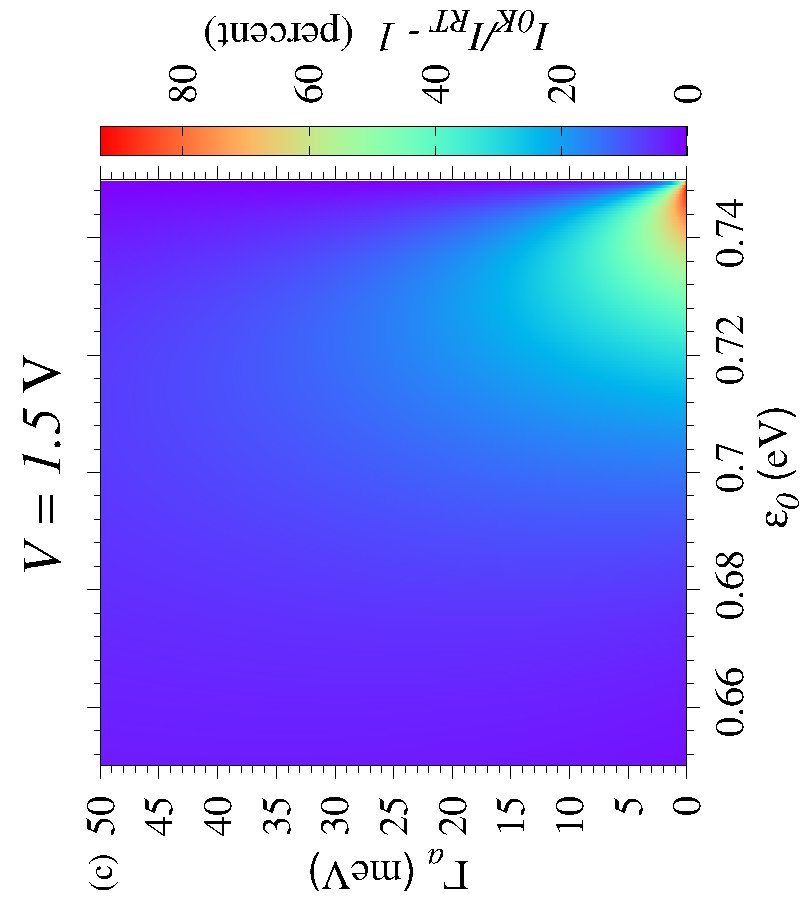}
    \includegraphics[width=0.22\textwidth,height=0.42\textwidth,angle=-90]{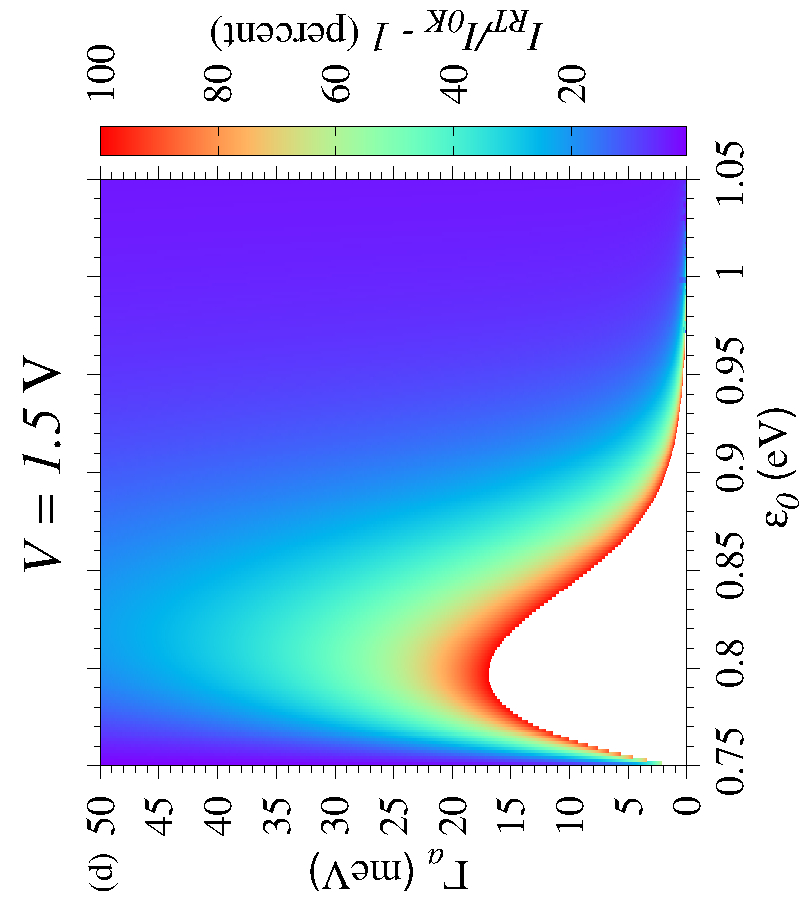}}
  \centerline{\includegraphics[width=0.22\textwidth,height=0.42\textwidth,angle=-90]{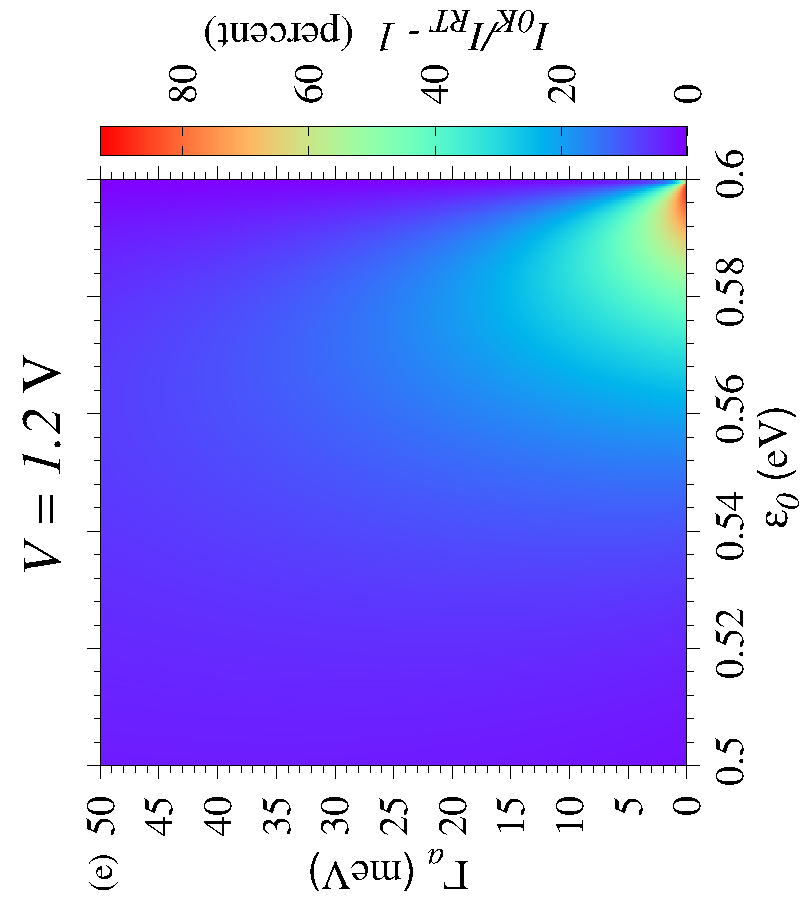}
    \includegraphics[width=0.22\textwidth,height=0.42\textwidth,angle=-90]{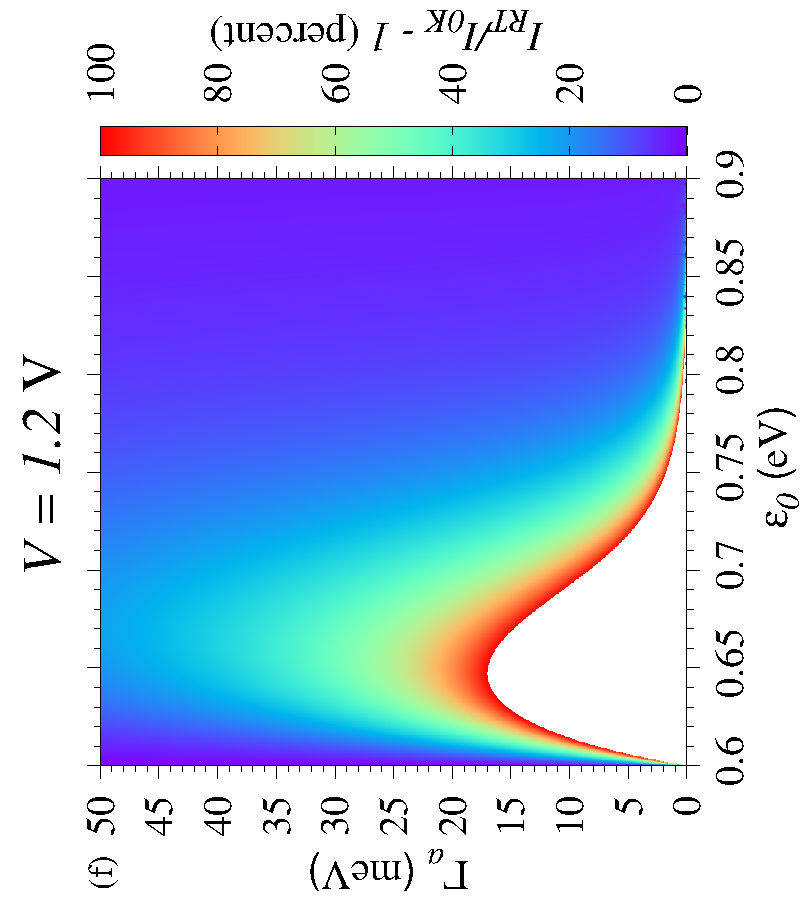}}
  \centerline{\includegraphics[width=0.22\textwidth,height=0.42\textwidth,angle=-90]{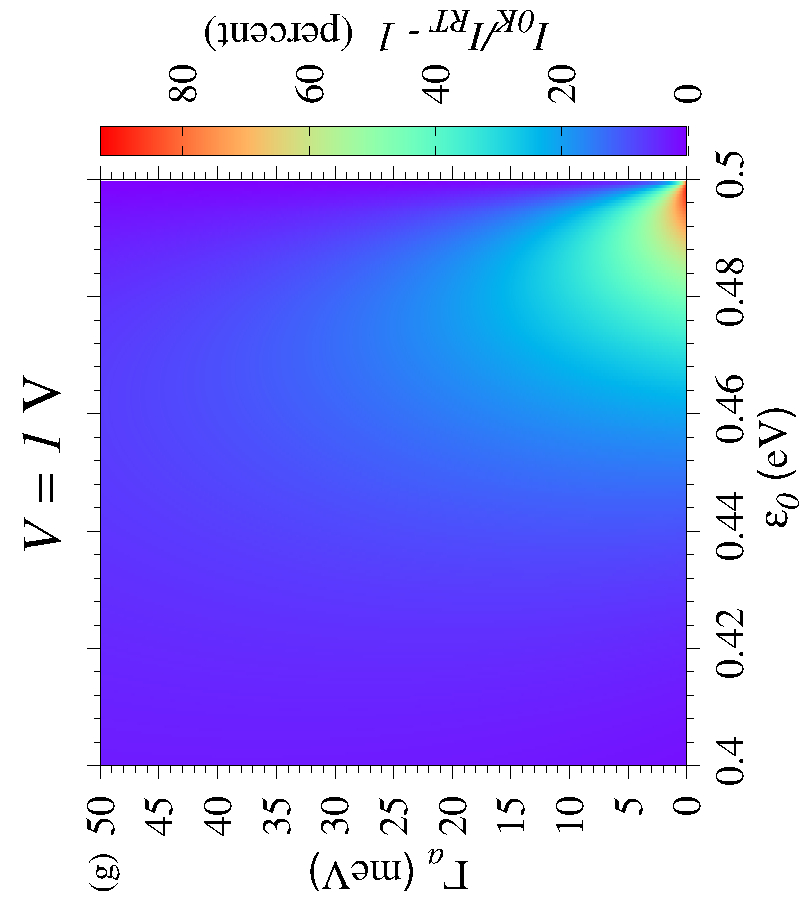}
    \includegraphics[width=0.22\textwidth,height=0.42\textwidth,angle=-90]{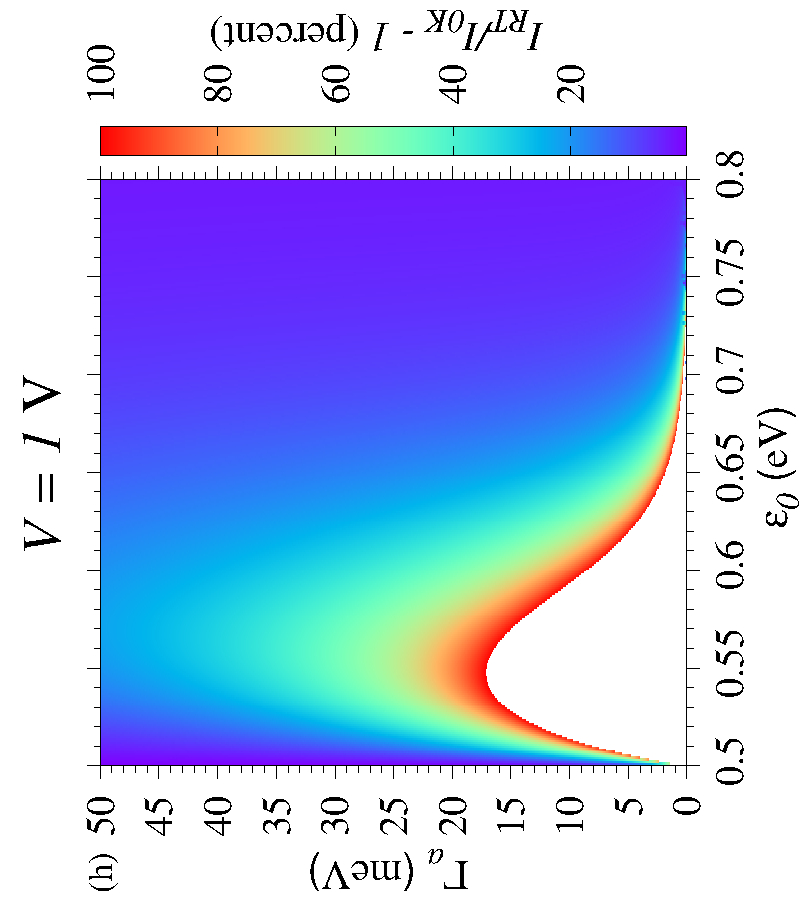}}
  \centerline{\includegraphics[width=0.22\textwidth,height=0.42\textwidth,angle=-90]{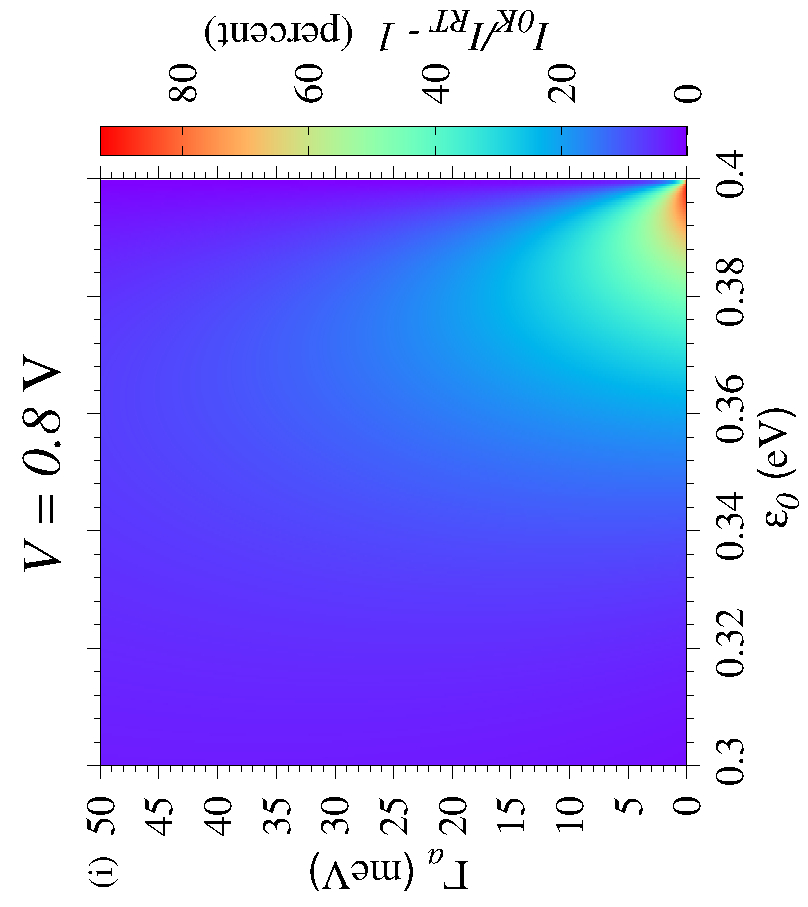}
    \includegraphics[width=0.22\textwidth,height=0.42\textwidth,angle=-90]{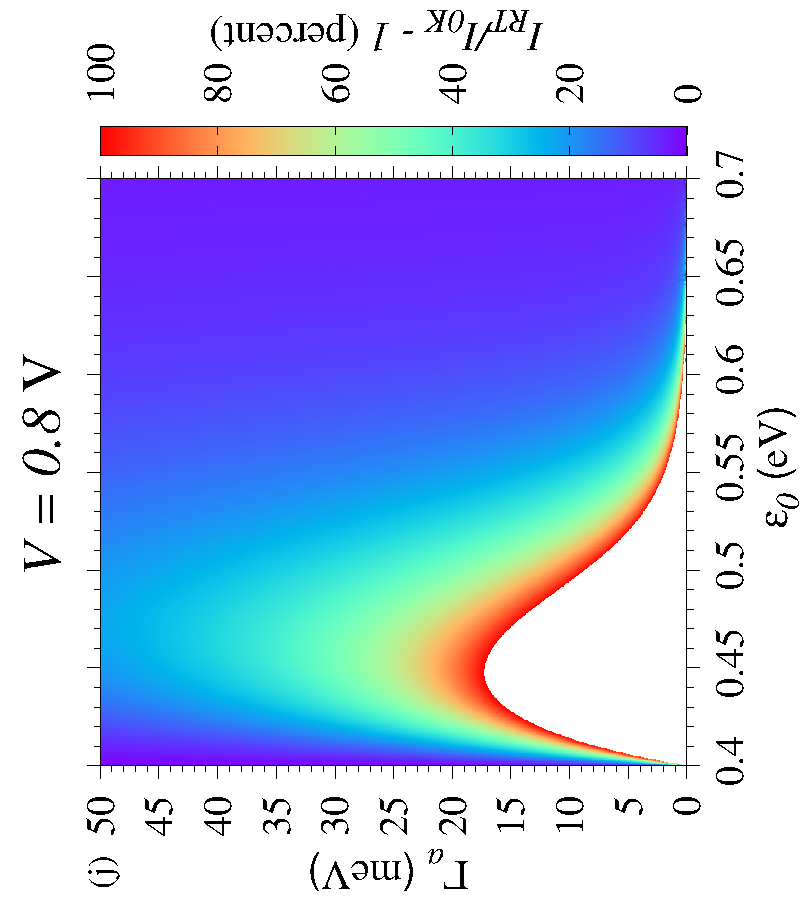}}
  \caption{Below resonance ($\vert e V\vert < 2 \left\vert\varepsilon_0 \right\vert$, right panels),
     thermal effects enhance the room temperature current $I_{RT}$
     (eqn~(\ref{eq-Iexact})) with respect to the zero-temperature current $I_{0K}$ (eqn~(\ref{eq-I0K})),
     while slightly reducing it ($I_{RT} < I_{0K}$) above resonance ($\vert e V\vert > 2 \left\vert\varepsilon_0 \right\vert$, left panels).
     For sufficiently large values of $\vert V \vert$($\agt 0.8$\,V), the location around resonance
     ($\vert e V\vert = 2 \left\vert\varepsilon_0 \right\vert$) of the
     regions of the ($\varepsilon_0, \Gamma_a$)-plane affected is nearly independent of $V$.
     Notice that all rightmost (leftmost) positions of the left (right) panels are aligned to resonance.
     In the white (empty) regions of the left panels the relative deviations exceed 100\%.}
  \label{fig:errors-V-aligned}
\end{figure*}

Complementary to the presentation in \secname\ref{sec:fixed-e0},
in \figsname\ref{fig:errors-1.5V} to \ref{fig:errors-V-aligned}
and \figsname\ref{fig:errors-1.0V-si} to \ref{fig:errors-V-g-aligned-si} of the {\esi}
we next show results depicting room temperature effects
in the plane ($\varepsilon_0, \Gamma_a$) for several values of the bias ranging from
the upper limit of biases which real molecular junctions can withstand ($V=1.5$\,V) down to
low biases ($V=0.1$\,V) typically chosen to experimentally estimate the ``zero-bias'' conductance; see
\figsname\ref{fig:errors-1.5V} and \ref{fig:errors-V-aligned}, and \figsname\ref{fig:errors-1.0V-si} to \ref{fig:errors-V-g-aligned-si}.

Again, these figures show that 
the current $I_{0K}$ computed at $T = 0$ deviates from the current at room temperature for
energy offsets around the resonance value $\varepsilon_0 \approx e V/2$.
Basically, \figsname\ref{fig:errors-1.5V}a and \ref{fig:errors-0.5V}a, and \ref{fig:errors-1.0V-si}a and \ref{fig:errors-0.1V-si}a depict
changes expected in view of the above considerations;
lowering the bias $V = 1.5; 1.0; 0.5; 0.1$\,V shifts 
the predominantly red region in these figures (corresponding to deviations up to 100\%) to the smaller $\varepsilon_0$
around the values $\varepsilon_0 (= e V/2) = 0.75; 0.5; 0.25; 0.05$\,eV.

To emphasize again the fact that current's thermal enhancement is directly related to the resonance condition
($\varepsilon_0 \approx e V/2$), we depicted in \figname\ref{fig:errors-e0-aligned-si} 
relative deviations of $I_{0K}$ from $I_{RT}$ computed for various biases
aligned to the same abscissa value ($\varepsilon_0 = e V/2$).
\figname\ref{fig:errors-e0-aligned-si} makes it clear that thermal effects around resonance
($\varepsilon_0 \approx e V/2$) are insensitive to
$V$, provided that the corresponding energy is reasonably large with respect to the thermic energy $k_B T$.
Consistent to \secname\ref{sec:fixed-e0}, thermal effects appear to be weaker above resonance ($\varepsilon_0 < 2 e V$)
than below resonance, where they are pronounced in the range
\begin{eqnarray}
  & & e V / 2 < \varepsilon_0 \alt e V/2 + \delta\varepsilon_0\left(\Gamma_a\right) \nonumber \\
  & & \delta \varepsilon_0 \left(\Gamma_a\right)
  < \left . \delta\varepsilon_0\right\vert_{\Gamma_a\to 0} \approx 0.25\, \mbox{eV} \approx 3 \pi k_B T_{RT}
  \label{eq-delta-e0}
\end{eqnarray}
Consistent with \secname\ref{sec:fixed-e0} and eqn~(\ref{eq-delta-V}),
$ \delta\varepsilon_0\left(\Gamma_a\right) \approx e \delta V\left(\Gamma_a\right)/2$ 
is seen to be nearly independent of $V$. Its value at small $\Gamma_a$ is
$ \left . \delta\varepsilon_0\right\vert_{\Gamma_a\to 0} \approx \left . e \delta V\right\vert_{\Gamma_a \to 0} / 2 \approx 0.25$\,eV
and decreases as $\Gamma_a$ becomes larger.

Along with deviations with respect to the exact current $I_{RT}$ at room
temperature of the current $I_{0K}$ computed at $T=0$
(Figs.~\ref{fig:errors-1.5V}a, \ref{fig:errors-1.0V-si}a, \ref{fig:errors-0.5V}a, and \ref{fig:errors-0.1V-si}a), in
Figs.~\ref{fig:errors-1.5V}b, \ref{fig:errors-1.0V-si}b, \ref{fig:errors-0.5V}b, and \ref{fig:errors-0.1V-si}b
we also show deviations of current $I_{0K,off}$ computed via eqn~(\ref{eq-I0Koff}).
Inspection of the aforementioned figures reveals that,
in the range compatible with eqn~(\ref{eq-1.4}) and (\ref{eq-Gamma-vs-e0}), it is very accurate.
At $V=0.5$\,V, deviations of $I_{0K,off}$ from $I_{RT}$ are significant only for model parameter values where
deviations of $I_{0K}$ from $I_{RT}$ are also significant. This fact is understandable: being a limiting case
of eqn~(\ref{eq-I0K}), eqn~(\ref{eq-I0Koff}) cannot be expected to perform better than eqn~(\ref{eq-I0K}).
$I_{0K,off}$ is less accurate than $I_{0K}$ only for values of the parameters violating
the conditions assumed in the deduction of eqn~(\ref{eq-I0Koff}). This is the case of the upper left corner of
Fig.~\ref{fig:errors-0.1V-si}b, wherein the small values of $\varepsilon_0$ and the large values of $\Gamma_a$
are incompatible with eqn~(\ref{eq-Gamma-vs-e0}).
\subsection{Thermal effects on the zero bias conductance}
\label{sec:g}
\begin{figure}[htb]
  \centerline{\includegraphics[width=0.22\textwidth,height=0.42\textwidth,angle=-90]{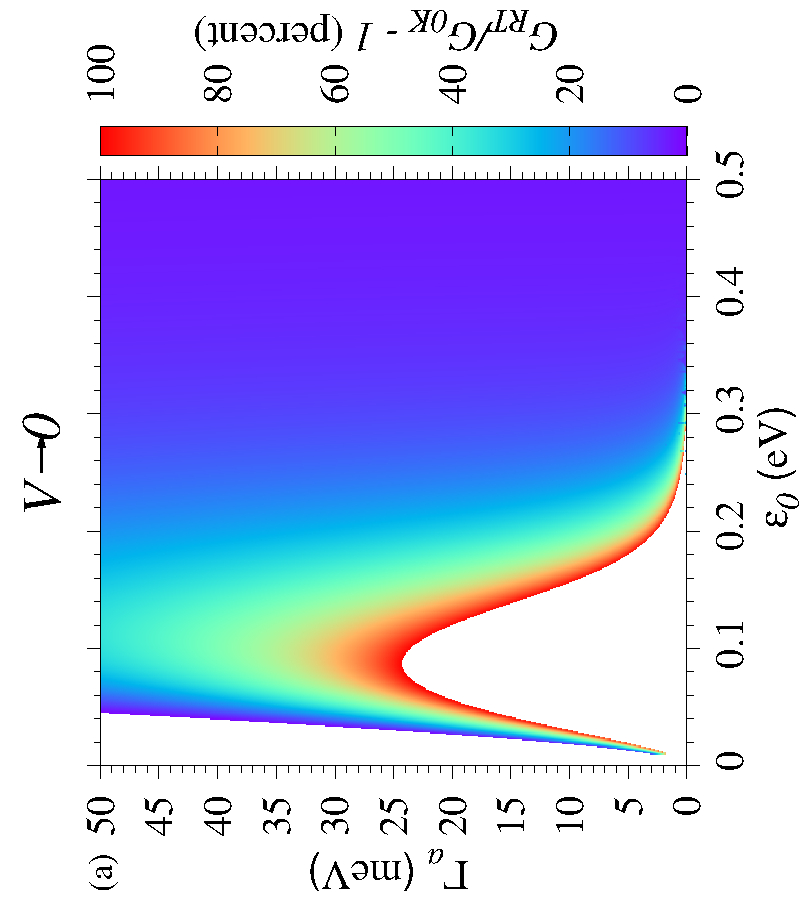}}
  \centerline{\includegraphics[width=0.22\textwidth,height=0.42\textwidth,angle=-90]{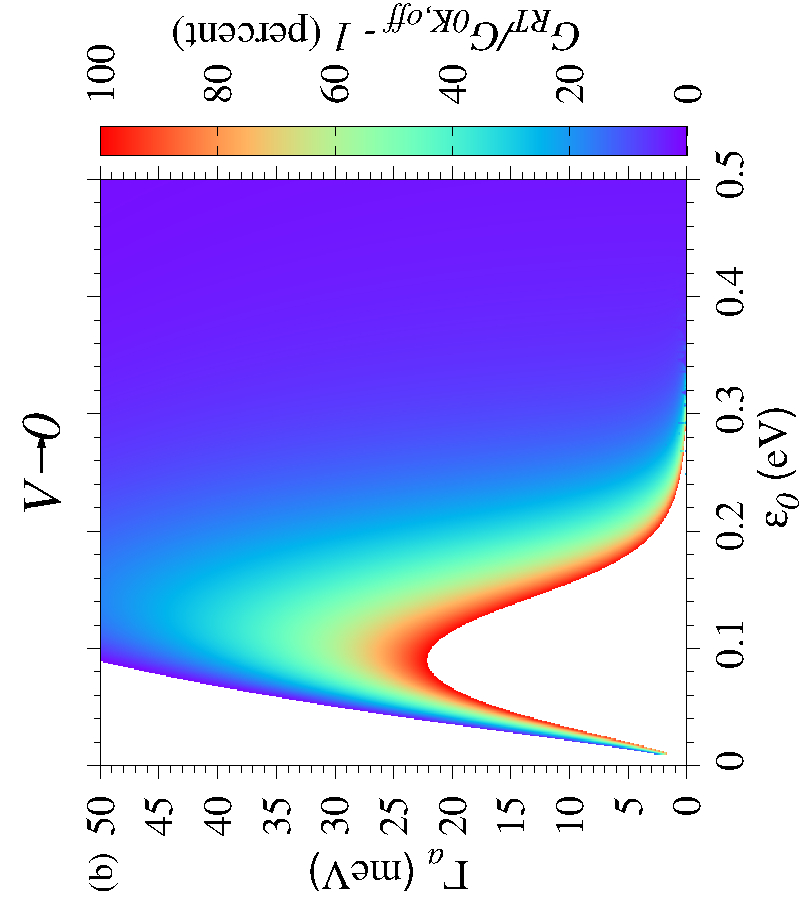}}
  \caption{Deviations in percent from the exact zero bias conductance $G_{exact} = G_{RT}$ 
    computed exactly at $T = 298.15$\,K (eqn~(\ref{eq-g-exact})) 
    of the ohmic conductance (a) $G_{0K}$ and (b) $G_{0K,off}$ 
    computed at $T=0$ using eqn~(\ref{eq-g0K}) and (\ref{eq-g0Koff}), respectively.
    Unless eqn~(\ref{eq-Gamma-vs-e0}) (see upper left corner), $G_{0K,off}$ is as accurate as $G_{0K}$.}
  \label{fig:errors-g0K-g0Koff}
\end{figure}

As an important case of a fixed value of $V$ (namely, $V\to 0$), in \figname\ref{fig:errors-g0K-g0Koff}
we depict deviations of the zero temperature values $G_{0K}$ and $G_{0K, off}$ (eqn~(\ref{eq-g0K}) and (\ref{eq-g0Koff}),
respectively) from the exact conductance $G_{exact}=G_{RT}$ (eqn~(\ref{eq-g-exact})).

In agreement with eqn~(\ref{eq-delta-e0}) at $V = 0$, significant departures of $G_{0K}$ from $G_{RT}$ occur 
in the range
\begin{equation}
  \label{eq-delta-e0-V=0}
  \varepsilon_0 < \delta \varepsilon_0\left(\Gamma_a\right) <
  \left . \delta\varepsilon_0\right\vert_{\Gamma_a\to 0} \approx 0.25\,\mbox{eV} \approx 3 \pi k_B T_{RT}
\end{equation}
whose width is the largest at $\Gamma_a \to 0$ ($\left . \delta\varepsilon_0\right\vert_{\Gamma_a \to 0} \approx 0.25\,\mbox{eV}$)
and becomes smaller as $\Gamma_a $ increases (\figname\ref{fig:errors-g0K-g0Koff}a).
As for $G_{0K, off}$, confirming the analysis of \secname\ref{sec:model},
it is as accurate as $G_{0K}$ unless the condition
$ \Gamma_a \ll \varepsilon_0$ is violated (cf.~eqn~(\ref{eq-g0K}) and (\ref{eq-g0Koff});
see the upper left corner of \figname\ref{fig:errors-g0K-g0Koff}b.

\subsection{An experimental digression}
\label{sec:exp}
To illustrate the above ideas with specific examples from real molecular electronics,
we will consider in this section three molecular junctions fabricated using different platforms:
single molecule junctions of 4,4'-bisnitrotolane (BNT) and gold electrodes
fabricated using mechanically controlled break
junction technique \cite{Zotti:10}, CP-AFM junctions fabricated with
perylene tetracarboxylic acid diimide (PDI) molecules and silver electrodes \cite{Baldea:2018a}, and large area 
junctions with EGaIn electrodes based on molecules of alkanethiolates functionalized with a ferrocene (Fc) unit \cite{Nijhuis:16b}.

What these junctions have in common is their comparable, small MO energy offset:
$\varepsilon_0 \simeq 0.27$\,eV \cite{Zotti:10}, $0.26$\,eV \cite{Baldea:2018a}, and $0.24$\,eV \cite{Baldea:2022j},
respectively. What makes the first junction different from the last two is the MO width:
$\Gamma_a \simeq 35$\,meV \cite{Zotti:10} much larger than
$\Gamma_a \simeq 6$\,meV \cite{Baldea:2018a} and $4.6$\,meV \cite{Baldea:2022j}, respectively.

\begin{figure*}[htb]
  \centerline{\includegraphics[width=0.45\textwidth]{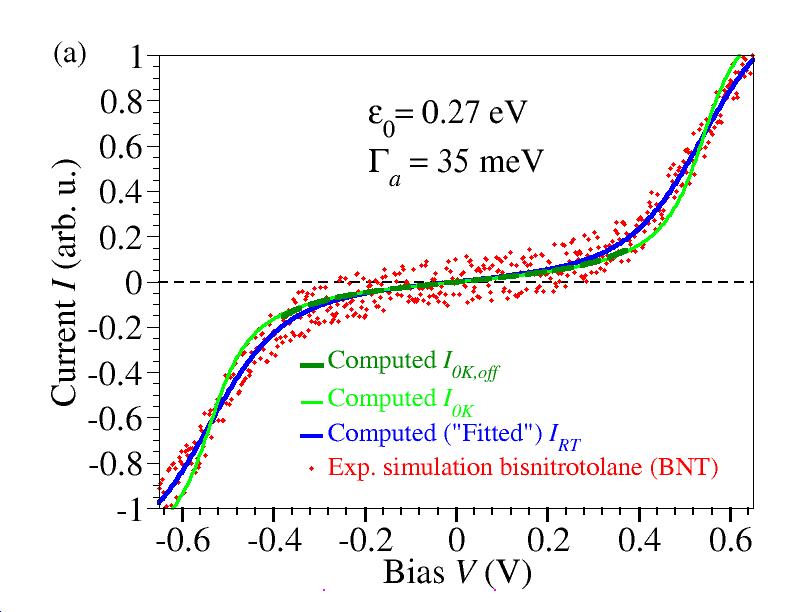}
    \includegraphics[width=0.45\textwidth]{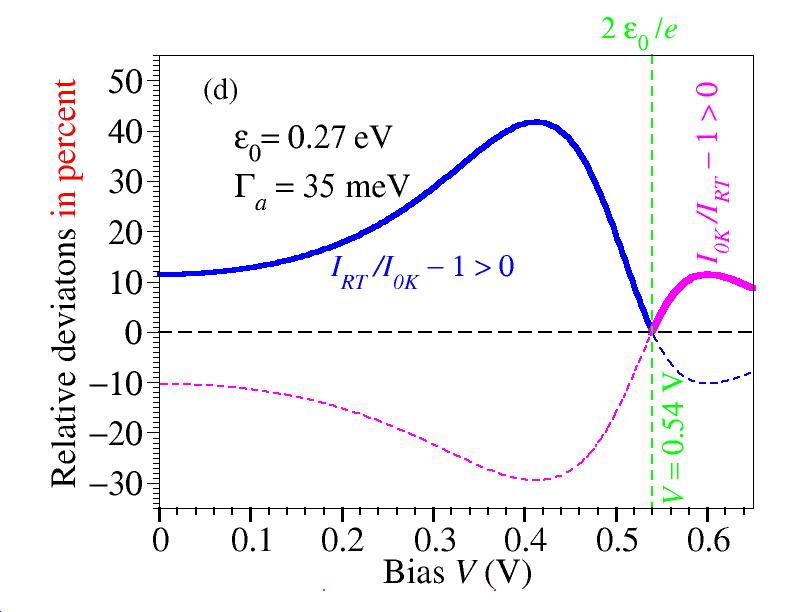}
  }
  \centerline{\includegraphics[width=0.45\textwidth]{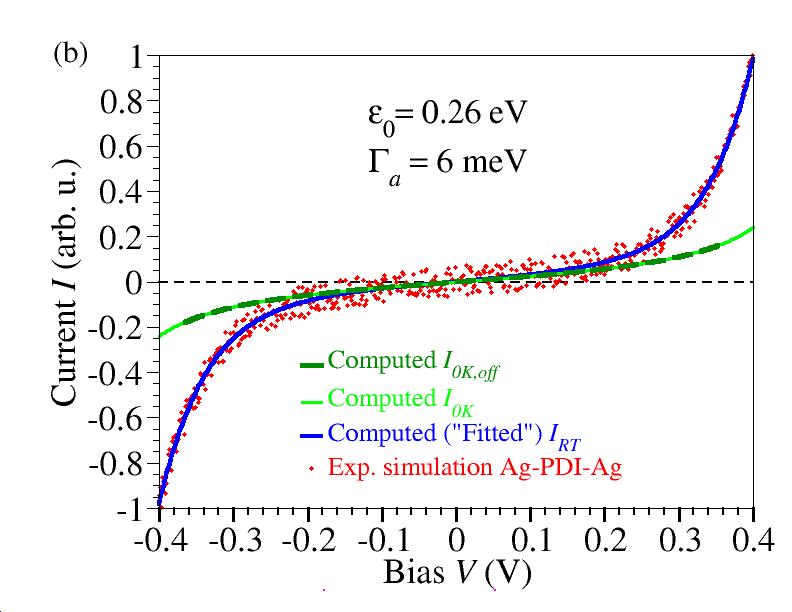}
    \includegraphics[width=0.45\textwidth]{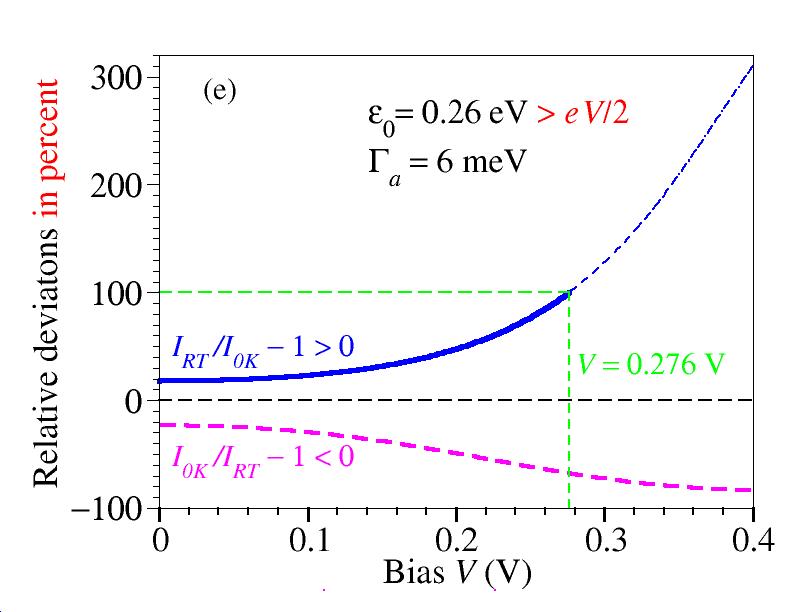}}
  \centerline{\includegraphics[width=0.45\textwidth]{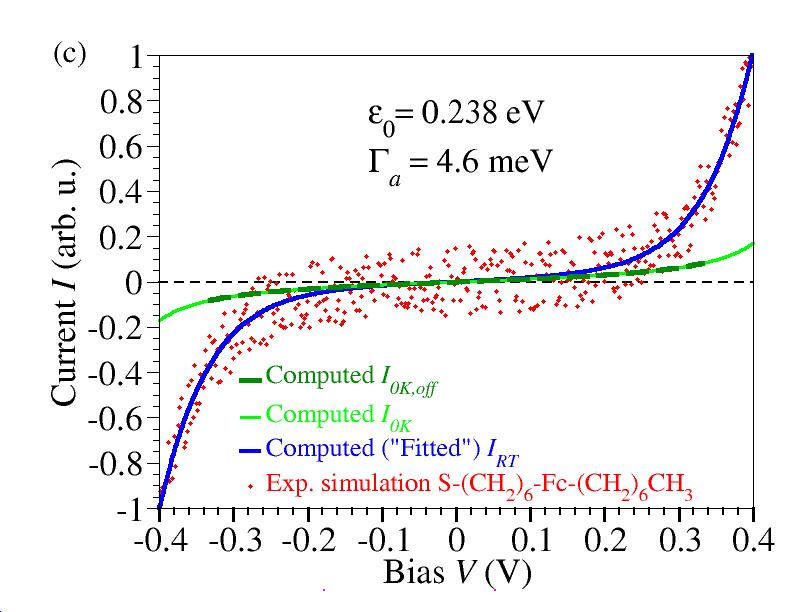}
    \includegraphics[width=0.45\textwidth]{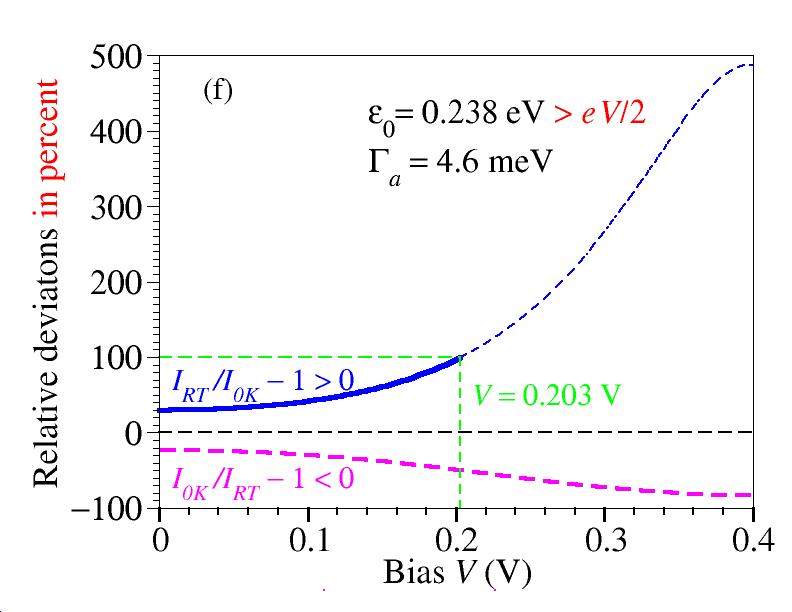}}
  \caption{Numerical simulations using literature model parameters \cite{Zotti:10,Baldea:2018a,Baldea:2022j}
    for (a, d) single-molecule BNT junctions \cite{Zotti:10}, (b, e) CP-AFM PDI junctions \cite{Baldea:2018a},
    and (c, f) large-area junctions with EGaIn electrodes based on molecules of alkanethiolates functionalized
    with a ferrocene (Fc) \cite{Nijhuis:16b}.
    Notwithstanding the similar values of the dominant MO energy offset $\varepsilon_0$, the much larger value of the MO width
    $\Gamma_a$ for BNT junctions makes the impact of temperature
    on current much weaker than for the other two junctions.}
  \label{fig:exp}
\end{figure*}

In \figname\ref{fig:exp} we used the above parameter values to simulate $I$-$V$ ``measurements'' (represented as red points)
by overimposing a bit disorder on the values of the current calculated via eqn~(\ref{eq-Iexact}) at $T=298.15$\,K. 
Re-fitting for self-consistency these ``experimental'' $I$-$V$ data using eqn~(\ref{eq-Iexact}) yielded the curves for $I_{RT}$
depicted by blue lines in \figname\ref{fig:exp}. 

The curves for $I_{0K}$ and $I_{0K,off}$ computed using
eqn~(\ref{eq-I0K}) and (\ref{eq-I0Koff}) and the same parameter values are depicted by the light green and dark green
lines, respectively. The difference between these curves (assuming $T = 0$) and the red curves (assuming $T=298.15$\,K)
is a temperature effect. 
Given the large value of $\Gamma_a$($=35$\,meV), notwithstanding the small value of $\varepsilon_0$,
this effect is weak for BNT (\figsname\ref{fig:exp}a and d). The temperature effect is considerably
more pronounced for the PDI junction having $\Gamma_a = 6$\,meV (cf.~\figsname\ref{fig:exp}b and e).
In agreement with \figsname\ref{fig:exp}f,
the thermal corrections $I_{RT}/I_{0K} - 1$ exceeding 100\% at $V=0.4$\,V and $V=0.3$\,V
lie in the empty (white) region in \figsname\ref{fig:errors-V-g-aligned-si}a and b.
The smaller values $\varepsilon_0 \simeq 0.24$\,eV versus  $\varepsilon_0 \simeq 0.26$\,eV and
$\Gamma_a = 6$\,meV versus $\Gamma_a = 4.6$\,meV make the temperature effect in the Fc-based junction (\figsname\ref{fig:exp}c and f)
even stronger than for PDI.
In the former case, the thermal corrections $I_{RT}/I_{0K} - 1$ exceed 100\% not only at  $V=0.4$\,V and $V=0.3$\,V but
also at the lower bias $V=0.2$\,V
(cf.~\figsname\ref{fig:exp}c and f and the empty (white) region in \figsname\ref{fig:errors-V-g-aligned-si}a to c).
\section{Applicability of the zero temperature approaches expressed by analytic inequalities}
\label{sec:applic}
The numerical results reported above allow us to 
indicate (more precisely than done in \secname\ref{sec:model}) the parameter ranges where
transport measurements performed at room temperature can be accurately analyzed using analytic formulas 
valid for $T=0$, which are more convenient for experimental data processing than those for $T\neq 0$.

As anticipated in \secname\ref{sec:model} and confirmed by the foregoing numerical simulations (\secname\ref{sec:results}),
the strongest thermal effects occur below resonance ($\vert e V\vert < 2\left\vert \varepsilon_0 \right\vert $).
Therefore, the analysis in this section will focus on such situations.
\subsection{$\mathbf{I_{0K} \approx I_{RT}}$. Applicability of eqn~(\ref{eq-I0K})}
\begin{figure*}[htb]
  \centerline{\includegraphics[width=0.22\textwidth,height=0.42\textwidth,angle=-90]{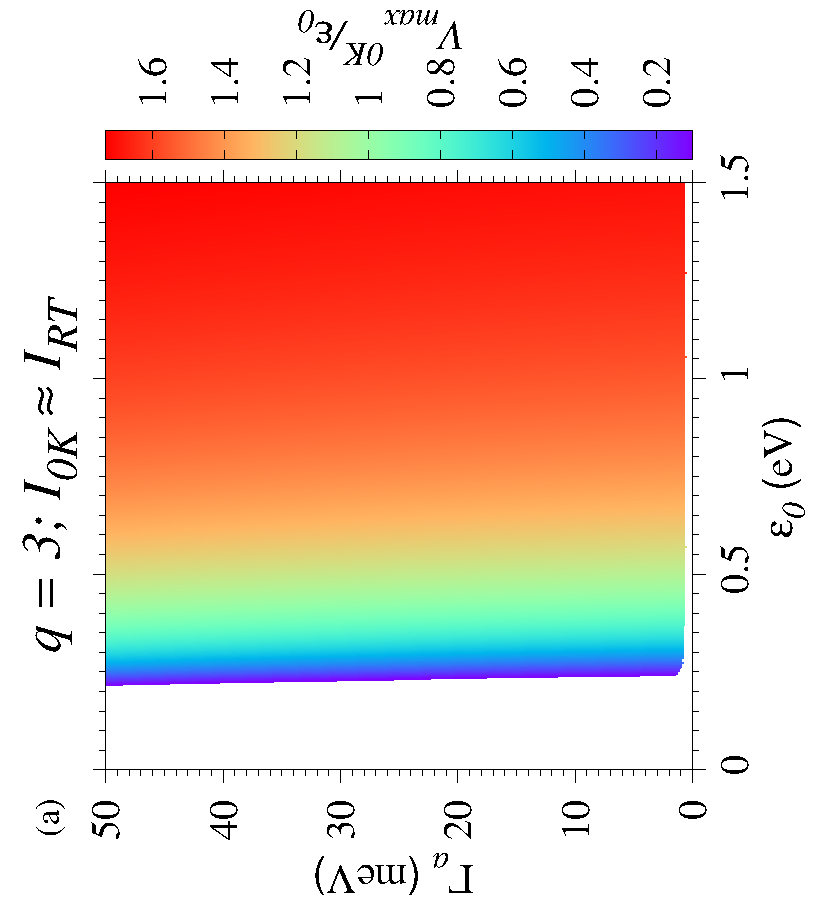}
    \includegraphics[width=0.22\textwidth,height=0.42\textwidth,angle=-90]{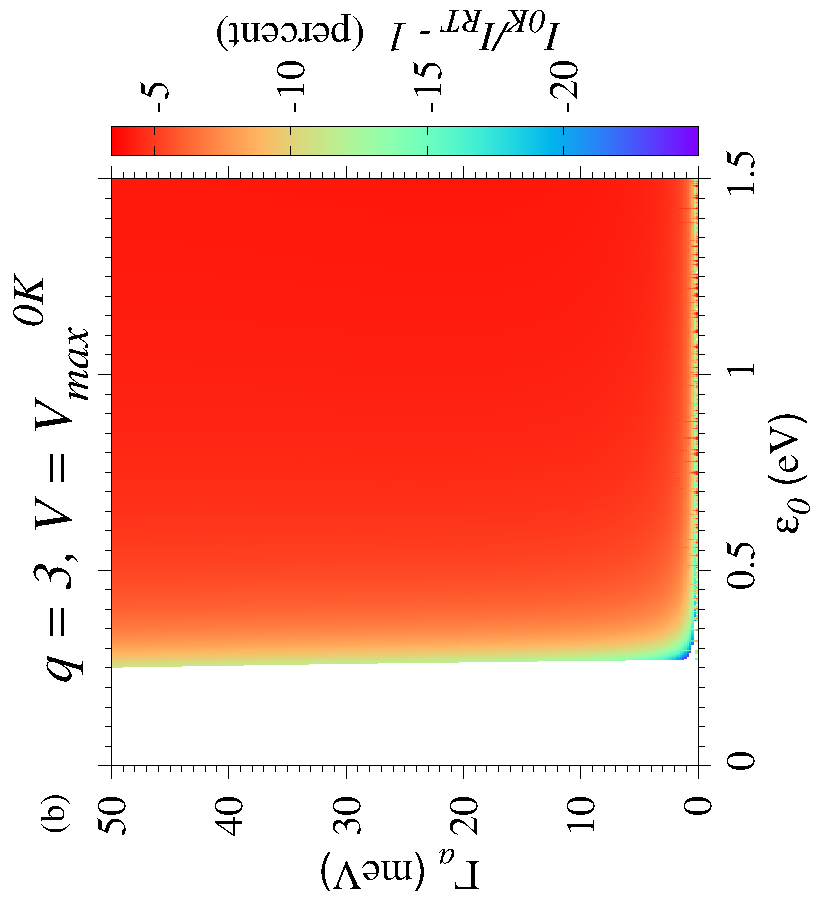}}
    \centerline{\includegraphics[width=0.22\textwidth,height=0.42\textwidth,angle=-90]{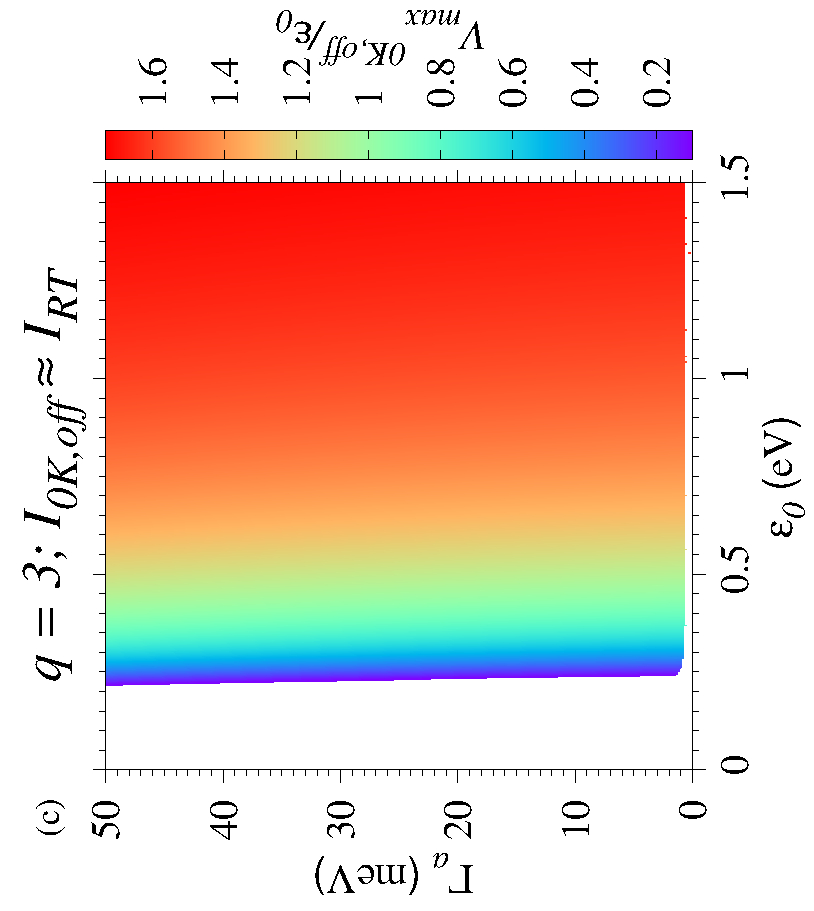}
    \includegraphics[width=0.22\textwidth,height=0.42\textwidth,angle=-90]{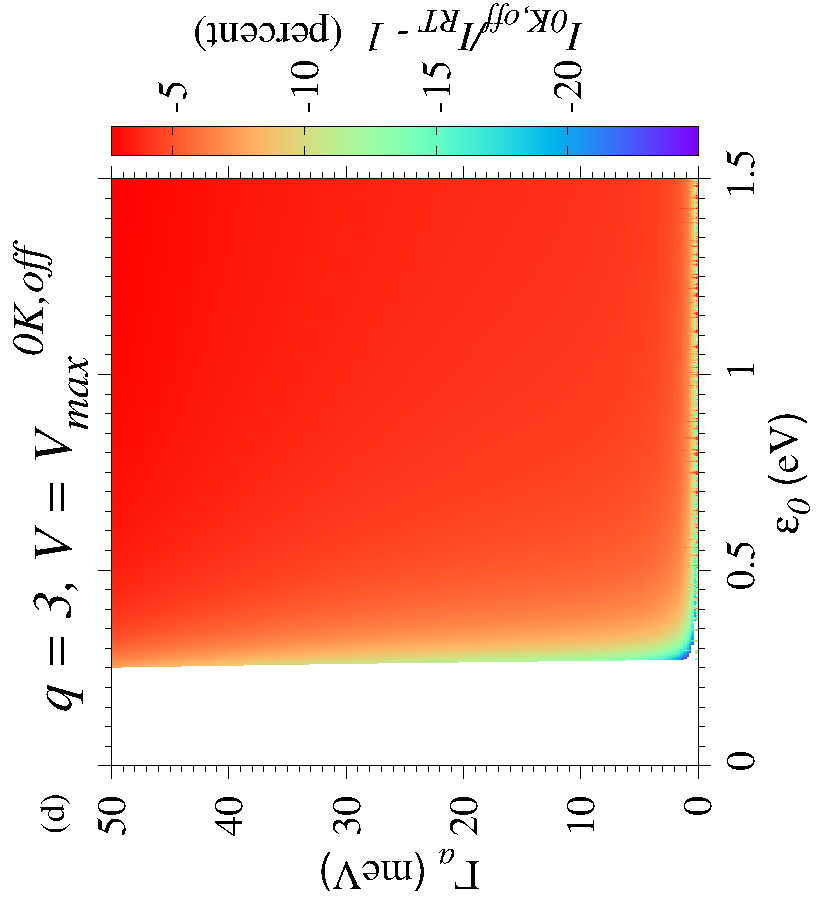}}
    \caption{The highest bias $V_{max}^{0K}$ and  $V_{max}^{0K,off}$ (panels a and b, respectively)
      at which eqn~(\ref{eq-I0K}) and eqn~(\ref{eq-I0Koff}) can reliably be applied for room temperature data processing.
      $V_{max}^{0K}$ and  $V_{max}^{0K,off}$ were computed from eqn~(\ref{eq-Vmax0K}) and eqn~(\ref{eq-extra-restriction-I0Koff}), respectively.
      They correspond to setting $q = 3$ in eqn~(\ref{eq-low-T-V-q}).
      The pertaining thermal corrections are presented in panels b and d, respectively.}
  \label{fig:q=3}
\end{figure*}

Corroborating the general considerations that led to eqn~(\ref{eq-low-T-V}) with the specific results expressed by
eqn~(\ref{eq-delta-V}) and (\ref{eq-delta-e0}), we arrive at concluding that the value of $q$ ``sufficiently'' larger
than two (cf.~\secname\ref{sec:model}) needs in fact not be very large.
Namely, to ensure that the description based on $I_{0K}$ is accurate (i.e., $I_{0K}\simeq I_{RT}$), it is sufficient to set $q = 3$.
The highest bias at which eqn~(\ref{eq-I0K}) applies
is defined by the value $V_{max}^{0K}$($ < 2 \left\vert\varepsilon_0\right\vert /e$) indicated below
\begin{subequations}
\label{eq-valid-I0K}
  \begin{equation}
  \label{eq-Vmax0K}
\sqrt{\left(\pi k_B T + \Gamma_a\right)^2 + \left(  \left\vert \varepsilon_0 \right\vert - e V_{max}^{0K} /2\right)^2} = 3 \pi k_B T  
\end{equation}

The diagram of $V_{max}^{0K} = \left . V_{max}^{0K}\left(\varepsilon_0, \Gamma_a\right)\right\vert_{q = 3}$ is depicted in \figname\ref{fig:q=3}a 
while the pertaining thermal corrections for current
(not exceeding a few percent for most realistic parameters) are shown in \figname\ref{fig:q=3}b.
Imposing larger values of $q$ (e.g., $q=3.5$, \figname\ref{fig:q=3.5} or $q=4$, \figname\ref{fig:q=4}) does not significantly decrease
the thermal corrections while artificially decreasing $V_{max}^{0K}$.

Notice that only values $\varepsilon_0 > \varepsilon_{0}^{min}$ larger than a certain minimum value are shown in \figname\ref{fig:q=3}.
This expresses the physical reality that, irrespective of bias ($e V <2\varepsilon_0$),
the low temperature limit does not apply for too small values of $\varepsilon_0$.
The smallest value $\varepsilon_{0}^{min}$, estimated by setting $q=3$ in eqn~(\ref{eq-low-T-V-q}),  
\begin{eqnarray}
  \left\vert \varepsilon_0 \right\vert >
  \varepsilon_{0}^{min} & = &
  \sqrt{\left( 3 \pi k_B T\right)^2 - \left(\pi k_B T + \Gamma_a\right)^2}
  \label{eq-e0-min}
\end{eqnarray}
very weakly depends on $\Gamma_a$. It is visualized by the line separating the white and colored
portions in \figname\ref{fig:q=3}.
\end{subequations}

Eqn~(\ref{eq-low-T-V-gg}) also makes it clear that, irrespective of the values of $\varepsilon_0$ and $V$,
the low temperature limit applies for $\Gamma_a$ ``sufficiently'' larger than $k_B T$.
Inspection of the previously analyzed figures reveals that, in fact, $\Gamma_a$ needs not be much larger than $k_B T$($ = 25.7$\,meV).
The low temperature limit $I_{RT} \approx I_{0K}$, $G_{RT} \approx G_{0K}$ is reasonably accurate at the largest value ($\Gamma_a = 50$\,meV)
shown in those figures.
Depicting values of the ratio $I_{0K}/I_{RT}$ close to unity both for $\Gamma_a = 50$\,meV and
for $\Gamma_a = \pi k_B T_{RT} = 80.7$\,meV, \figname\ref{fig:errors-fixed-Gamma}
additionally emphasizes this aspect.

To sum up, eqn~(\ref{eq-I0K}) is applicable for biases $\vert V\vert $ smaller than
$V_{max}^{0K}$ (eqn~(\ref{eq-Vmax0K})) and
$\left\vert \varepsilon_0\right\vert$ larger than $\varepsilon_{0}^{min}$ (eqn~(\ref{eq-e0-min})).
\subsection{$\mathbf{I_{0K,off} \approx I_{RT}}$. Applicability of eqn~(\ref{eq-I0Koff})}
\label{sec:inequalities}
Let us now discuss the applicability of eqn~(\ref{eq-I0Koff}).  
As re-emphasized recently \cite{baldea:comment}, this equation should be applied only for biases sufficiently
below resonance \cite{Baldea:2012a}. Eqn~(\ref{eq-I0Koff}) should by no means be applied above resonance
($ \vert e V \vert \geq 2 \left\vert \varepsilon_0\right\vert $) where the denominator becomes
negative and, completely nonphysically, current and bias would have opposite directions \cite{baldea:comment}.

Derived as a limiting case of eqn~(\ref{eq-I0K}) \cite{Baldea:2012a},
eqn~(\ref{eq-I0Koff}) is implicitly subject to the low temperature
restrictions expressed by eqn~(\ref{eq-valid-I0K}). In addition, the highest bias to which eqn~(\ref{eq-I0Koff})
applies has to satisfy eqn~(\ref{eq-arctan}), that is
\begin{subequations}
  \begin{equation}
  \vert V\vert < V_{max}^{off} ; \
  e V_{max}^{off} < 2 \left\vert\varepsilon_0\right\vert - 5.856 \Gamma_a \ \left(< 2\left\vert \varepsilon_0\right\vert / e \right)
\end{equation}
The above condition ensures that $I_{0K,off}$ does not differ from $I_{0K}$ by more than 2\% (\figname\ref{fig:arctan}).
Corroborating with (eqn~(\ref{eq-Vmax0K})), this yields
the highest bias at which eqn~(\ref{eq-I0Koff}) is accurate
\begin{equation}
  \label{eq-extra-restriction-I0Koff}
  \vert V \vert < V_{max}^{0K,off} \equiv \min\left(V_{max}^{0K}, V_{max}^{off}\right) 
\end{equation}
\end{subequations}
Diagrams for $V_{max}^{0K,off}$ along with the current deviations $I_{0K,off}/I_{RT} - 1$
at this bias ($V = V_{max}^{0K,off}$) are depicted in
panels c and d of \figsname\ref{fig:q=3}, \ref{fig:q=3.5}, and \ref{fig:q=4}.
Notice the very close similarity of these panels (relying on $I_{0K,off}$) to panels a and b
(relying on  $I_{0K}$) of the same figures. This confirms the analysis of \secname\ref{sec:model}.
In off-resonance cases (which do represent the main focus of most experiments on molecular junctions), 
the description based on eqn~(\ref{eq-I0Koff}) is essentially as good as that based on eqn~(\ref{eq-I0K})
while applicable up to an upper bias which is basically the same.

To conclude, eqn~(\ref{eq-I0Koff}) is accurate at biases $\vert V\vert $ smaller than
$V_{max}^{0K,off}$ (eqn~(\ref{eq-extra-restriction-I0Koff})) and
$\left\vert \varepsilon_0\right\vert$ larger than $\varepsilon_{0}^{min}$ (eqn~(\ref{eq-e0-min})).
\subsection{Interactive data fitting using eqn~(\ref{eq-I0Koff})}
\label{sec:interactive}
The various figures presented above revealed that eqn~(\ref{eq-I0Koff}) used for biases $\vert V\vert < V_{1.4}$
(eqn~(\ref{eq-1.4})) 
is reliable in broad area of the model parameters $\varepsilon_0$ and $\Gamma_a$. Therefore, we recommend
to use this method first for $I$-$V$ data fitting,
as the fitting parameters $\varepsilon_0$ thus obtained were validated through additional ultraviolet photoelectron spectroscopy
(UPS) studies on benchmark molecular junctions \cite{Frisbie:11,Baldea:2019d,Baldea:2019h}.

When we suggested $V_{1.4}$ of eqn~(\ref{eq-1.4}) as upper bias 
for the applicability of eqn~(\ref{eq-I0Koff}),
we had in mind a pragmatic reason (see ref.~\citenum{baldea:comment} and citations therein): most molecular junctions
currently fabricated possess a conductance $G/G_0 <\alt 0.01$ obeying eqn~(\ref{eq-Gamma-vs-e0}).
\figname\ref{fig:I0Koff-1.4-1.0-10perc}a depicts parameter ranges wherein at $V = V_{1.4}$ $I_{0K,off} \approx I_{RT}$ holds
within 10\% (=``typical'' experimental accuracy).
The model parameters characterizing benchmark molecular junctions with alkyl
\cite{Baldea:2019h} and oligophenylene \cite{Baldea:2019d} backbones 
deduced from data fitting using eqn~(\ref{eq-I0Koff}) for biases $ \vert V \vert < V_{1.4} $
fall in the parameter ranges depicted in \figname\ref{fig:I0Koff-1.4-1.0-10perc}a.

For junctions having a normalized conductance $G/G_0$ larger than 0.01,
narrower bias ranges should be employed 
for reliably extracting the model parameters $\varepsilon_0$ and $\Gamma_a$
from data fitting based on eqn~(\ref{eq-I0Koff}).

\figname\ref{fig:I0Koff-1.4-1.0-10perc} may help to illustrate this idea.
Suppose we investigate a molecular junction having $\varepsilon_0 = 0.5$\,eV and
$\Gamma_a = 1.5$\,meV (obviously, values not known a priori) and can collect experimental $I$-$V$ data in the range 
$-0.7\,\mbox{V} < V < 0.7\,\mbox{V}$. 
To exploit the full experimental information available, we use eqn~(\ref{eq-I0Koff})
for data fitting in the entire range $V_{fit} = 0.7$\,V, $-V_{fit} < V < V_{fit}$.
This yields certain best fit parameters $\tilde{\varepsilon}_{0}$ and $\tilde{\Gamma}_{a}$,.
We insert these parameter values in eqn~(\ref{eq-extra-restriction-I0Koff}) and (\ref{eq-e0-min})
and compute $V_{max}^{0K,off} \to \tilde{V}_{max}^{0K,off}$ and $\varepsilon_{0}^{min} \to \tilde{\varepsilon}_{0}^{min}$.
Because the point 
($\varepsilon_0 = 0.5$\,eV, $\Gamma_a = 1.5$\,meV) lies in the empty (white) part of the diagram
in \figname\ref{fig:I0Koff-1.4-1.0-10perc}a depicted for $ e V = 1.4\,\varepsilon_0 $, we will have to conclude
that our values of $V_{fit}$ and $\tilde{\varepsilon}_{0}$ fail to satisfy at least one
of the two conditions requested
($V_{fit} < V_{max}^{0K,off}$, $\tilde{\varepsilon}_{0} > \varepsilon_{0}^{min}$).
We narrow the fitting range and arrive (possibly after several trials and errors)
at selecting the smaller value $V_{fit} = 0.5$\,V. With the new best fit parameters $\tilde{\varepsilon}_{0} = 0.5$\,eV and $\tilde{\Gamma}_{a} = 1.5$\,meV
and the new pertaining values $\tilde{V}_{max}^{0K,off}$ and $\tilde{\varepsilon}_{0}^{min}$, we
check that eqn~(\ref{eq-extra-restriction-I0Koff}) and eqn~(\ref{eq-e0-min}) are simultaneously satisfied.
Indeed, the point ($\varepsilon_0 = 0.5$\,eV, $\Gamma_a = 1.5$\,meV) belong to the ``allowed'' zone 
in \figname\ref{fig:I0Koff-1.4-1.0-10perc}b depicted for $\varepsilon_0 = e V $.
\section{Conclusion}
\label{sec:conclusion}
Notwithstanding impressive computational facility currently available,
experimentalists continue to prefer simple theoretical models to process the data they measure.
Representing ``by definition'' a simplified description of the real world,
a model cannot be blindly utilized ignoring the conditions of applicability.
Theory should make these conditions as transparent as possible.

By combining insight gained from a qualitative analysis of the relevant equations with extensive
numerical simulations, in the present we were able to provide the experimentalists not only with numerous diagrams
wherein they can presumably identify the specific case of their interest, but also with
simple mathematical inequalities that they can straightforwardly use to check whether
processing transport data measured on molecular junctions at room temperature 
using a zero temperature formalism is adequate or not.

Irrespective whether or not the fitting curves acceptably reproduced the measured $I$-$V$
traces, model parameters extracted by using eqn~(\ref{eq-I0K})
for data fitting can be trusted only if they satisfy
eqn~(\ref{eq-Vmax0K})) and (\ref{eq-e0-min}).
Likewise, model parameters extracted by using eqn~(\ref{eq-I0Koff}) are reliable only if
they obey eqn~(\ref{eq-extra-restriction-I0Koff}) and (\ref{eq-e0-min}).

Should this be not the case,
one can next try to obtain reliable parameters by gradually narrowing the bias range used for
data fitting according to the interactive procedure described in \secname\ref{sec:interactive}.
Should this attempt also fail, employing the less convenient eqn~(\ref{eq-Iexact}) is the last attempt
to be done before concluding that either electron (or hole) tunneling does not occur via a single level (MO),
that transmission is not Lorentzian \cite{Baldea:2017d},
or that, e.g., hopping rather than tunneling is at work in the envisaged junction.
\section*{Acknowledgments}
Financial support from the German Research Foundation
(DFG Grant No. BA 1799/3-2) in the initial stage of this work and computational support by the
state of Baden-W\"urttemberg through bwHPC and the German Research Foundation through
Grant No.~INST 40/575-1 FUGG
(bwUniCluster 2.0, bwForCluster/MLS\&WISO 2.0/HELIX, and JUSTUS 2.0 cluster) are gratefully acknowledged.
\renewcommand\refname{Notes and references}
\footnotesize{
  \providecommand*{\mcitethebibliography}{\thebibliography}
\csname @ifundefined\endcsname{endmcitethebibliography}
{\let\endmcitethebibliography\endthebibliography}{}

}
\pagestyle{empty}
\twocolumn[
  \begin{@twocolumnfalse}
\centerline{\LARGE{\textbf{Supplementary Information}}}
    \noindent\LARGE{\textbf{Can room temperature data for tunneling molecular junctions be analyzed within a theoretical framework assuming zero temperature? 
    }}
\vspace{0.6cm}

\noindent\large{\textbf{Ioan B\^aldea 
\textit{$^{a \ast}$}
}}\vspace{0.5cm}

\noindent

  {{\bf Keywords}:
molecular electronics, nanojunctions, single level model, thermal effects}
\vspace{0.5cm}
  \end{@twocolumnfalse}
]

\footnotetext{\textit{$^{a}$~Theoretical Chemistry, Heidelberg University, Im Neuenheimer Feld 229, D-69120 Heidelberg, Germany}}
\footnotetext{$^\ast$~E-mail: ioan.baldea@pci.uni-heidelberg.de
}
\renewcommand{\theequation}{S\arabic{equation}}
\setcounter{equation}{0}
\renewcommand{\thefigure}{S\arabic{figure}}
\setcounter{figure}{0}
\renewcommand{\thetable}{S\arabic{table}}
\setcounter{table}{0}
\renewcommand{\thesection}{S\arabic{section}}
\setcounter{section}{0}
\renewcommand{\thefootnote}{\alph{footnote}}
\begin{figure}[htb]
  \centerline{\includegraphics[width=0.45\textwidth,angle=0]{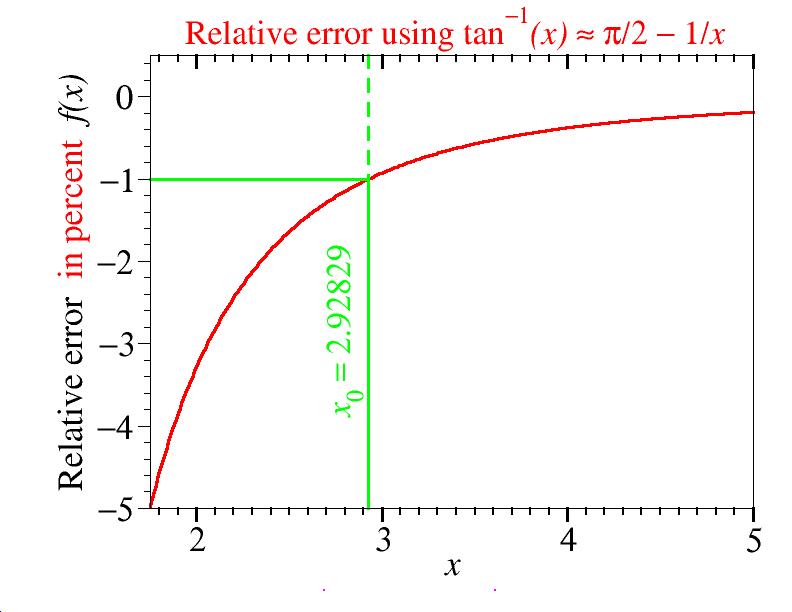}}
  \caption{Function $f(x) \equiv 100 \left[\left(\pi/2 - 1/x\right)/\tan^{-1} x - 1\right]$
    visualizing that the relative error in percent implied by using eqn~(\ref{eq-arctan}) is negligible.
  }
  \label{fig:arctan}
\end{figure}
\begin{figure*}[htb]
  \centerline{\includegraphics[width=0.22\textwidth,height=0.42\textwidth,angle=-90]{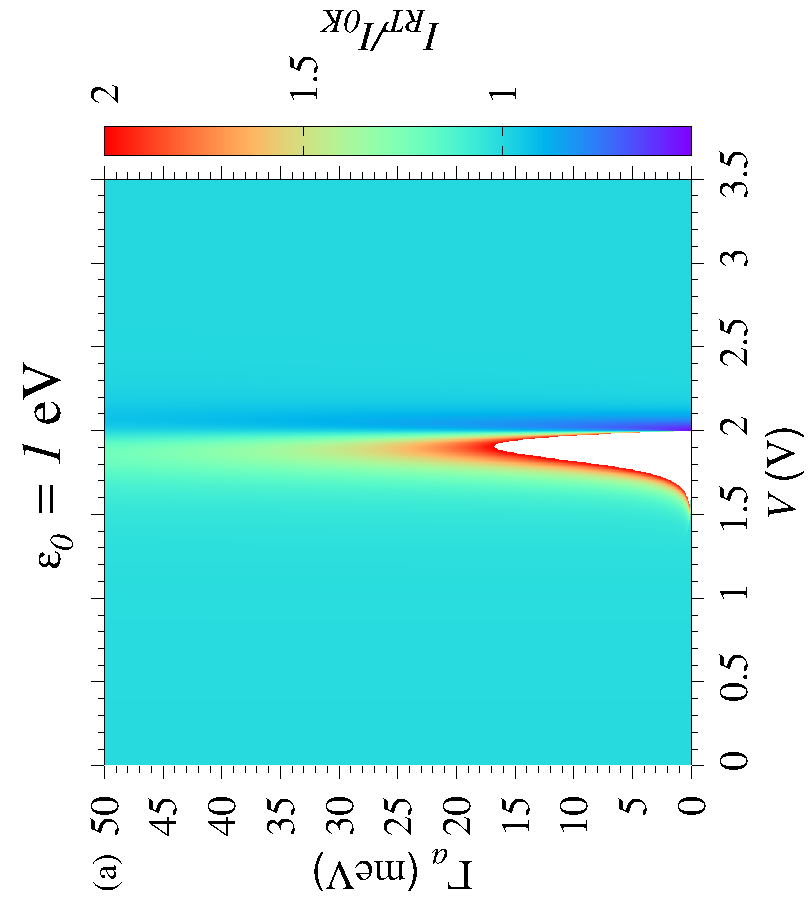}
  \includegraphics[width=0.22\textwidth,height=0.42\textwidth,angle=-90]{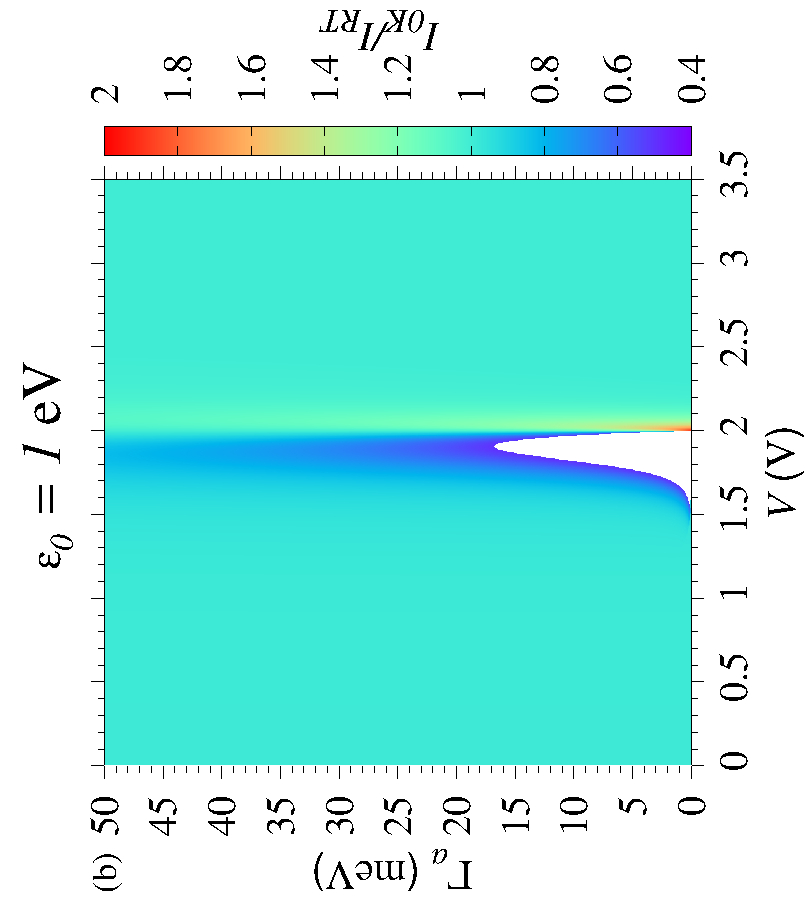}}
  \centerline{\includegraphics[width=0.22\textwidth,height=0.42\textwidth,angle=-90]{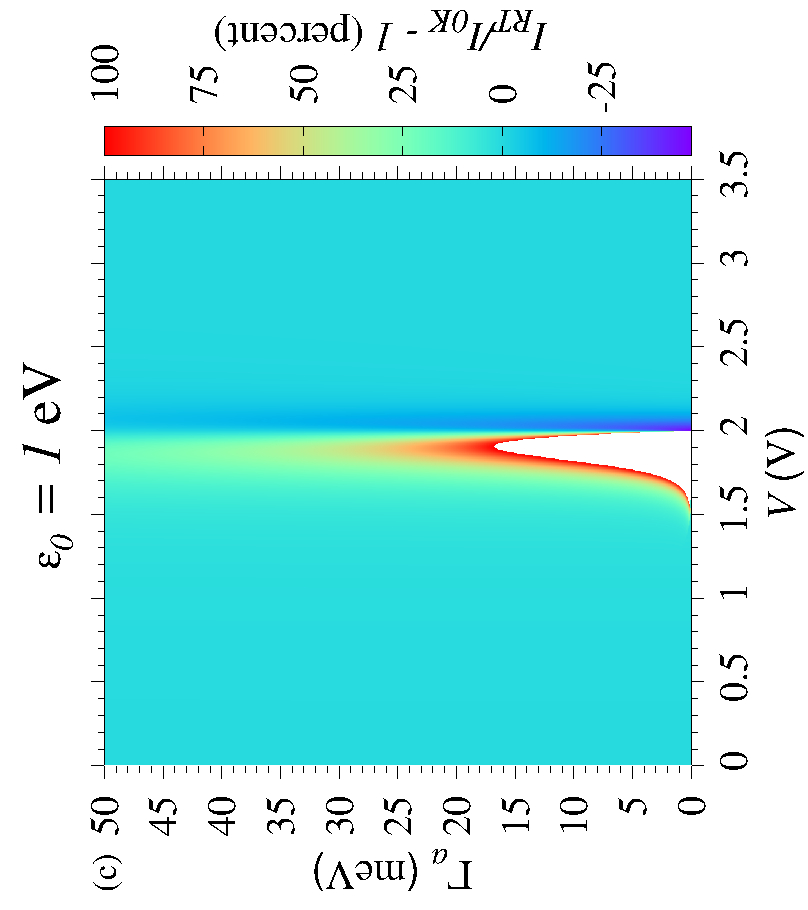}
  \includegraphics[width=0.22\textwidth,height=0.42\textwidth,angle=-90]{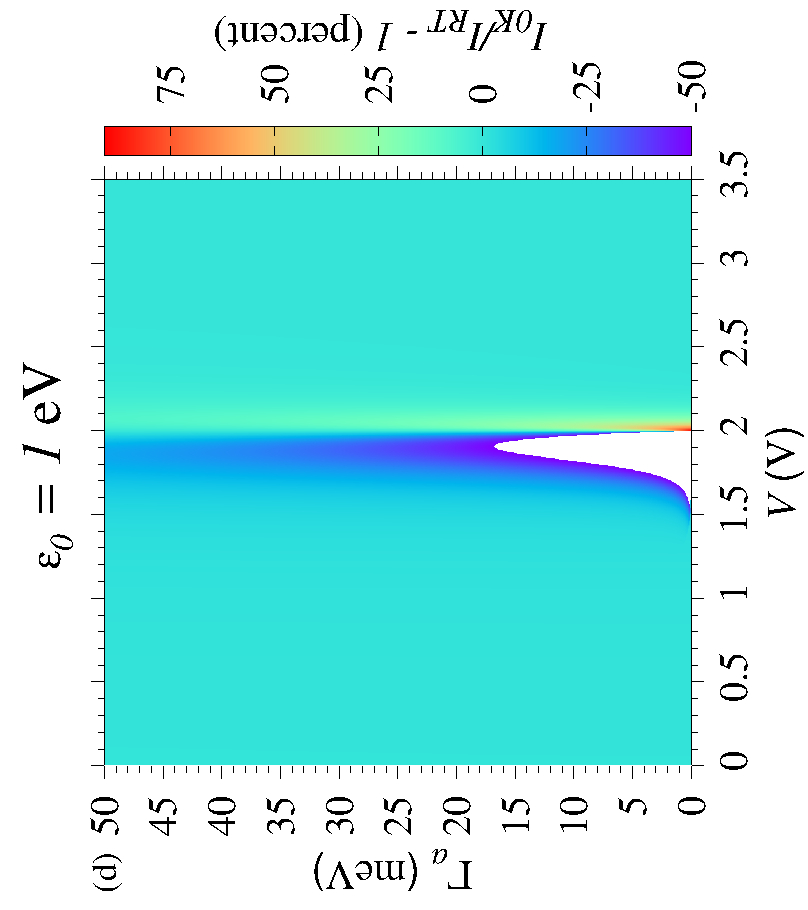}}
  \caption{Basically, the information presented in this figure on the deviations of the current
    $I_{0K}$ computed at zero temperature using eqn~(\ref{eq-I0K})
    from the room temperature $I_{RT}$ computed via eqn~(\ref{eq-Iexact})
    is the same as that of \figname\ref{fig:errors-e0-1.0}a and b.
    We prefer the latter manner of presentation because we find it is easier to understand.}
  \label{fig:presentation}
\end{figure*}
\begin{figure}[htb]
  \centerline{\includegraphics[width=0.45\textwidth,angle=0]{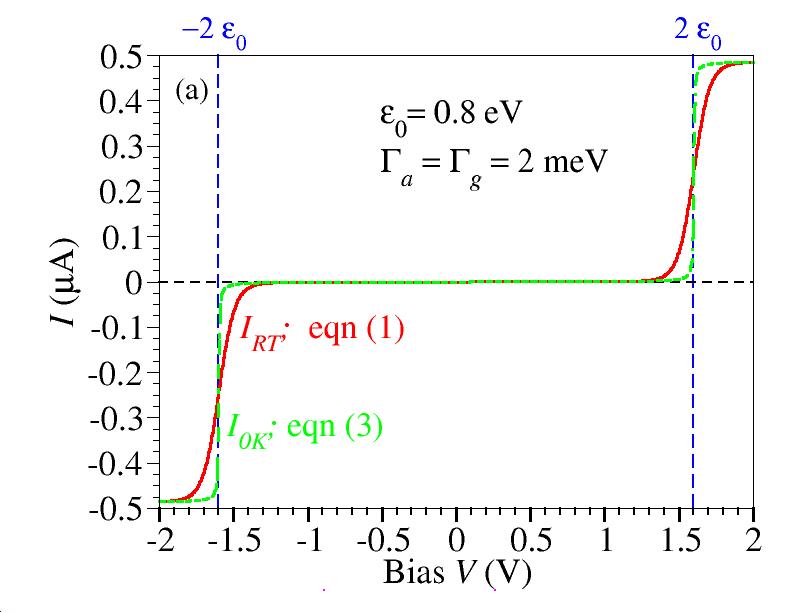}}
  \centerline{\includegraphics[width=0.45\textwidth,angle=0]{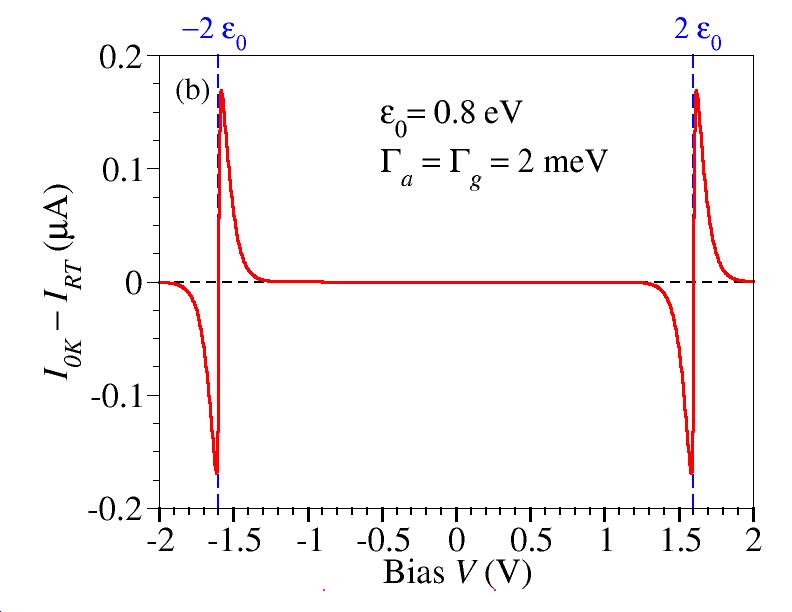}
  }
  \caption{(a) $I$-$V$ curves computed using eqn~(\ref{eq-Iexact}) and eqn~(\ref{eq-I0K}) illustrating
    that the thermal effect (b) enhances the current below resonance
    ($\left\vert I_{RT} \right\vert > \left\vert I_{0K} \right\vert$ for $ \vert e V\vert < 2 \left\vert\varepsilon_0\right\vert $)
    while reducing it above resonance
    ($\left\vert I_{RT} \right\vert < \left\vert I_{0K} \right\vert$ for $ \vert e V\vert > 2 \left\vert\varepsilon_0\right\vert $).}
  \label{fig:iv}
\end{figure}
\begin{figure*}[htb]
  \centerline{\includegraphics[width=0.22\textwidth,height=0.42\textwidth,angle=-90]{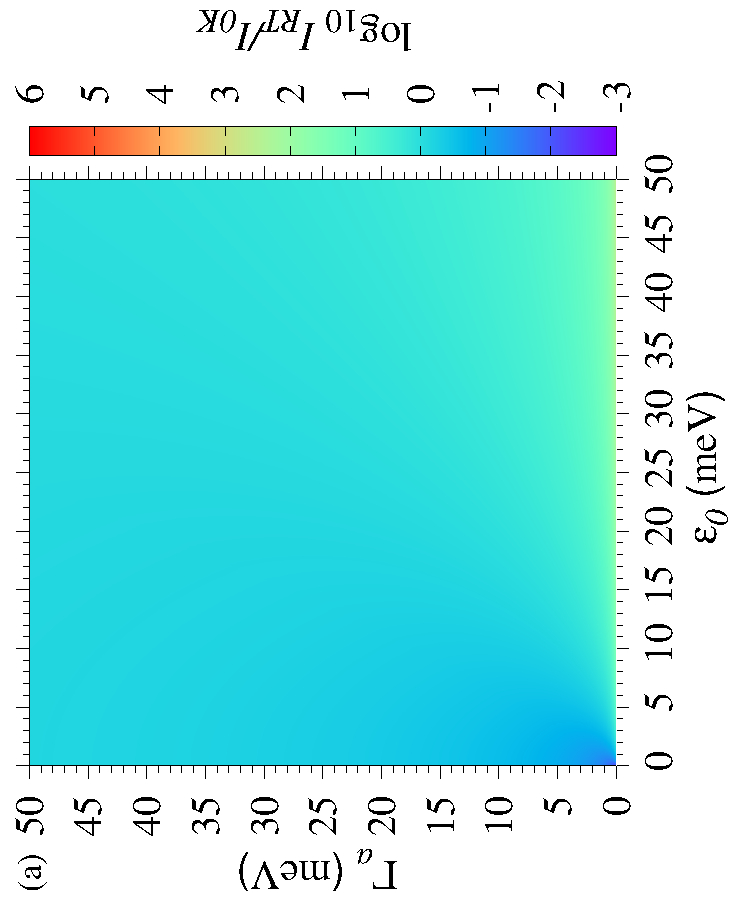}
        \includegraphics[width=0.22\textwidth,height=0.42\textwidth,angle=-90]{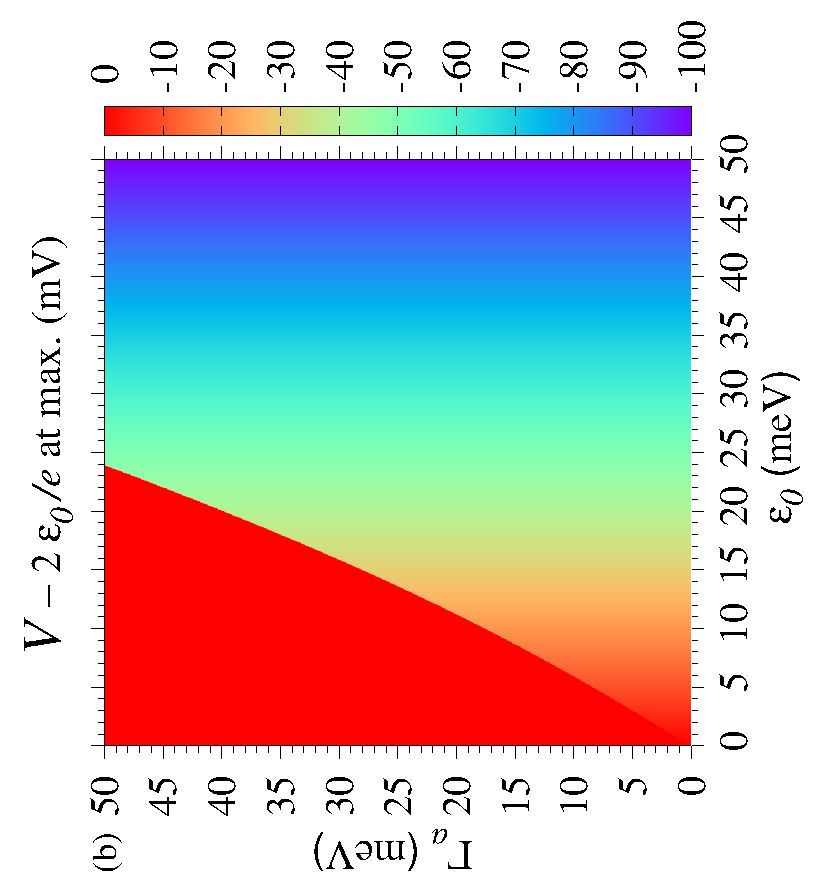}}  
  \centerline{\includegraphics[width=0.22\textwidth,height=0.42\textwidth,angle=-90]{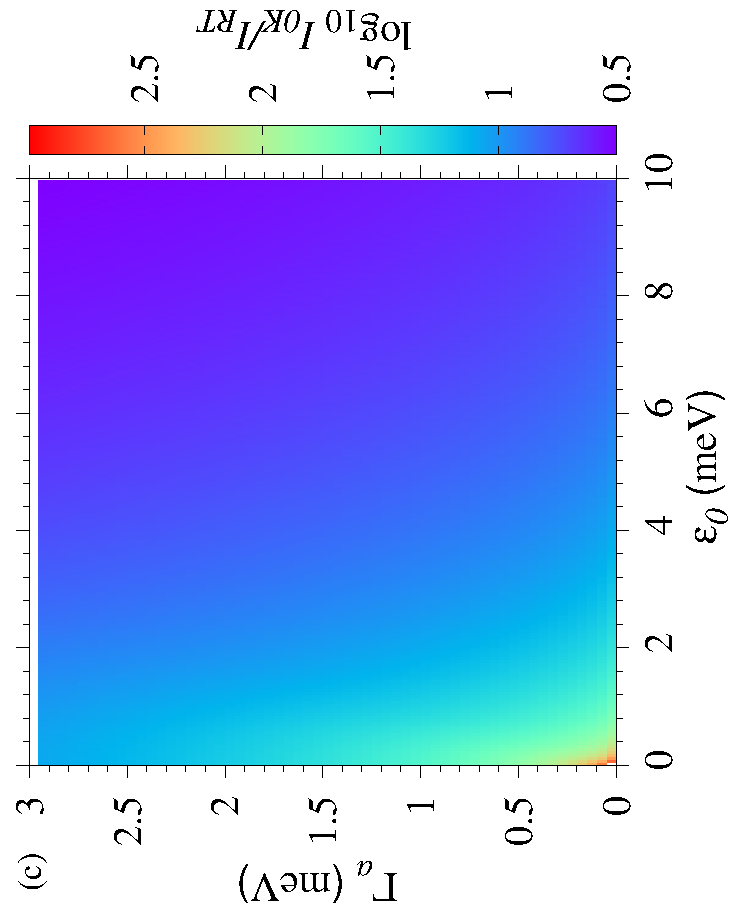}
        \includegraphics[width=0.22\textwidth,height=0.42\textwidth,angle=-90]{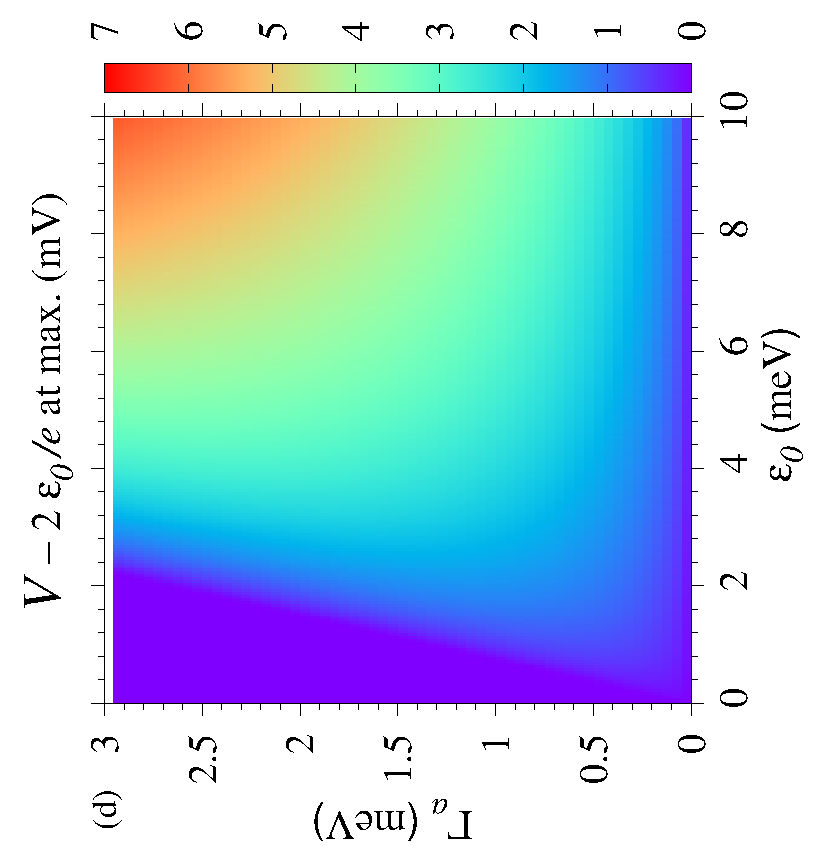}}  
  \caption{Results at very small values of the MO offset $\varepsilon_0$ showing an opposite behavior to that at (reasonably) large
   $\varepsilon_0$. Panel (a) depicts situations below resonance wherein $I_{RT} < I_{0K}$. Likewise, panel c shows situations above resonance wherein
   $I_{0K} < I_{RT}$. The values of panel a (panel c) were computed at the biases $V_m$ that maximize the ratio $I_{RT} / I_{0K}$ ($I_{0} / I_{RT}$).
   The corresponding differences from resonance $V_m - 2 \varepsilon_0 / e$ are presented in panels b and d, respectively.} 
  \label{fig:reversal-very-small-e0}
\end{figure*}
\begin{figure}[htb]
  \centerline{\includegraphics[width=0.22\textwidth,height=0.42\textwidth,angle=-90]{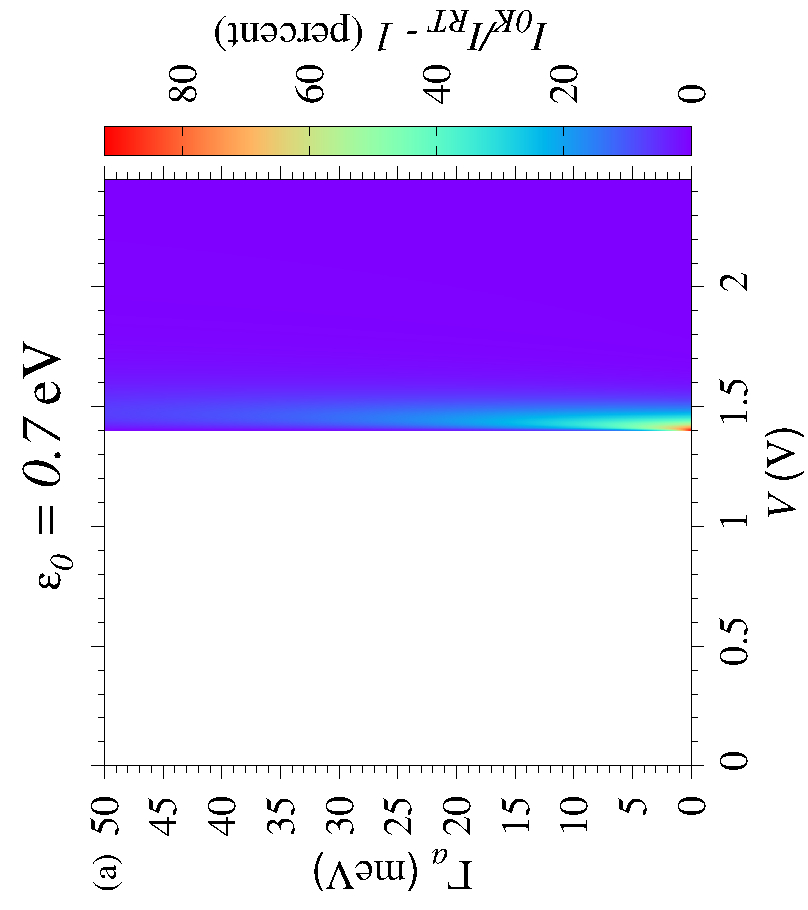}}
  \centerline{\includegraphics[width=0.22\textwidth,height=0.42\textwidth,angle=-90]{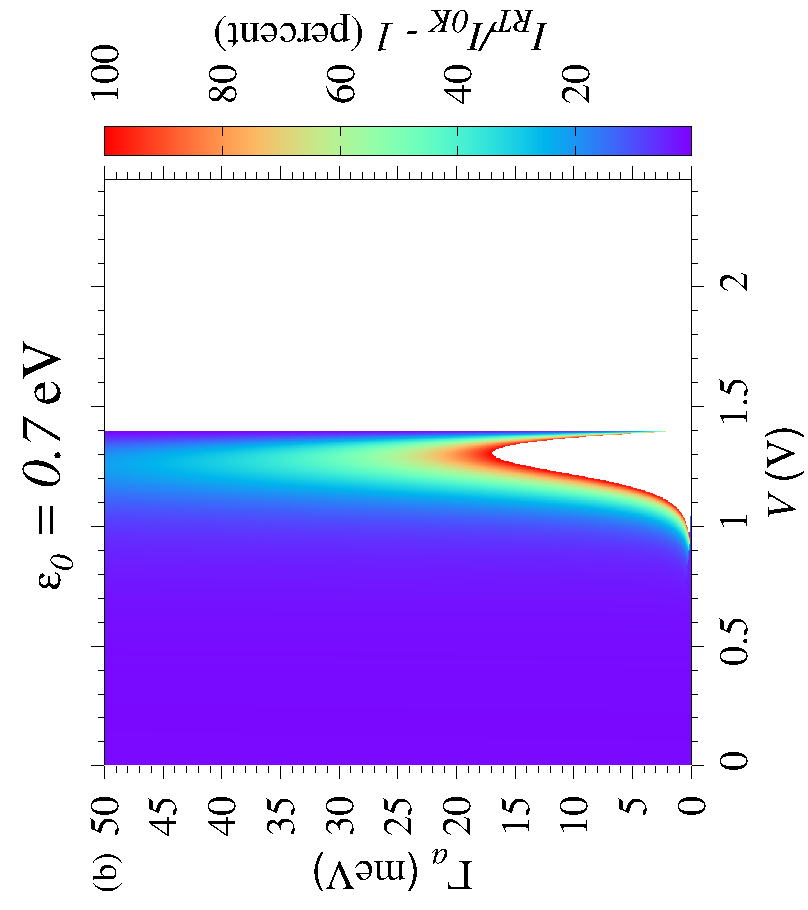}}
  \centerline{\includegraphics[width=0.22\textwidth,height=0.42\textwidth,angle=-90]{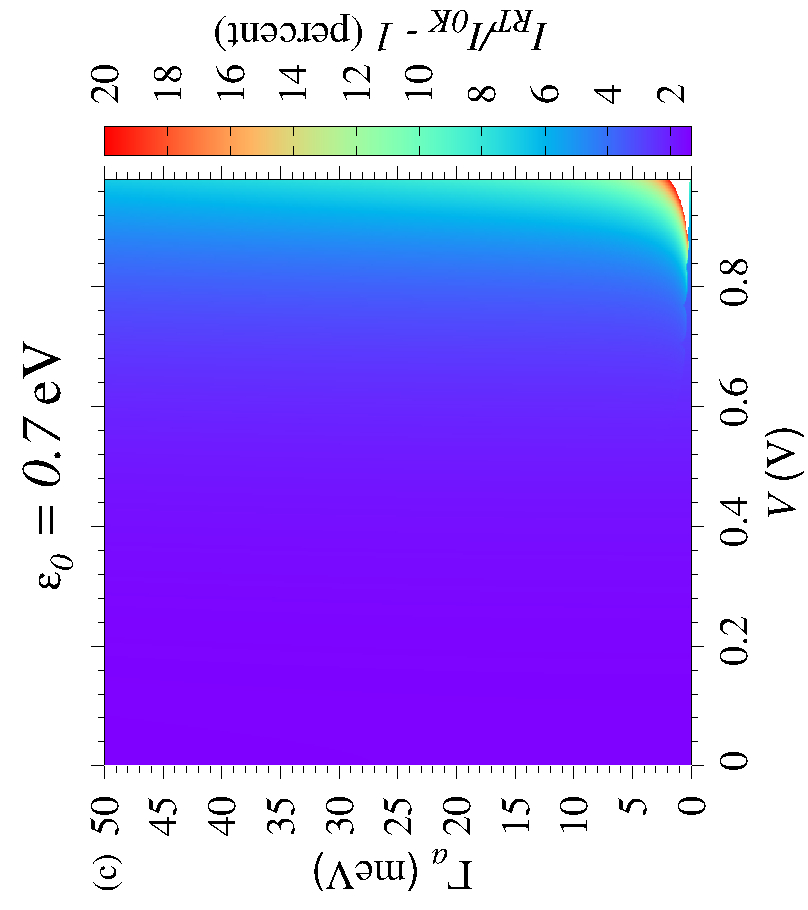}}
  \centerline{\includegraphics[width=0.22\textwidth,height=0.42\textwidth,angle=-90]{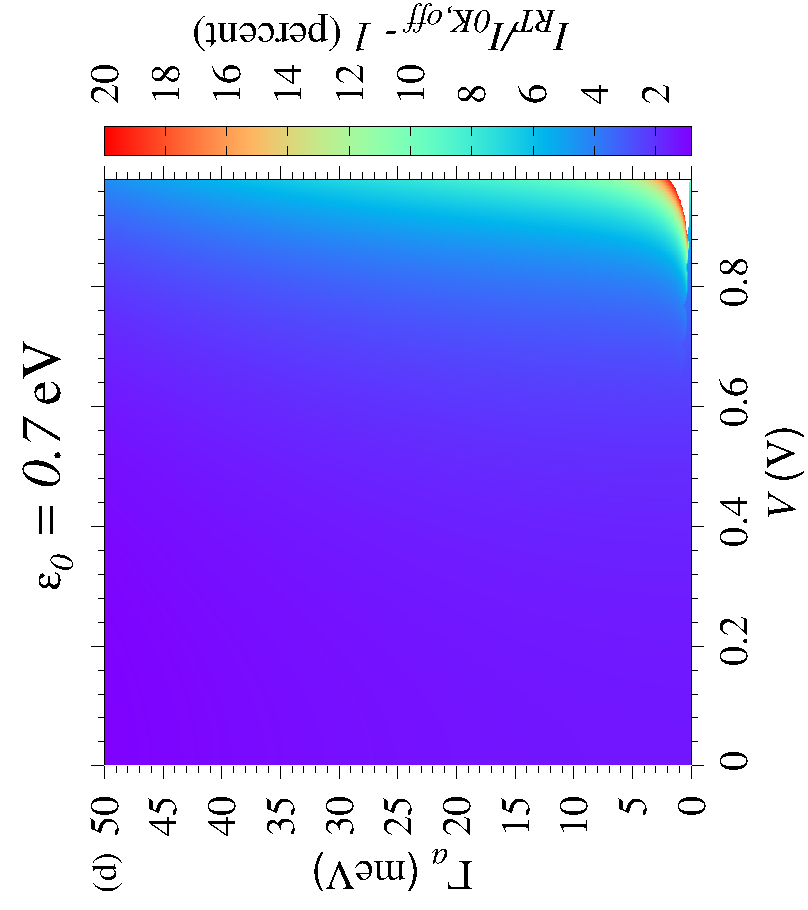}}
  \caption{The colored regions in the plane ($V, \Gamma_a$) depict situations where,
    at the fixed value of the MO energy offset indicated ($\varepsilon_0 = 0.7$\,eV), 
    the current $I_{0K}$ computed at $T=0$ using eqn~(\ref{eq-I0K})
    is larger ($\vert eV\vert > 2 \left\vert\varepsilon_0 \right\vert$, panel a) or smaller ($\vert eV\vert < 2 \left\vert\varepsilon_0 \right\vert$, panel b)
    than the exact current $I_{RT}$ computed from eqn~(\ref{eq-Iexact}) at room temperature ($T = 298.15$\,K).
    For parameter values compatible with eqn~(\ref{eq-1.4}) and (\ref{eq-Gamma-vs-e0}),
    the current $I_{0K,off}$ computed using eqn~(\ref{eq-I0Koff}) is very accurate (panel d);
    it is as accurate as $I_{0K}$ (panel c). Relative deviations (shown only when not exceeding 100\%) are indicated in the color box.
    To facilitate comparison between $I_{0K,off}$ and $I_{0K}$, abscissas in panel c depicting $I_{0K}$ are restricted to those in panel d.
    Notice that the $z$-range in panels (c) and (d) is different from panel b.}
  \label{fig:errors-e0-0.7-si}
\end{figure}
\begin{figure}[htb]
  \centerline{\includegraphics[width=0.22\textwidth,height=0.42\textwidth,angle=-90]{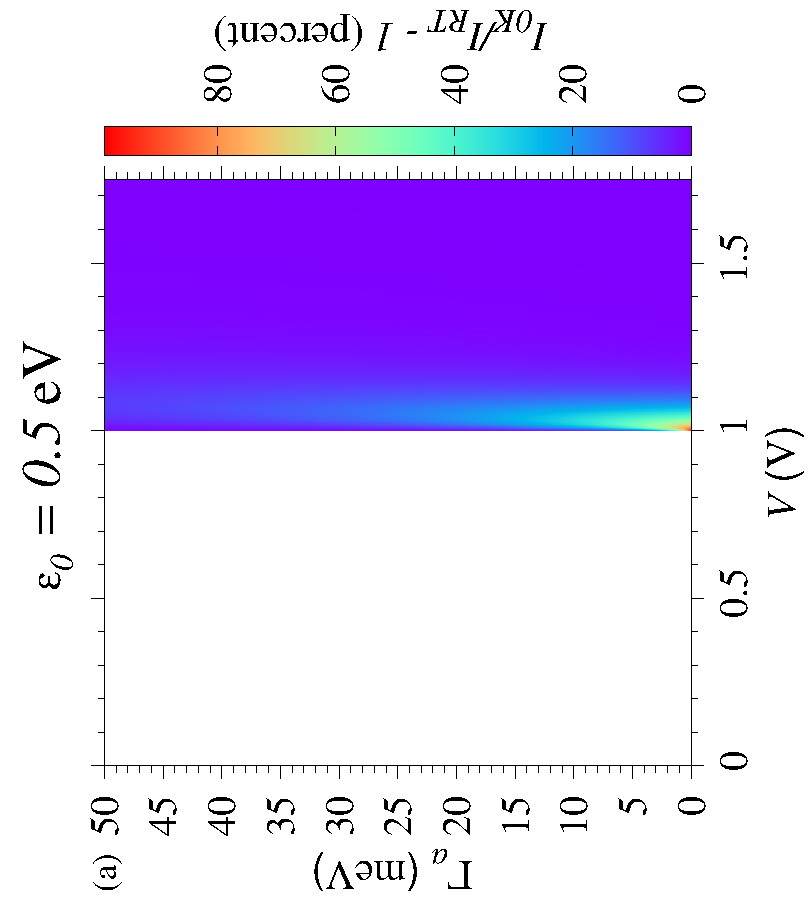}}
  \centerline{\includegraphics[width=0.22\textwidth,height=0.42\textwidth,angle=-90]{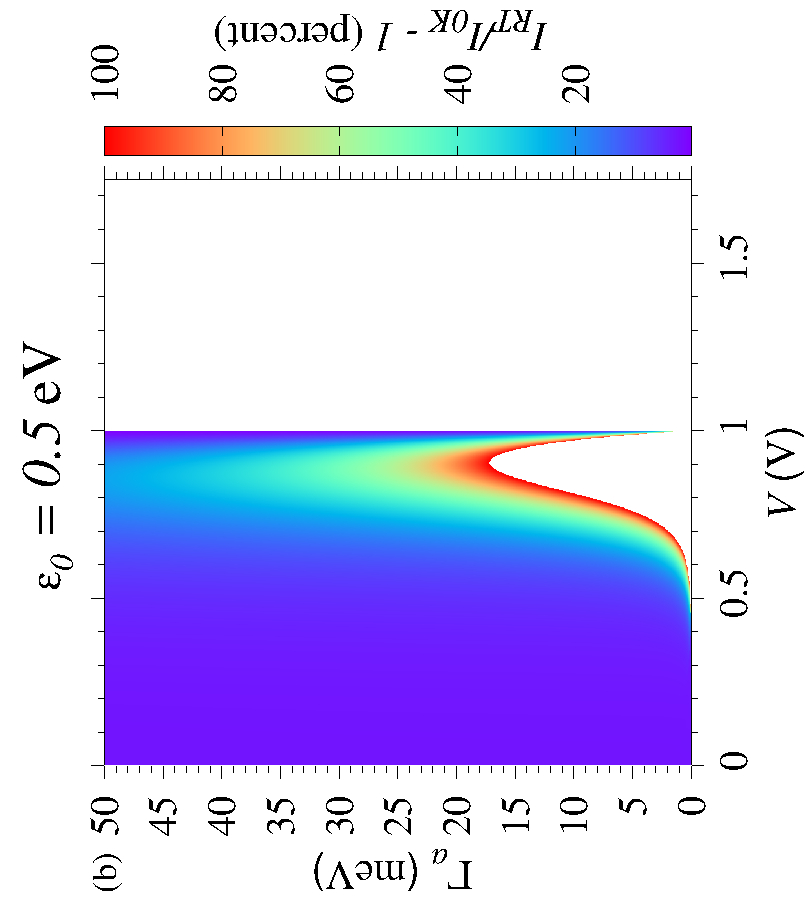}}
  \centerline{\includegraphics[width=0.22\textwidth,height=0.42\textwidth,angle=-90]{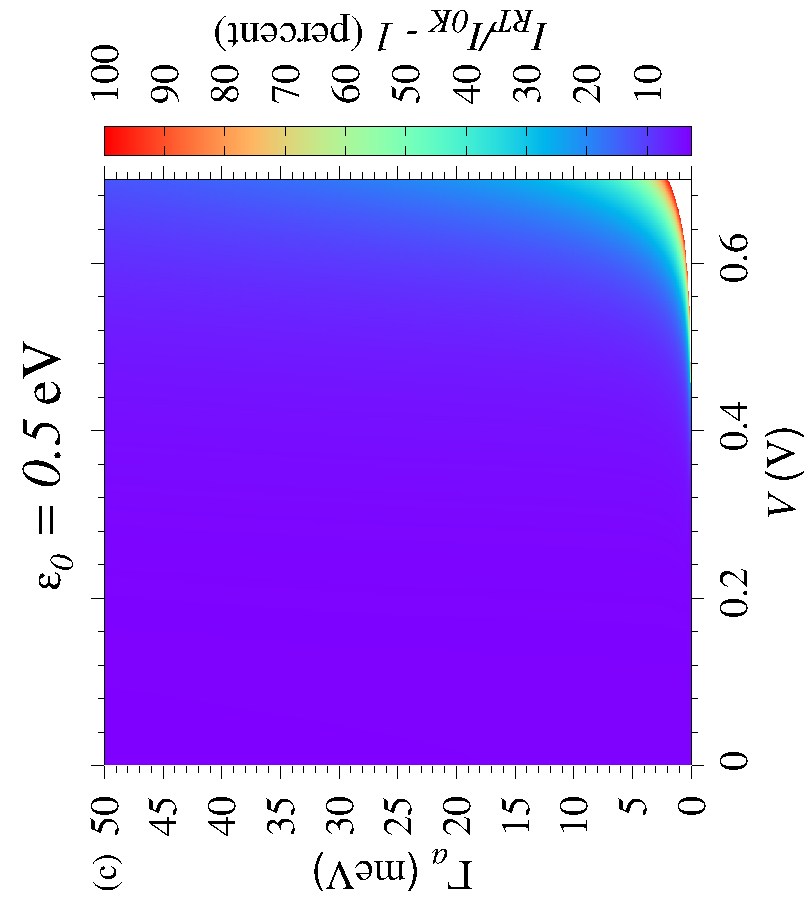}}
  \centerline{\includegraphics[width=0.22\textwidth,height=0.42\textwidth,angle=-90]{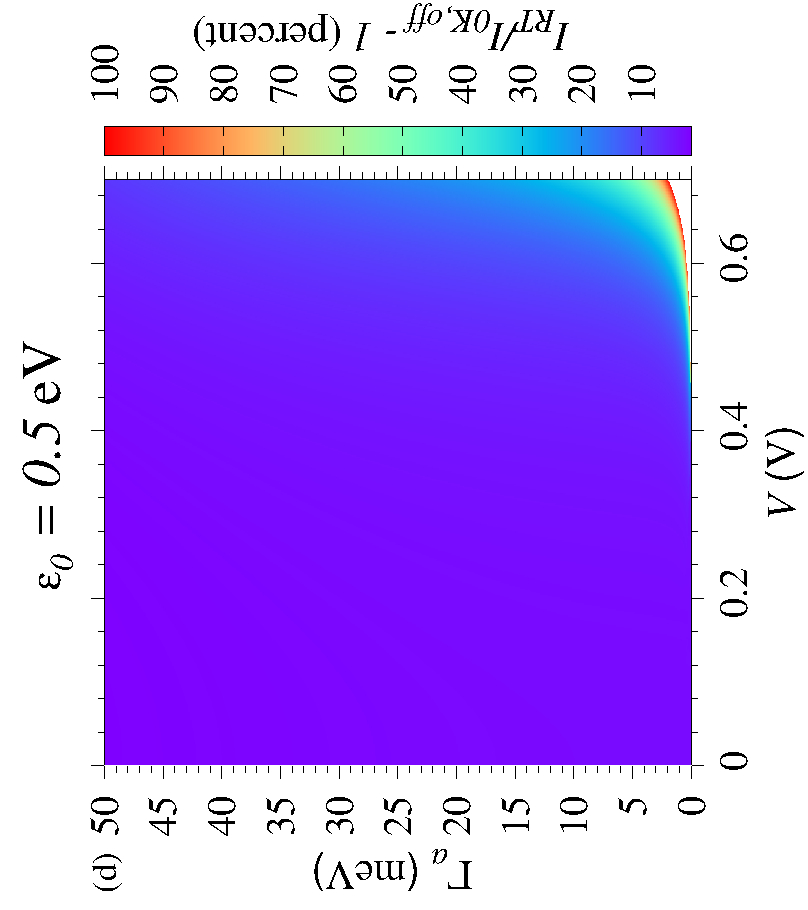}}
  \caption{The colored regions in the plane ($V, \Gamma_a$) depict situations where,
    at the fixed value of the MO energy offset indicated ($\varepsilon_0 = 0.5$\,eV), 
    the current $I_{0K}$ computed at $T=0$ using eqn~(\ref{eq-I0K})
    is larger ($\vert eV\vert > 2 \left\vert\varepsilon_0 \right\vert$, panel a) or smaller ($\vert eV\vert < 2 \left\vert\varepsilon_0 \right\vert$, panel b)
    than the exact current $I_{RT}$ computed from eqn~(\ref{eq-Iexact}) at room temperature ($T = 298.15$\,K).
    For parameter values compatible with eqn~(\ref{eq-1.4}) and (\ref{eq-Gamma-vs-e0}),
    the current $I_{0K,off}$ computed using eqn~(\ref{eq-I0Koff}) is very accurate (panel d);
    it is as accurate as $I_{0K}$ (panel c). Relative deviations (shown only when not exceeding 100\%) are indicated in the color box.
    To facilitate comparison between $I_{0K,off}$ and $I_{0K}$, abscissas in panel c depicting $I_{0K}$ are restricted to those in panel d.
    Notice that the $z$-range in panels (c) and (d) is different from panel b.}
  \label{fig:errors-e0-0.5-si}
\end{figure}
\begin{figure}[htb]
  \centerline{\includegraphics[width=0.22\textwidth,height=0.42\textwidth,angle=-90]{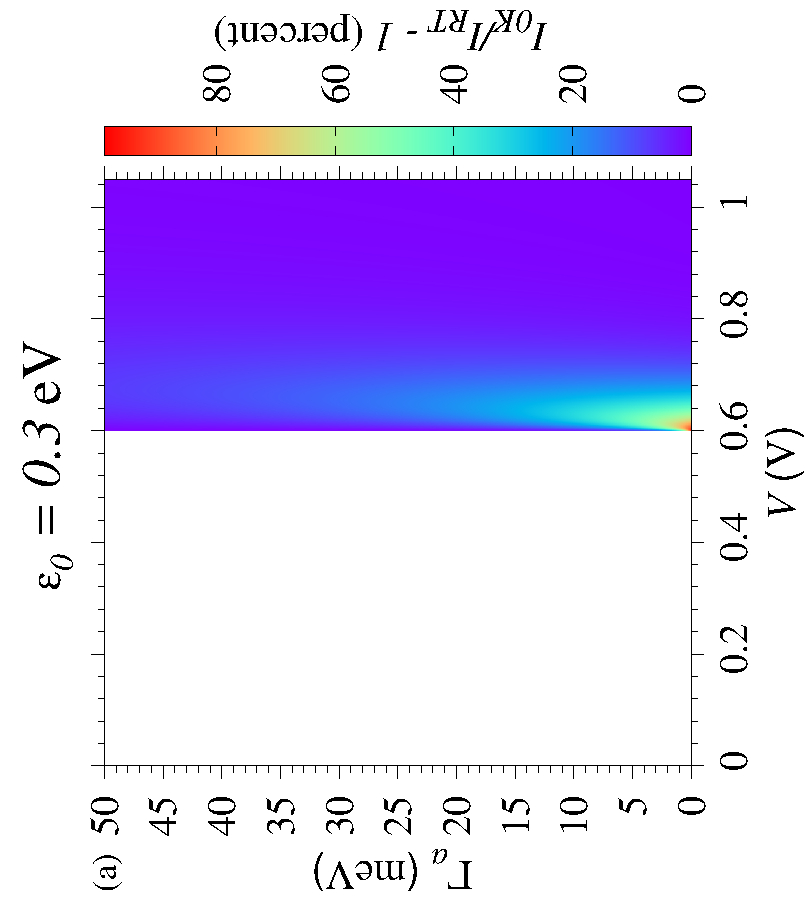}}
  \centerline{\includegraphics[width=0.22\textwidth,height=0.42\textwidth,angle=-90]{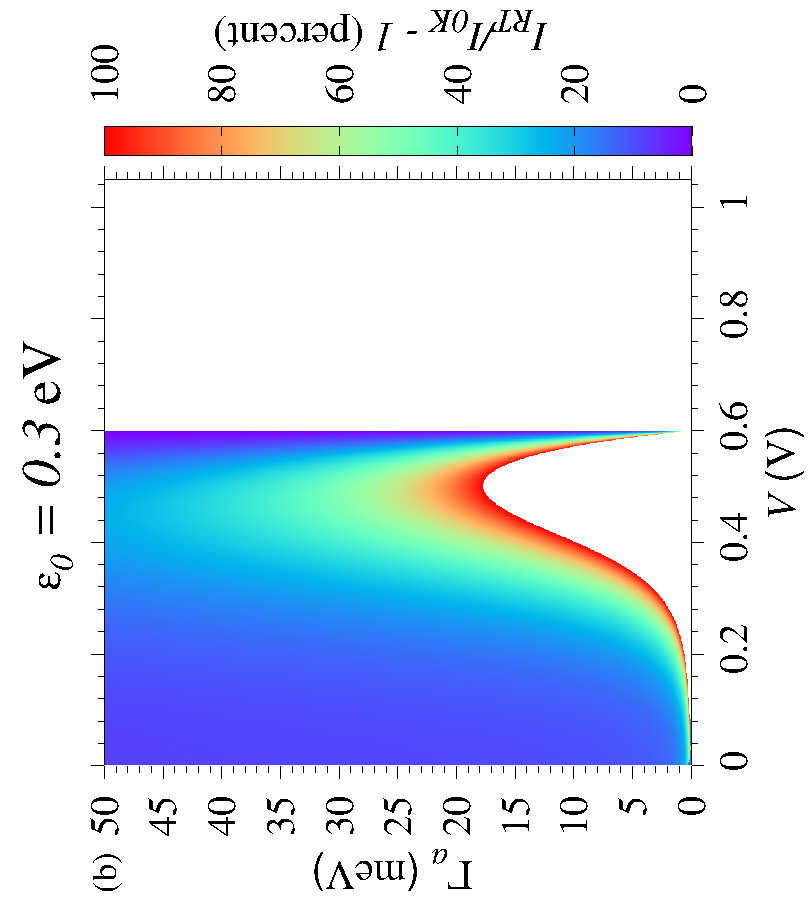}}
  \centerline{\includegraphics[width=0.22\textwidth,height=0.42\textwidth,angle=-90]{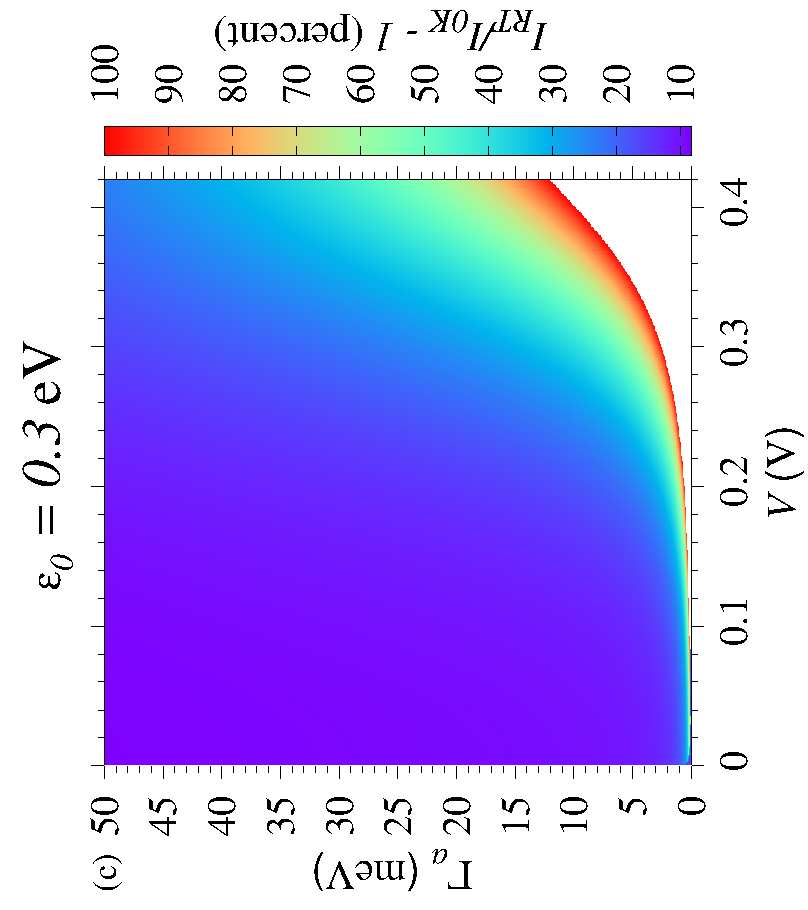}}
  \centerline{\includegraphics[width=0.22\textwidth,height=0.42\textwidth,angle=-90]{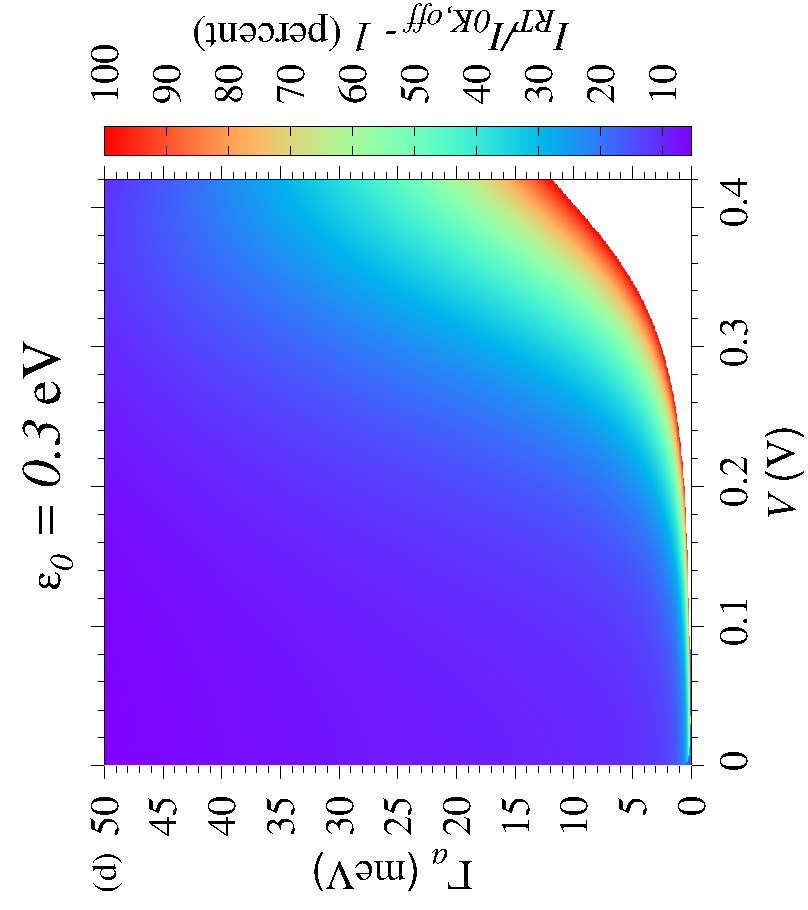}}
  \caption{The colored regions in the plane ($V, \Gamma_a$) depict situations where,
    at the fixed value of the MO energy offset indicated ($\varepsilon_0 = 0.3$\,eV), 
    the current $I_{0K}$ computed at $T=0$ using eqn~(\ref{eq-I0K})
    is larger ($\vert eV\vert > 2 \left\vert\varepsilon_0 \right\vert$, panel a) or smaller ($\vert eV\vert < 2 \left\vert\varepsilon_0 \right\vert$, panel b)
    than the exact current $I_{RT}$ computed from eqn~(\ref{eq-Iexact}) at room temperature ($T = 298.15$\,K).
    For parameter values compatible with eqn~(\ref{eq-1.4}) and (\ref{eq-Gamma-vs-e0}),
    the current $I_{0K,off}$ computed using eqn~(\ref{eq-I0Koff}) (panel d)
    is as accurate as $I_{0K}$ (panel c). Relative deviations (shown only when not exceeding 100\%) are indicated in the color box.
    To facilitate comparison between $I_{0K,off}$ and $I_{0K}$, abscissas in panel c depicting $I_{0K}$ are restricted to those in panel d.
    Notice that the $z$-range in panels (c) and (d) is different from panel b.}
  \label{fig:errors-e0-0.3-si}
\end{figure}
\begin{figure}[htb]
  \centerline{\includegraphics[width=0.22\textwidth,height=0.42\textwidth,angle=-90]{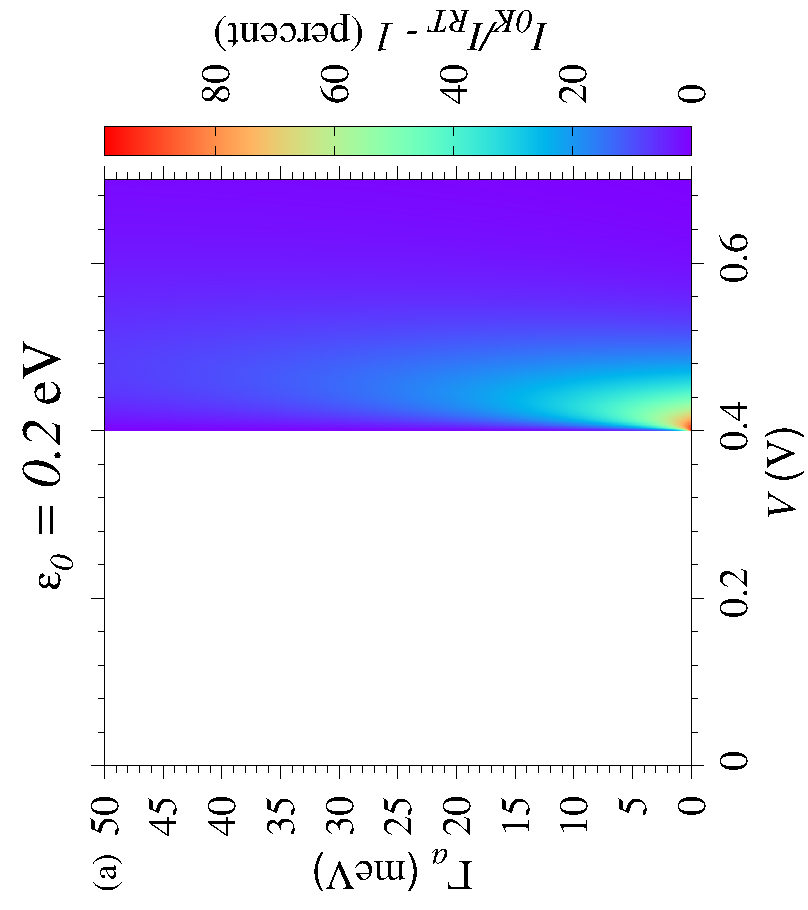}}
  \centerline{\includegraphics[width=0.22\textwidth,height=0.42\textwidth,angle=-90]{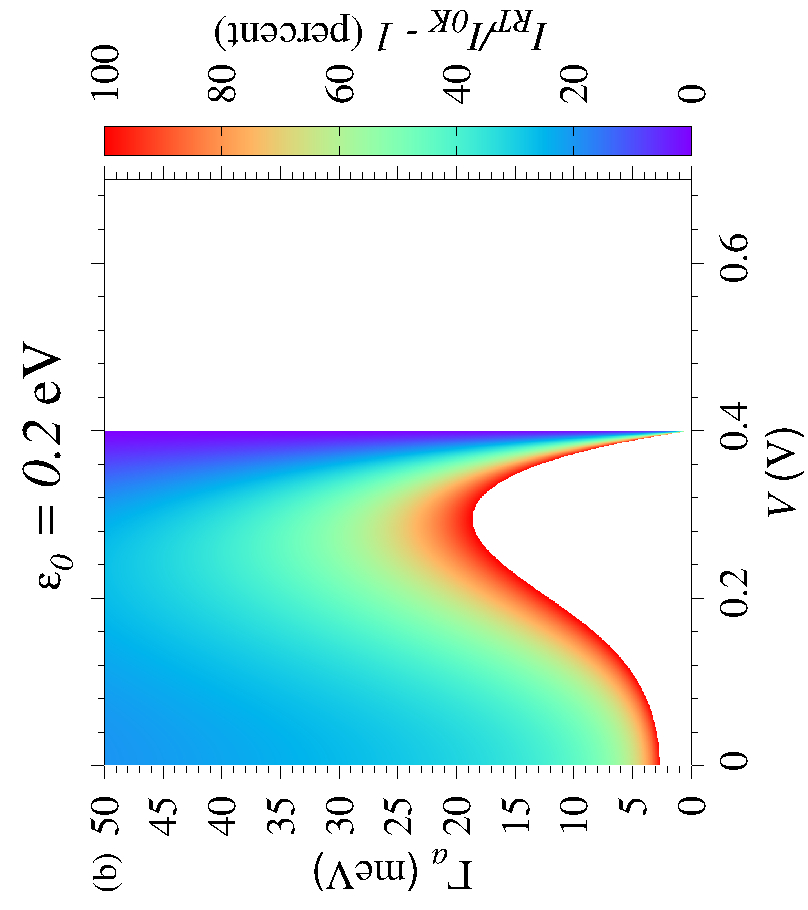}}
  \centerline{\includegraphics[width=0.22\textwidth,height=0.42\textwidth,angle=-90]{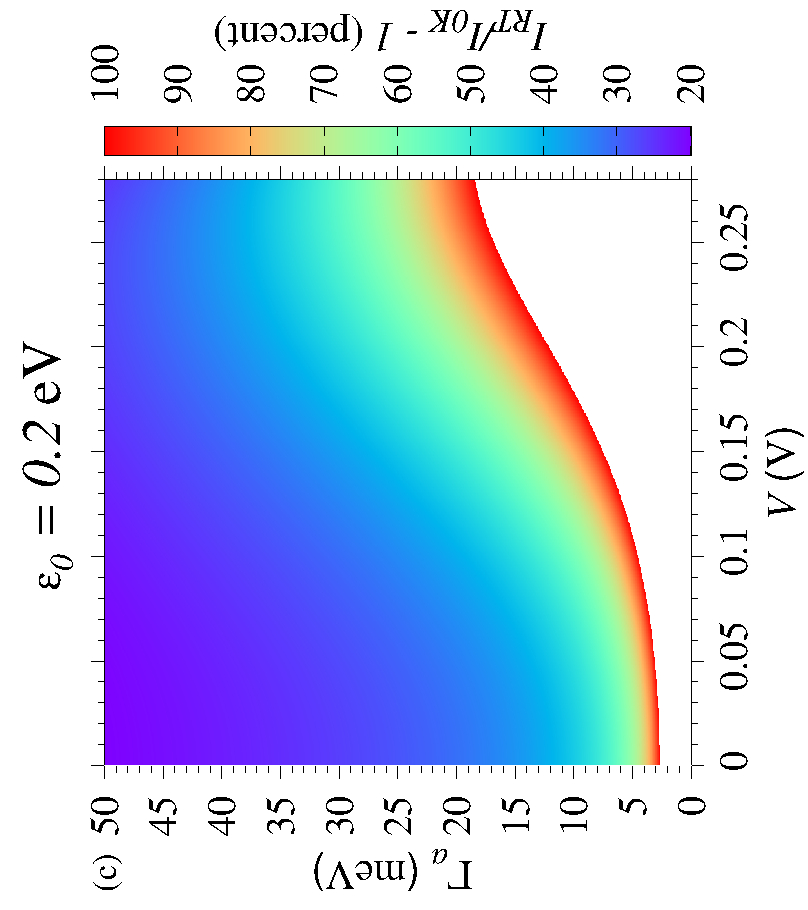}}
  \centerline{\includegraphics[width=0.22\textwidth,height=0.42\textwidth,angle=-90]{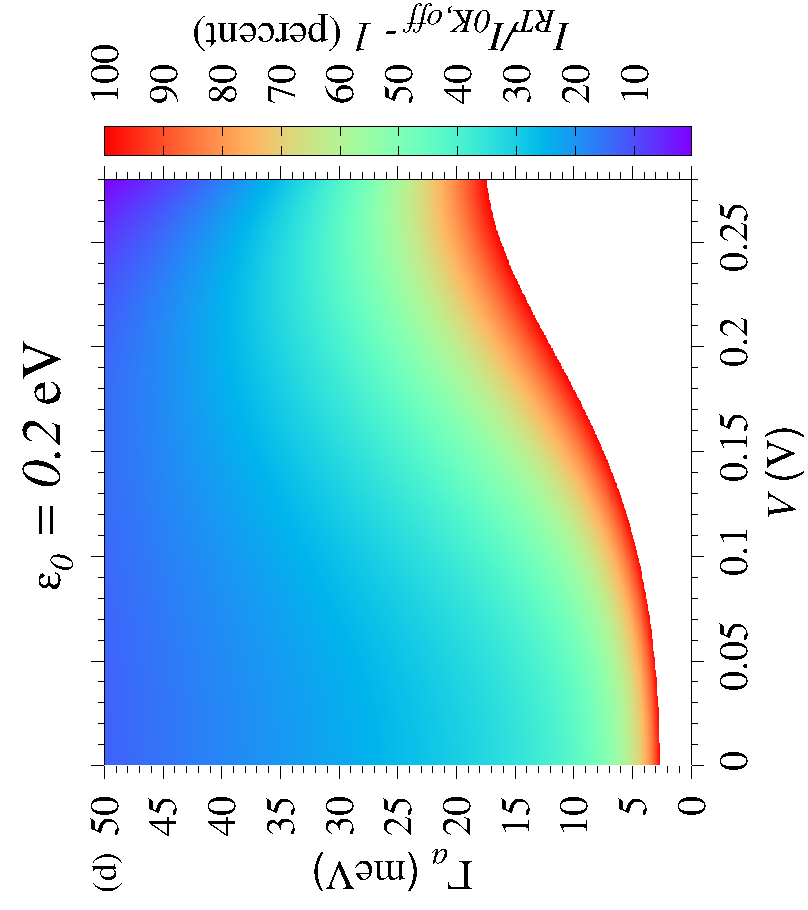}}
  \caption{The colored regions in the plane ($V, \Gamma_a$) depict situations where,
    at the fixed value of the MO energy offset indicated ($\varepsilon_0 = 0.2$\,eV), 
    the current $I_{0K}$ computed at $T=0$ using eqn~(\ref{eq-I0K})
    is larger ($\vert eV\vert > 2 \left\vert\varepsilon_0 \right\vert$, panel a) or smaller ($\vert eV\vert < 2 \left\vert\varepsilon_0 \right\vert$, panel b)
    than the exact current $I_{RT}$ computed from eqn~(\ref{eq-Iexact}) at room temperature ($T = 298.15$\,K).
    For parameter values compatible with eqn~(\ref{eq-1.4}) and (\ref{eq-Gamma-vs-e0}),
    the current $I_{0K,off}$ computed using eqn~(\ref{eq-I0Koff}) (panel d)
    is as accurate as $I_{0K}$ (panel c). Relative deviations (shown only when not exceeding 100\%) are indicated in the color box.
    To facilitate comparison between $I_{0K,off}$ and $I_{0K}$, abscissas in panel c depicting $I_{0K}$ are restricted to those in panel d.
    Notice that the $z$-range in panels (c) and (d) is different from panel b.}
  \label{fig:errors-e0-0.2-si}
\end{figure}
\begin{figure}[htb]
  \centerline{\includegraphics[width=0.22\textwidth,height=0.42\textwidth,angle=-90]{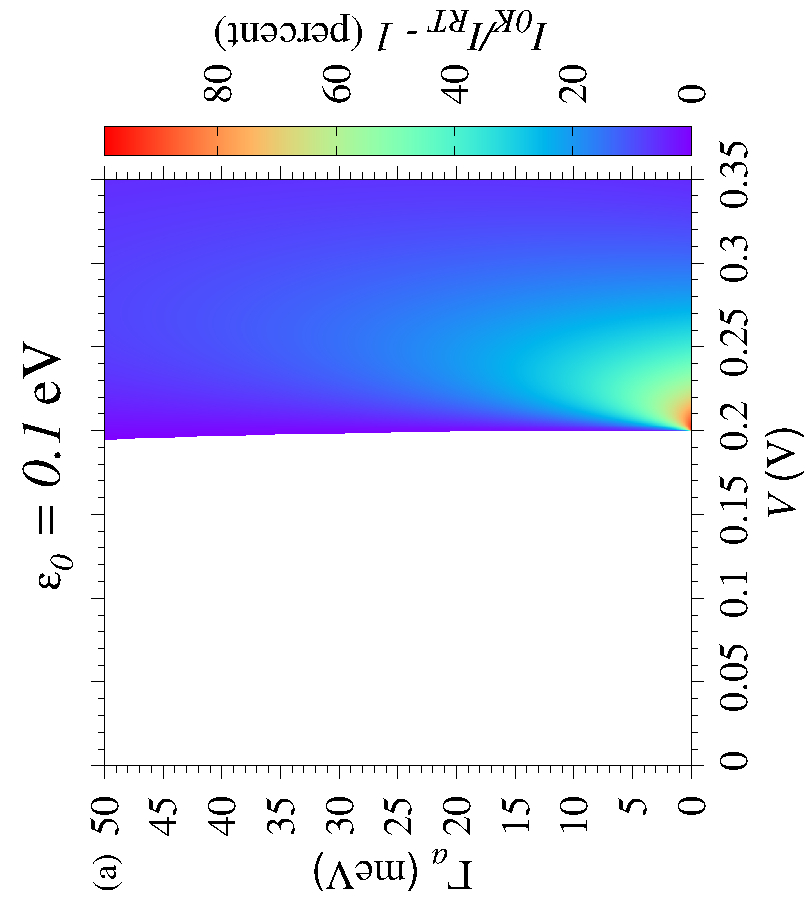}}
  \centerline{\includegraphics[width=0.22\textwidth,height=0.42\textwidth,angle=-90]{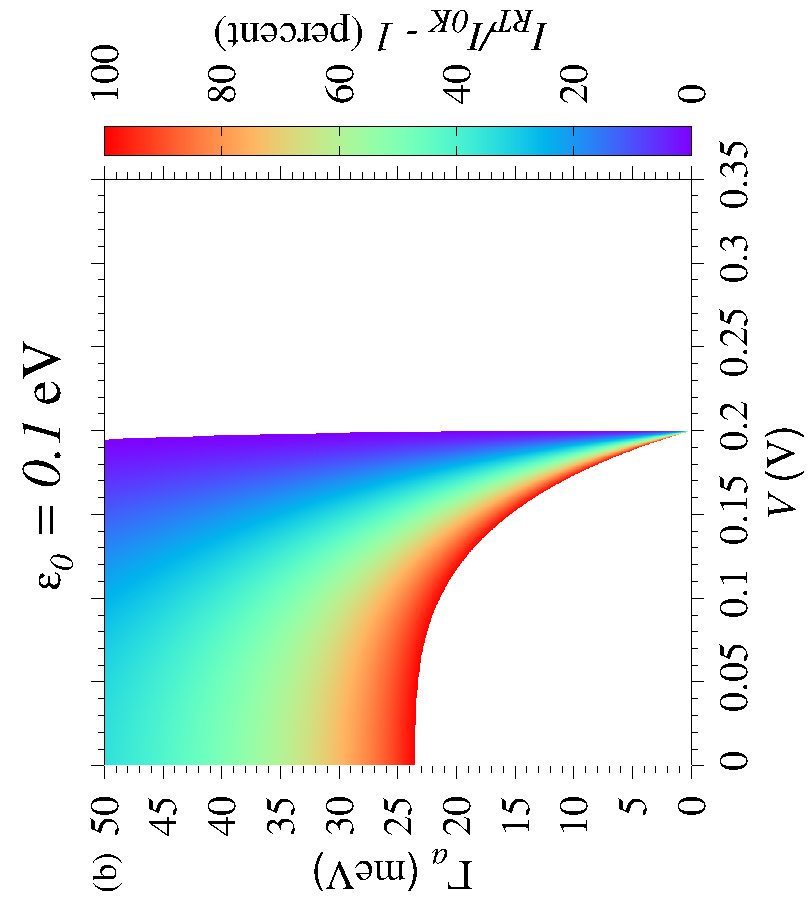}}
  \centerline{\includegraphics[width=0.22\textwidth,height=0.42\textwidth,angle=-90]{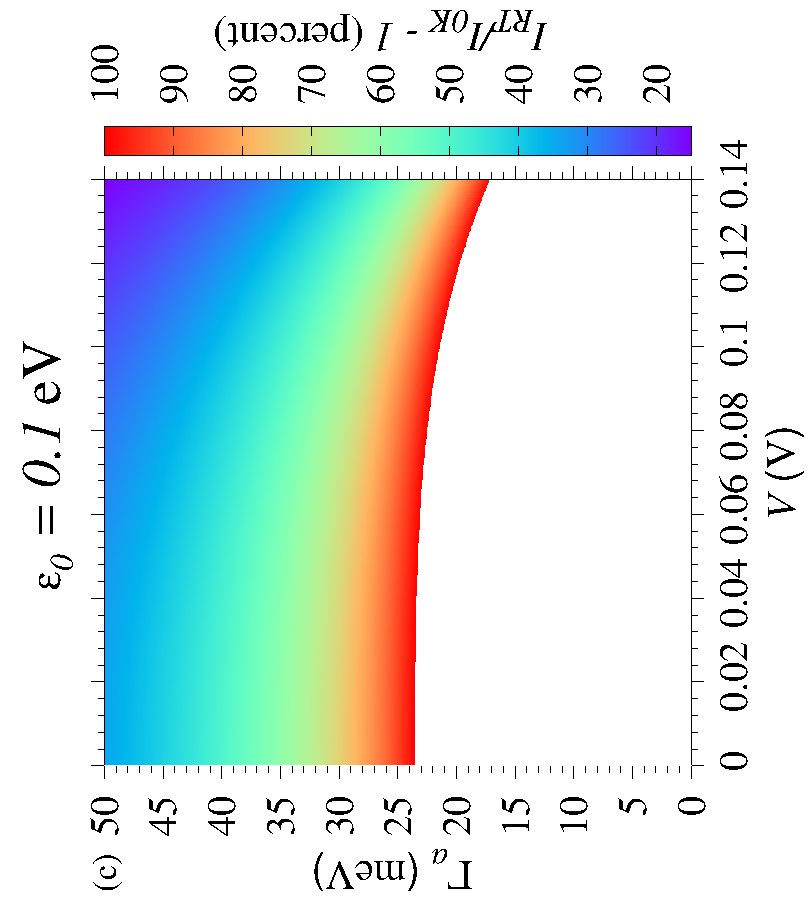}}
  \centerline{\includegraphics[width=0.22\textwidth,height=0.42\textwidth,angle=-90]{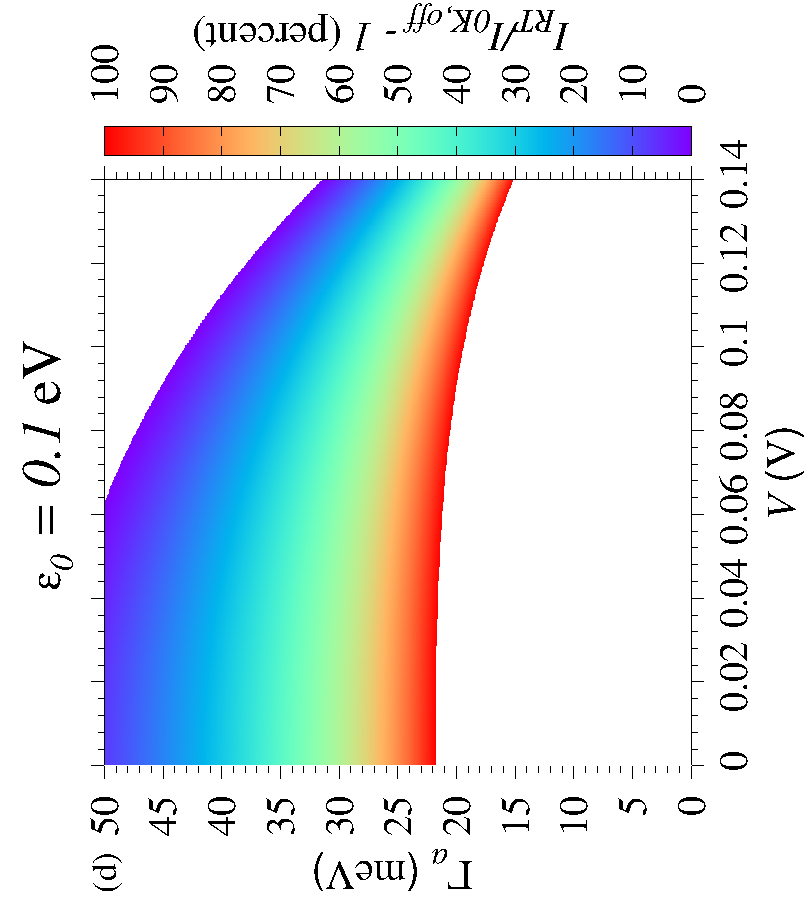}}
  \caption{The colored regions in the plane ($V, \Gamma_a$) depict situations where,
    at the fixed value of the MO energy offset indicated ($\varepsilon_0 = 0.1$\,eV), 
    the current $I_{0K}$ computed at $T=0$ using eqn~(\ref{eq-I0K})
    is larger ($\vert eV\vert > 2 \left\vert\varepsilon_0 \right\vert$, panel a) or smaller ($\vert eV\vert < 2 \left\vert\varepsilon_0 \right\vert$, panel b)
    than the exact current $I_{RT}$ computed from eqn~(\ref{eq-Iexact}) at room temperature ($T = 298.15$\,K).
    For situations violating eqn~(\ref{eq-Gamma-vs-e0}),
    the current $I_{0K,off}$ computed using eqn~(\ref{eq-I0Koff}) (panel d) stronger departs from $I_{RT}$ than $I_{0K}$ (panel d).
    Relative deviations (shown only when not exceeding 100\%) are indicated in the color box.
    To facilitate comparison between $I_{0K,off}$ and $I_{0K}$, abscissas in panel c depicting $I_{0K}$ are restricted to those in panel d.
    Notice that the $z$-range in panels (c) and (d) is different from panel b.}
  \label{fig:errors-e0-0.1-si}
\end{figure}
\begin{figure*}[htb]
  \centerline{\includegraphics[width=0.22\textwidth,height=0.42\textwidth,angle=-90]{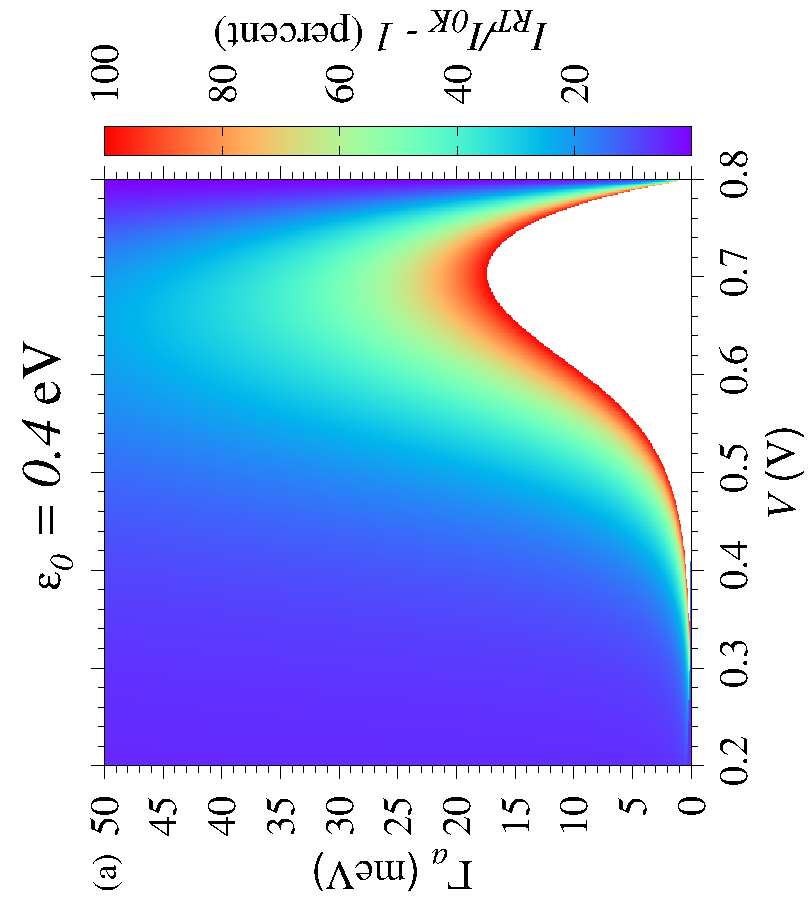}
    \includegraphics[width=0.22\textwidth,height=0.42\textwidth,angle=-90]{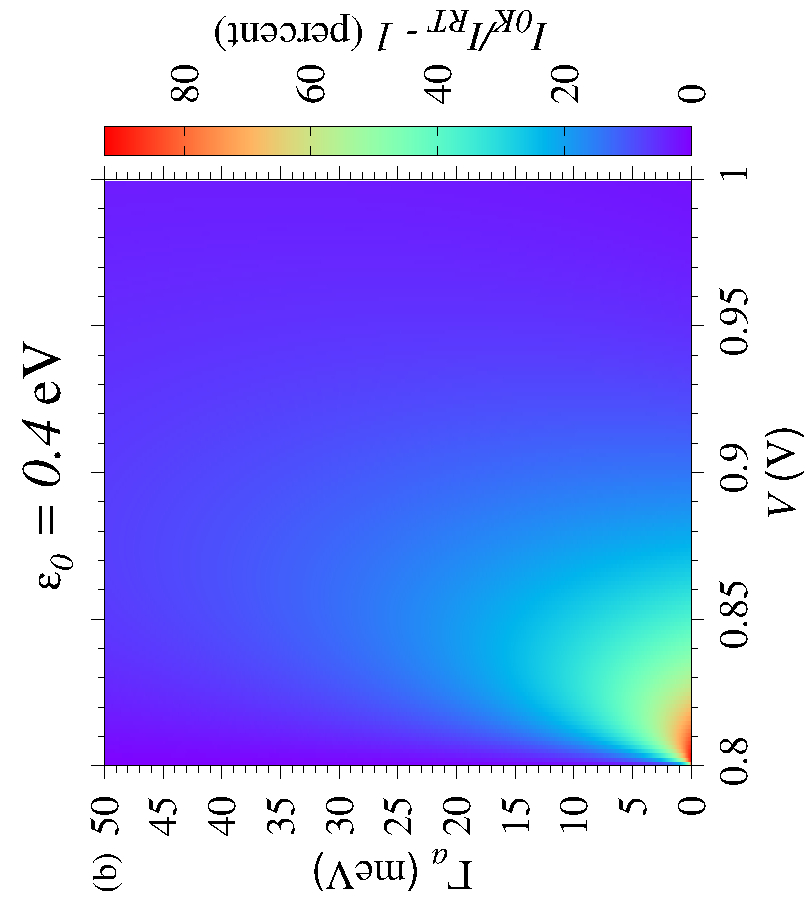}}
    \centerline{\includegraphics[width=0.22\textwidth,height=0.42\textwidth,angle=-90]{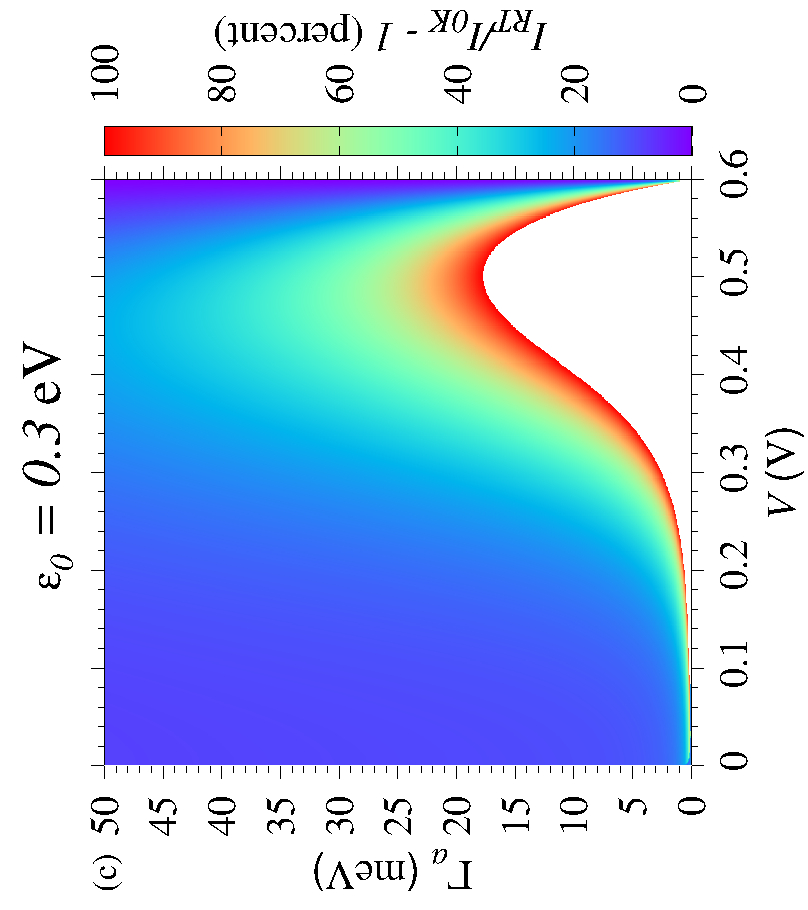}
    \includegraphics[width=0.22\textwidth,height=0.42\textwidth,angle=-90]{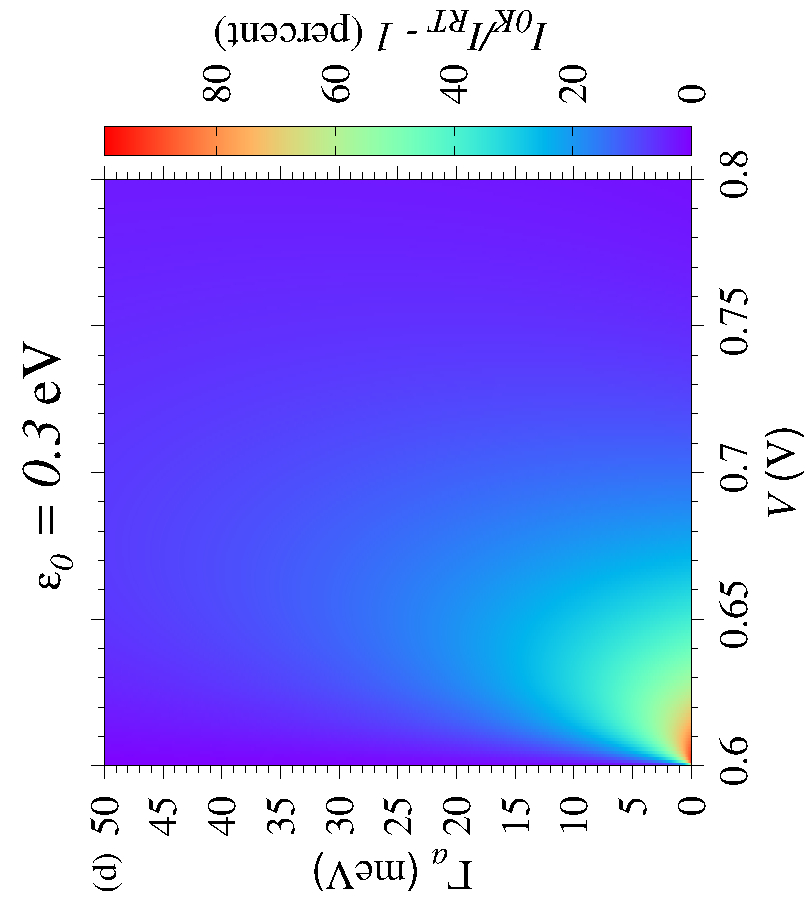}}
  \centerline{\includegraphics[width=0.22\textwidth,height=0.42\textwidth,angle=-90]{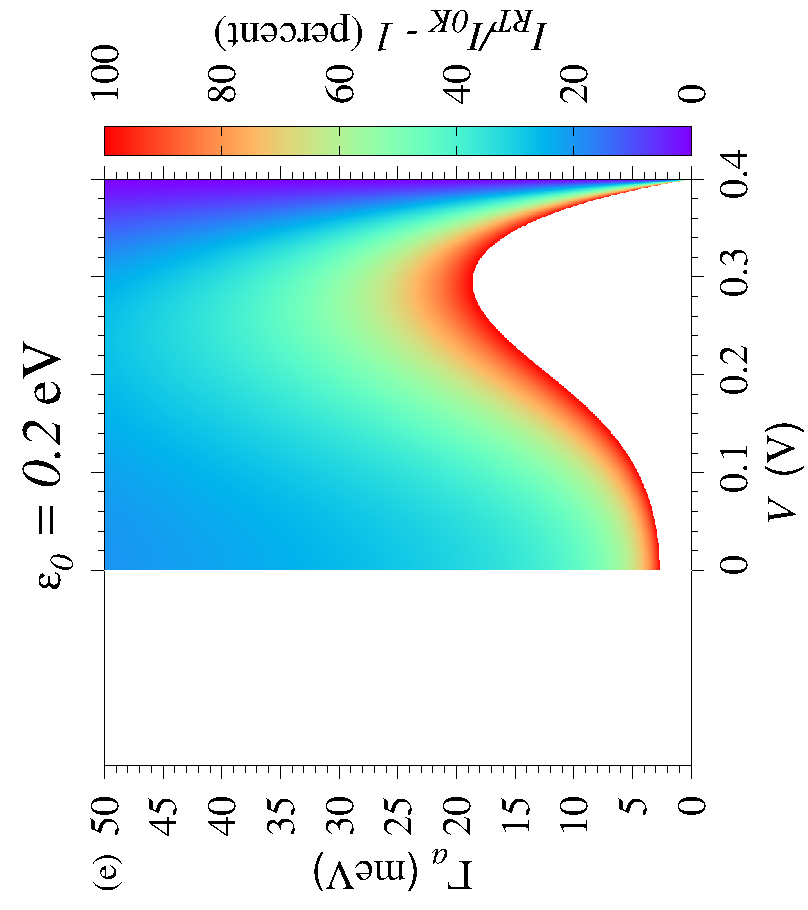}
    \includegraphics[width=0.22\textwidth,height=0.42\textwidth,angle=-90]{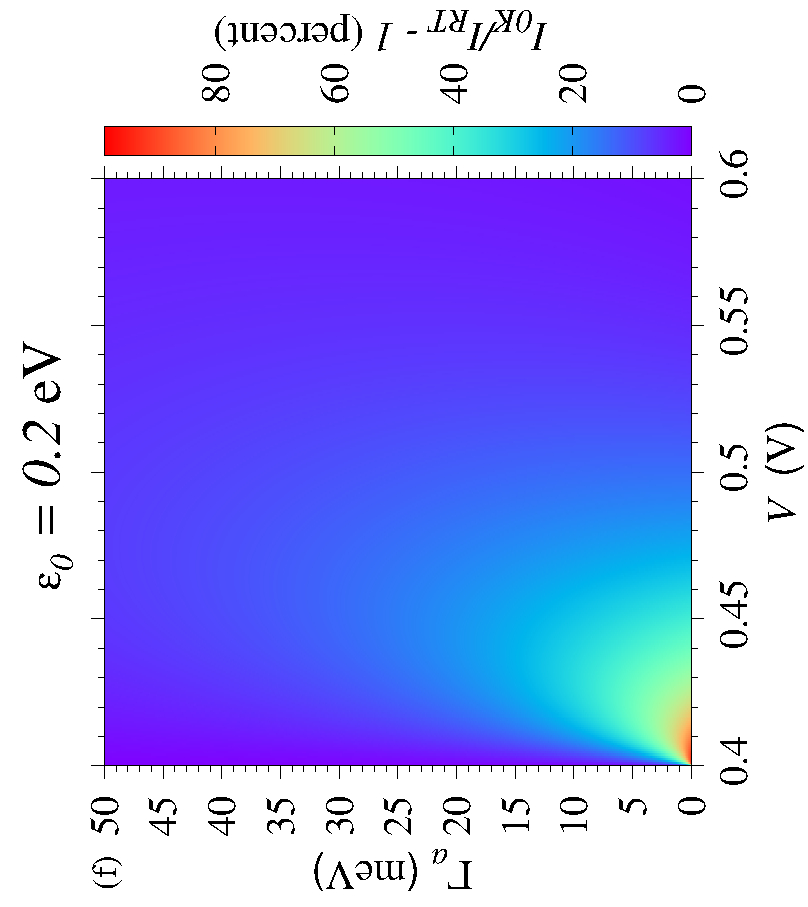}}
    \centerline{\includegraphics[width=0.22\textwidth,height=0.42\textwidth,angle=-90]{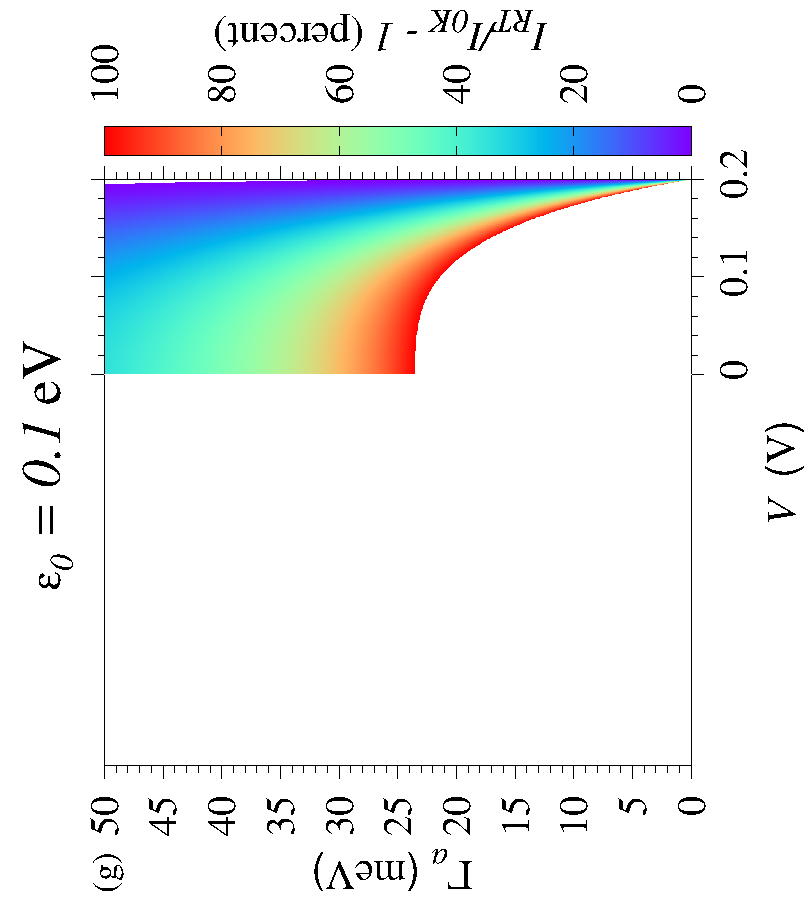}
    \includegraphics[width=0.22\textwidth,height=0.42\textwidth,angle=-90]{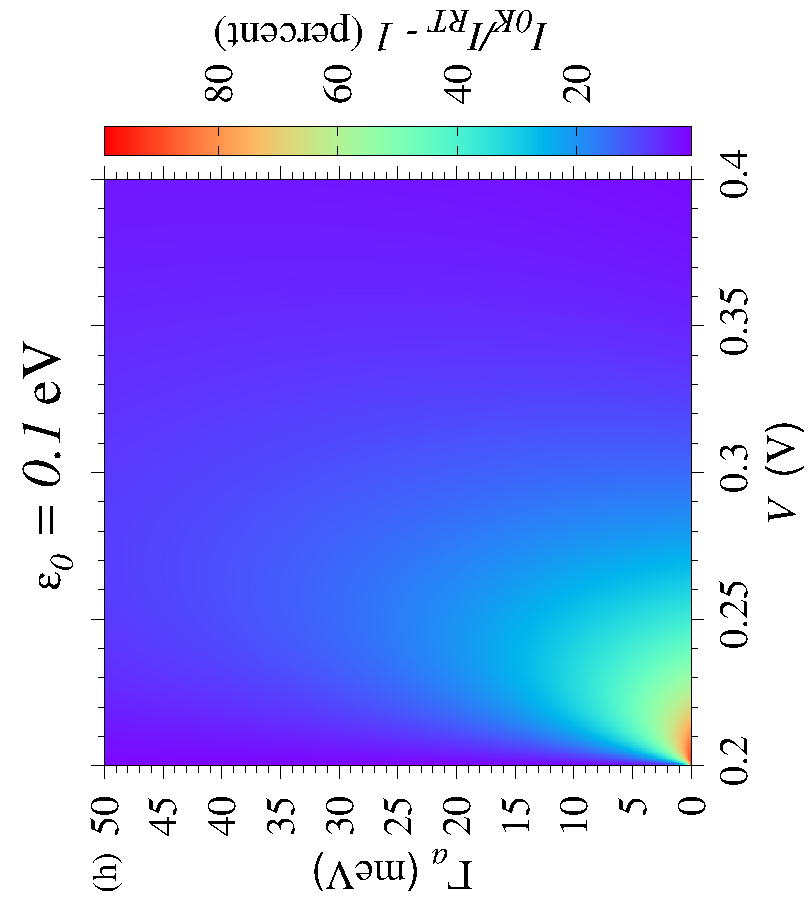}}
    \caption{Results for lower biases illustrating the current enhancement below resonance
      ($\vert e V\vert < 2 \left\vert\varepsilon_0 \right\vert$, left panels)
      and current reduction above resonance
      ($\vert e V\vert > 2 \left\vert\varepsilon_0 \right\vert$, right panels).
      As $\varepsilon_0$ decreases (downwards), the white (empty) region (wherein the relative deviations exceed 100\%)
      in the left panels extends upwards to larger $\Gamma_a$ and comprises a broader bias range.
      Notice that all rightmost (leftmost) positions of the left (right) panels are aligned to resonance.}
  \label{fig:errors-e0-aligned-si}
\end{figure*}
\begin{figure*}[htb]
  \centerline{\includegraphics[width=0.22\textwidth,height=0.42\textwidth,angle=-90]{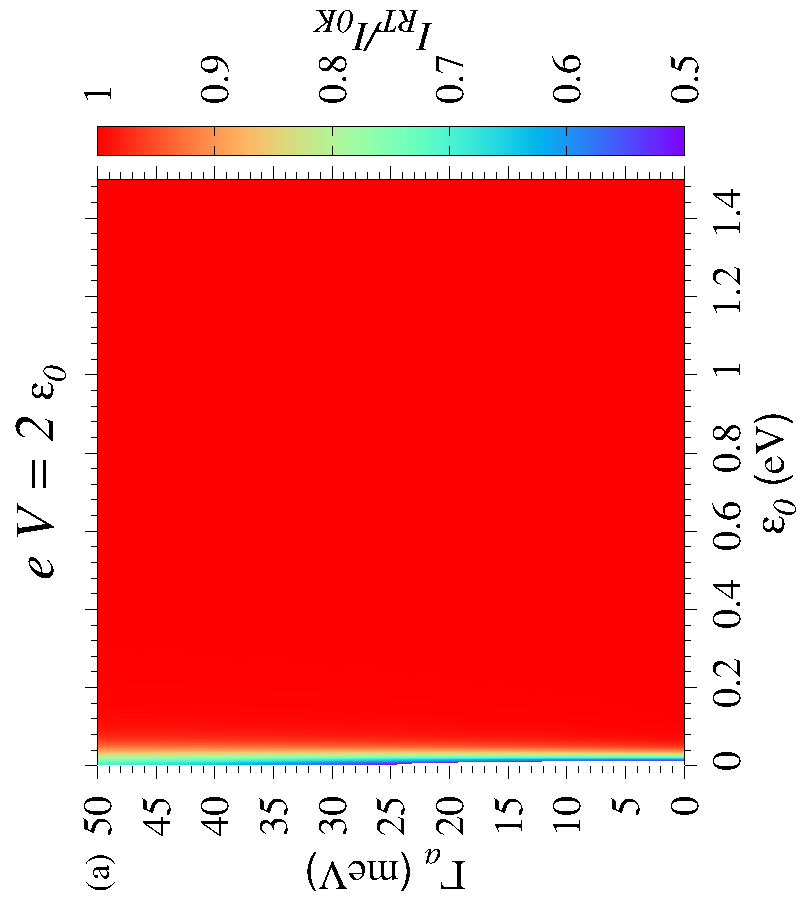}
    \includegraphics[width=0.22\textwidth,height=0.42\textwidth,angle=-90]{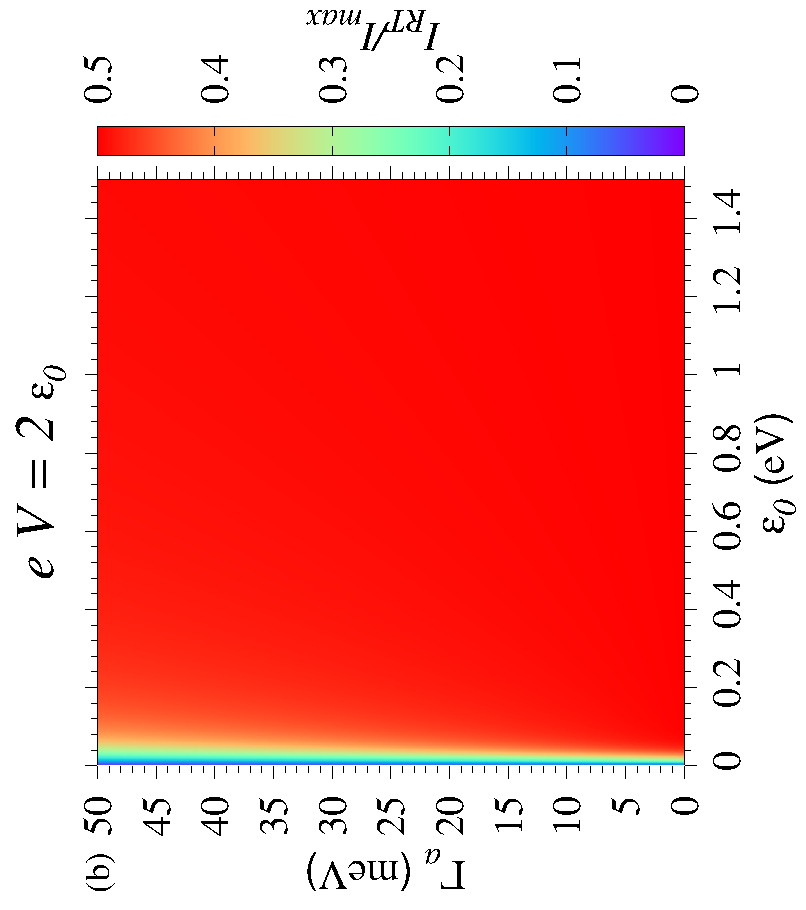}}
  \centerline{\includegraphics[width=0.45\textwidth,angle=0]{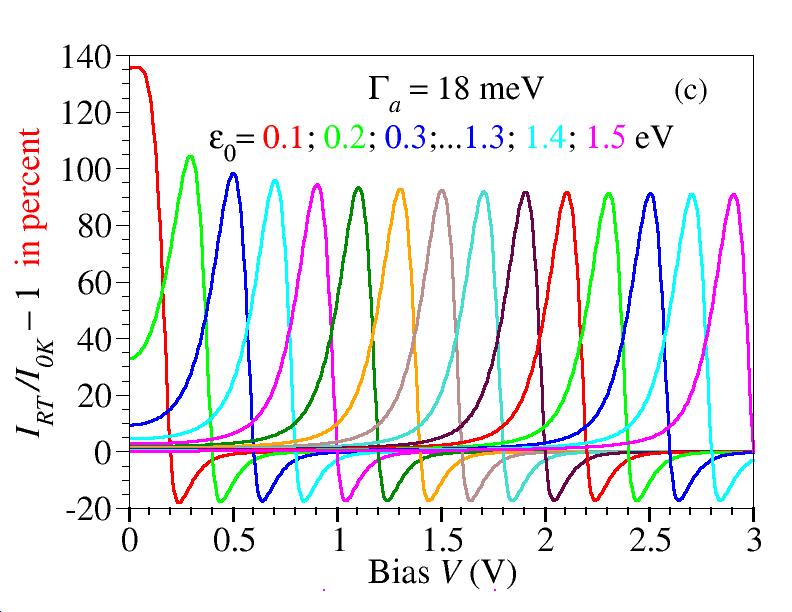}
    \includegraphics[width=0.45\textwidth,angle=0]{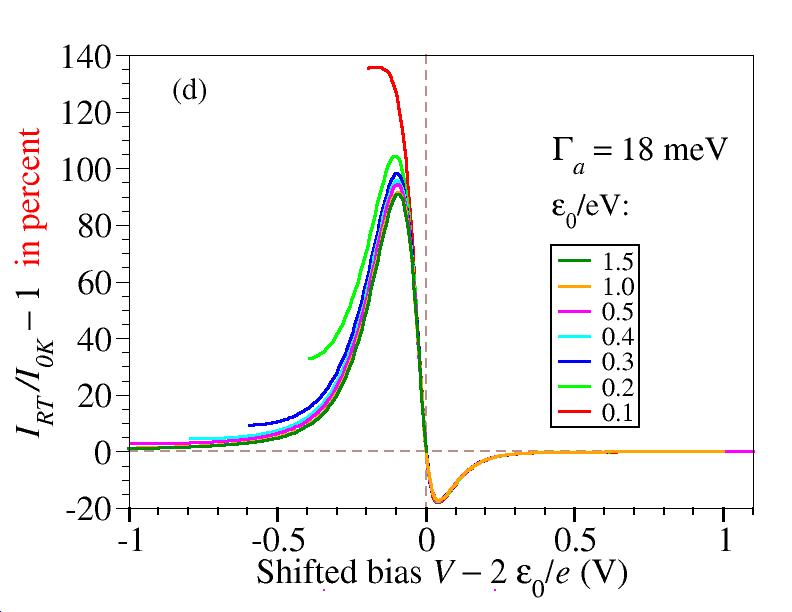}}
  \caption{(a,b) Strictly on resonance, the temperature impact on the current is negligible.
    (c,d) Except for small values of the MO energy offset $\varepsilon_0$, the thermal enhancement of the current occurs around resonance
    and is quite insensitive to $\varepsilon_0$.}
  \label{fig:shift}
\end{figure*}
\begin{figure}[htb]
  \centerline{\includegraphics[width=0.22\textwidth,height=0.42\textwidth,angle=-90]{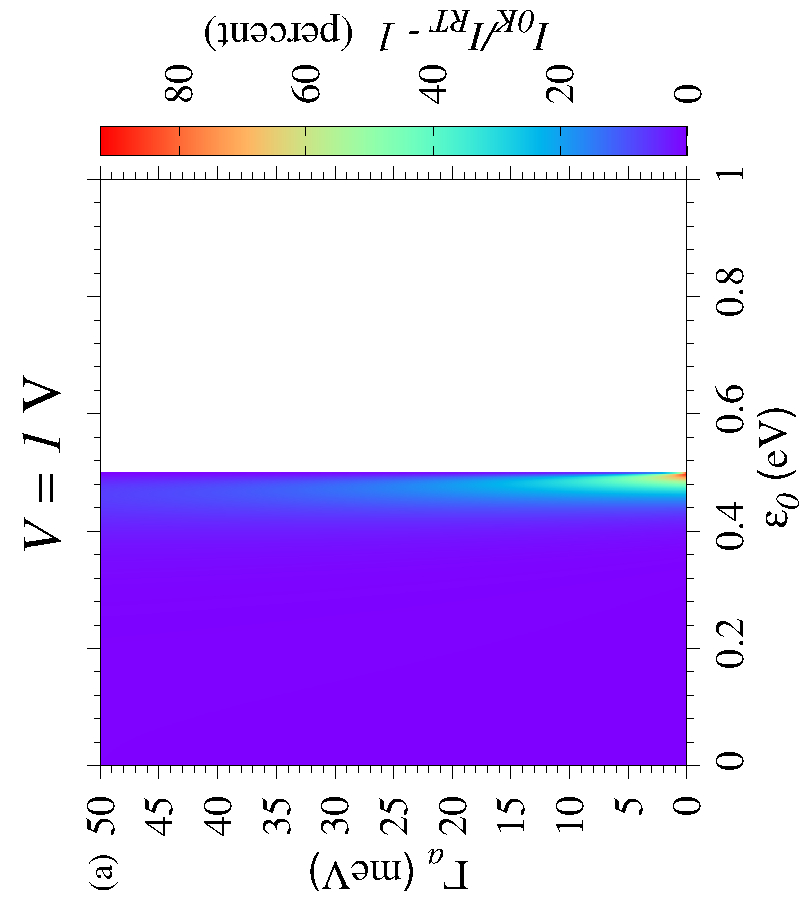}}
  \centerline{\includegraphics[width=0.22\textwidth,height=0.42\textwidth,angle=-90]{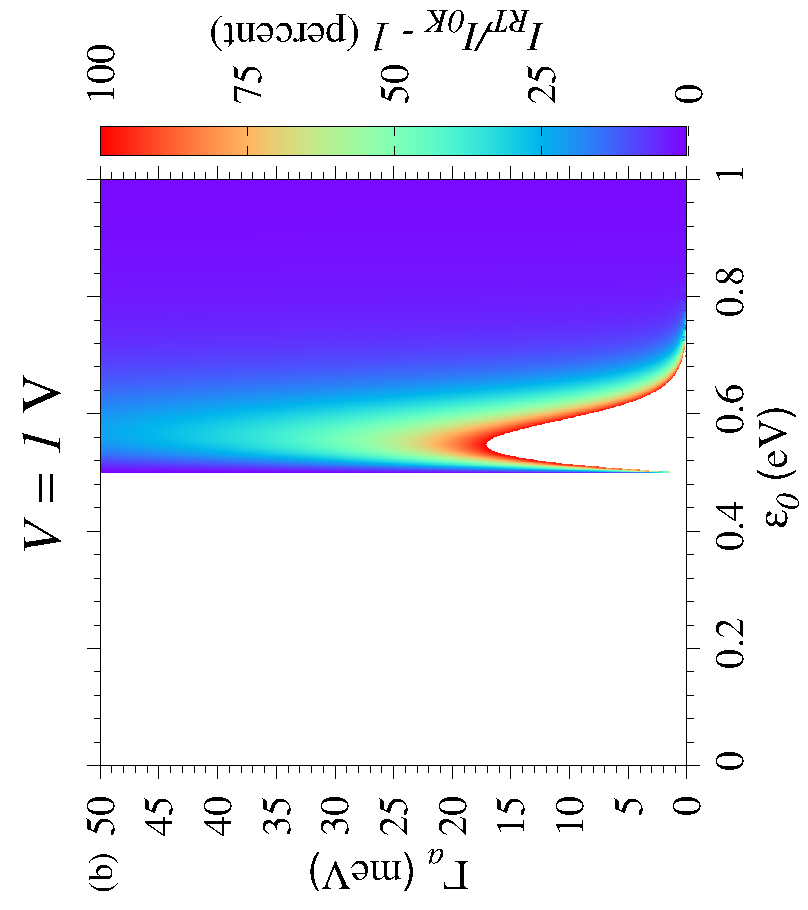}}
  \centerline{\includegraphics[width=0.22\textwidth,height=0.42\textwidth,angle=-90]{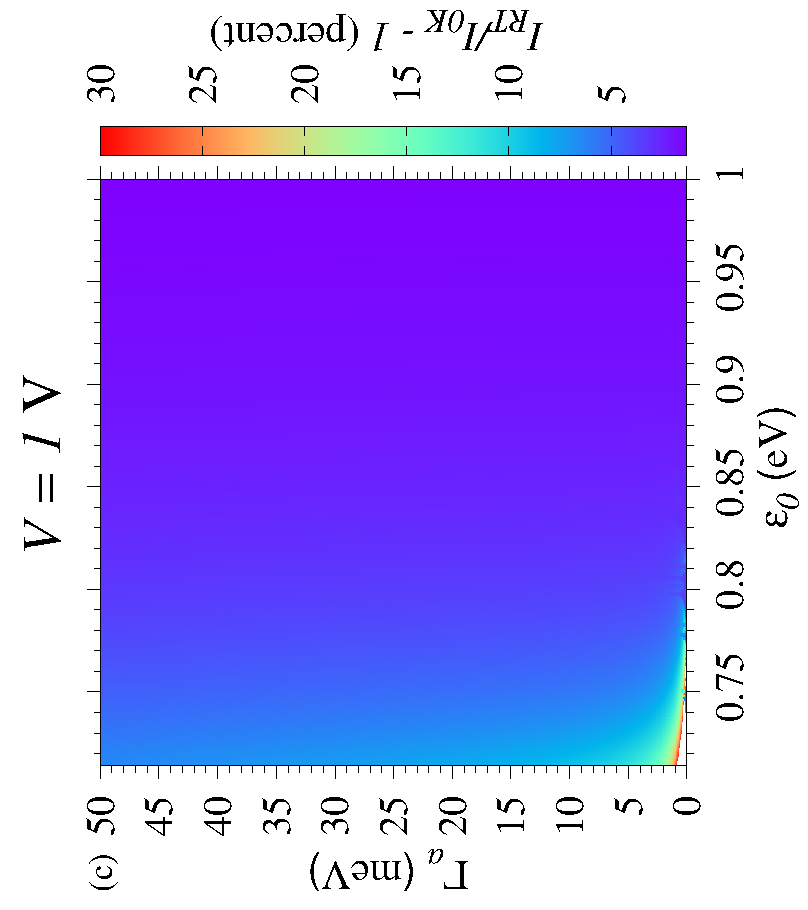}}
  \centerline{\includegraphics[width=0.22\textwidth,height=0.42\textwidth,angle=-90]{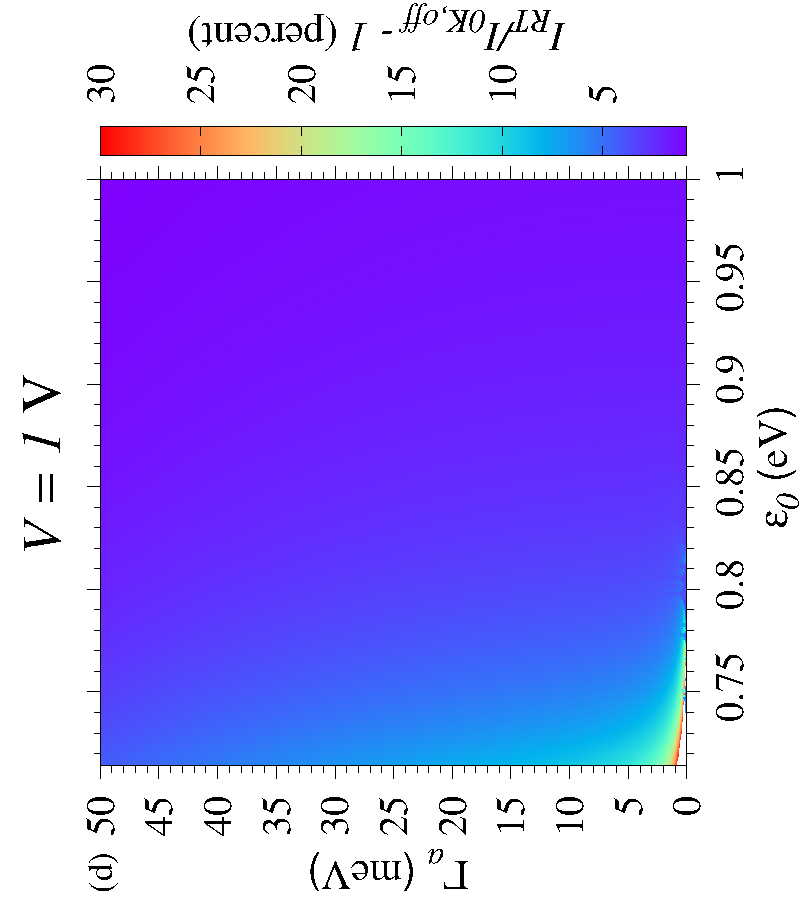}}
  \caption{The colored regions in the plane ($\varepsilon_0, \Gamma_a$) depict situations where,
    at the fixed bias indicated ($V = 1$\,V), 
    the current $I_{0K}$ computed at $T=0$ using eqn~(\ref{eq-I0K})
    is larger ($\vert eV\vert > 2 \left\vert\varepsilon_0 \right\vert$, panel a) or smaller ($\vert eV\vert < 2 \left\vert\varepsilon_0 \right\vert$, panel b)
    than the exact current $I_{RT}$ computed from eqn~(\ref{eq-Iexact}) at room temperature ($T = 298.15$\,K).
    For parameter values compatible with eqn~(\ref{eq-1.4}) and (\ref{eq-Gamma-vs-e0}),
    the current $I_{0K,off}$ computed using eqn~(\ref{eq-I0Koff}) is very accurate (panel d);
    it is as accurate as $I_{0K}$ (panel c). Relative deviations (shown only when not exceeding 100\%) are indicated in the color box.
    To facilitate comparison between $I_{0K,off}$ and $I_{0K}$, abscissas in panel c depicting $I_{0K}$ are restricted to those in panel d.
    Notice that the $z$-range in panels (c) and (d) is different from panel b.
   }
  \label{fig:errors-1.0V-si}
\end{figure}
\begin{figure}[htb]
  \centerline{\includegraphics[width=0.22\textwidth,height=0.42\textwidth,angle=-90]{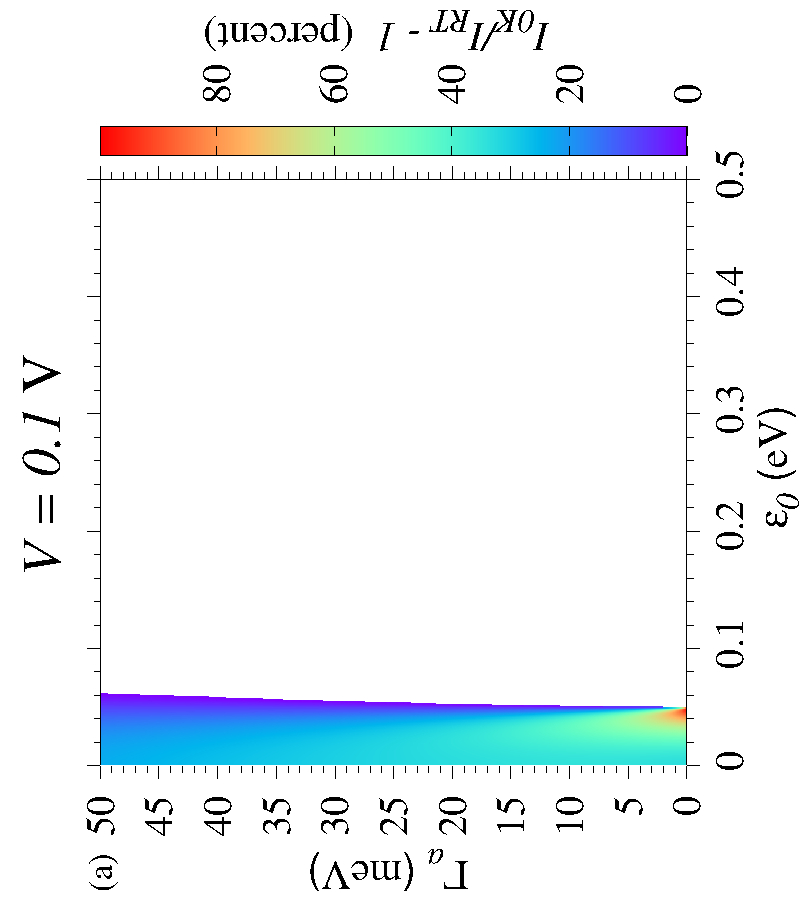}}
  \centerline{\includegraphics[width=0.22\textwidth,height=0.42\textwidth,angle=-90]{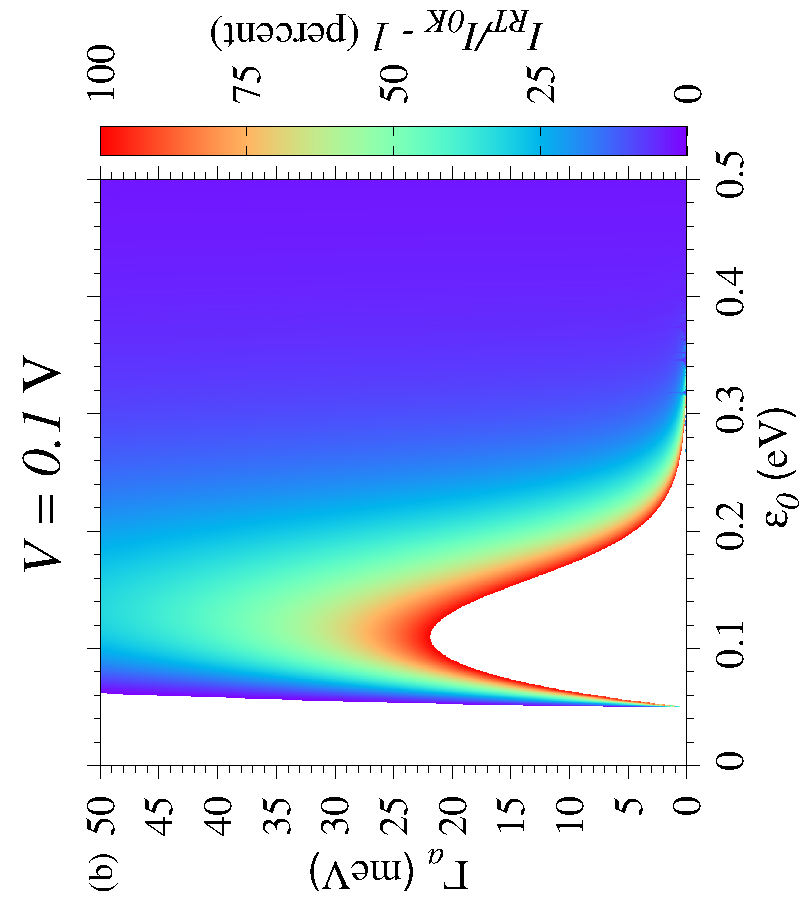}}
  \centerline{\includegraphics[width=0.22\textwidth,height=0.42\textwidth,angle=-90]{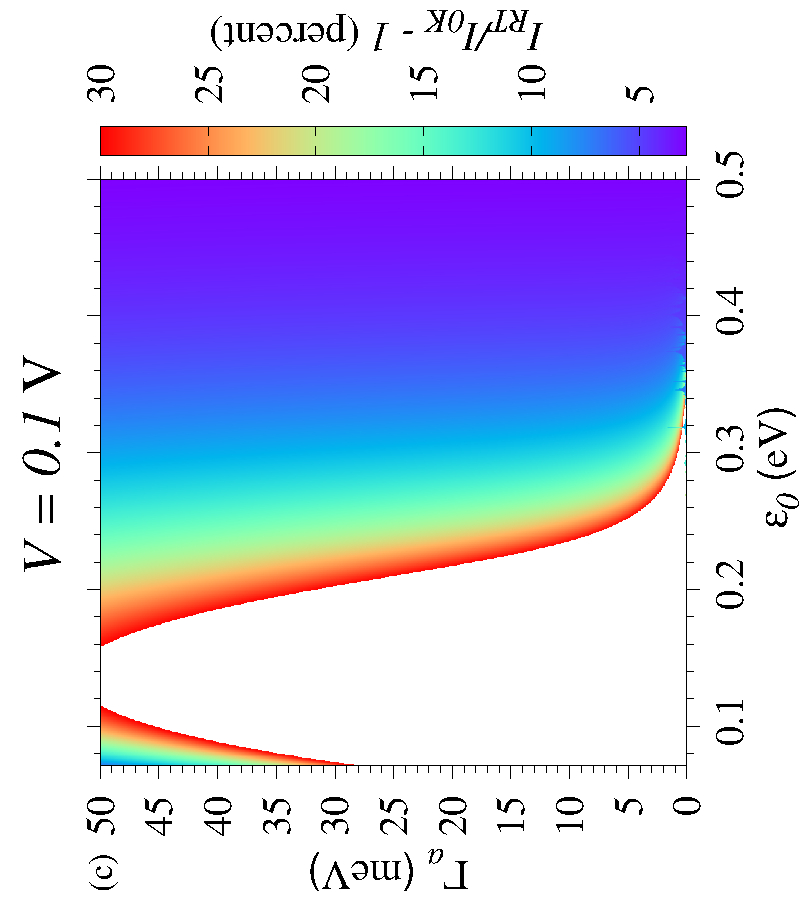}}
  \centerline{\includegraphics[width=0.22\textwidth,height=0.42\textwidth,angle=-90]{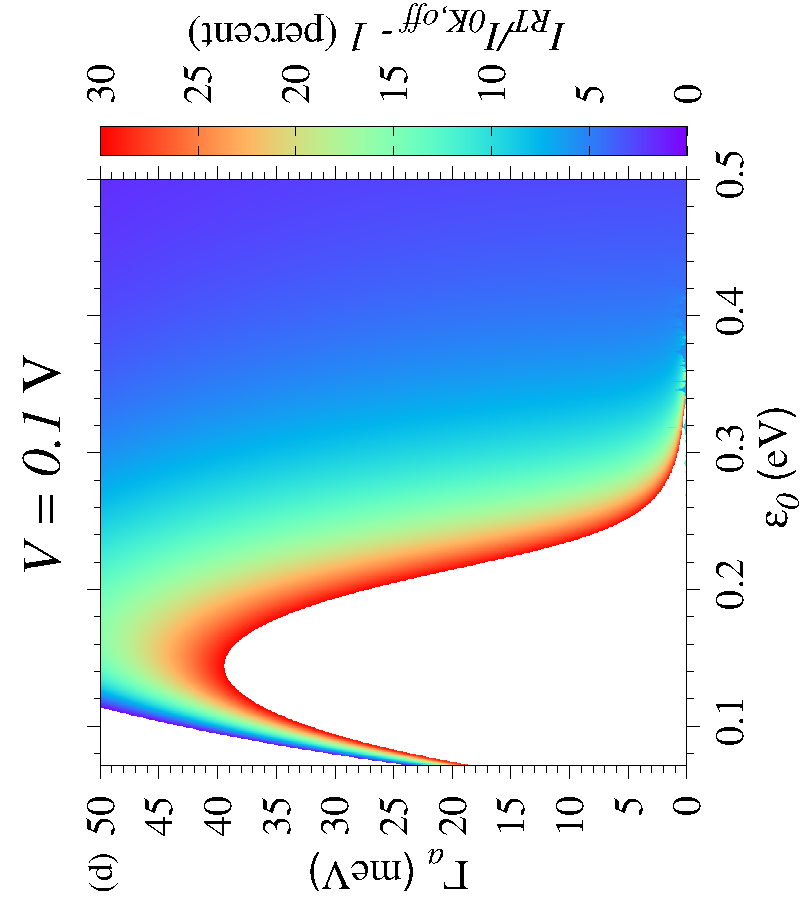}}
  \caption{The colored regions in the plane ($\varepsilon_0, \Gamma_a$) depict situations where,
    at the fixed bias indicated ($V = 0.1$\,V), 
    the current $I_{0K}$ computed at $T=0$ using eqn~(\ref{eq-I0K})
    is larger ($\vert eV\vert > 2 \left\vert\varepsilon_0 \right\vert$, panel a) or smaller ($\vert eV\vert < 2 \left\vert\varepsilon_0 \right\vert$, panel b)
    than the exact current $I_{RT}$ computed from eqn~(\ref{eq-Iexact}) at room temperature ($T = 298.15$\,K).
    The fact that in this case, paradoxically, the current $I_{0K,off}$ computed using eqn~(\ref{eq-I0Koff}) (panel d) is closer
    to $I_{RT}$ than $I_{0K}$ (panel c) is an error compensation effect.
    Relative deviations (shown only when not exceeding 100\%) are indicated in the color box.
    To facilitate comparison between $I_{0K,off}$ and $I_{0K}$, abscissas in panel c depicting $I_{0K}$ are restricted to those in panel d.
    Notice that the $z$-range in panels (c) and (d) is different from panel b.
   }
  \label{fig:errors-0.1V-si}
\end{figure}
\begin{figure}[htb]
  \centerline{\includegraphics[width=0.22\textwidth,height=0.42\textwidth,angle=-90]{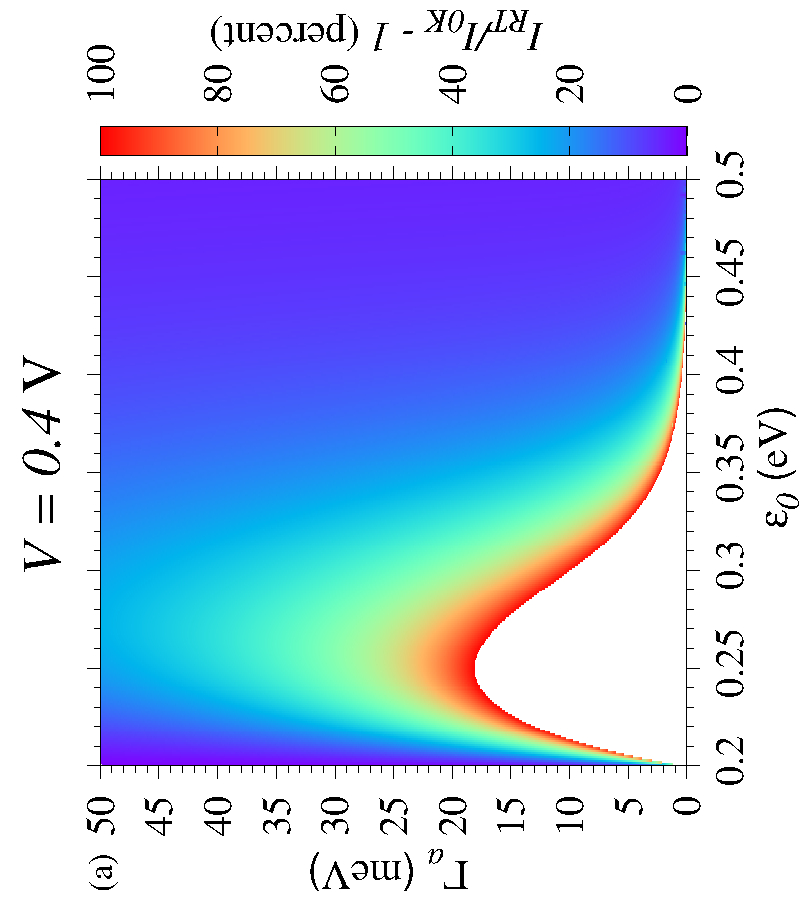}}
  \centerline{\includegraphics[width=0.22\textwidth,height=0.42\textwidth,angle=-90]{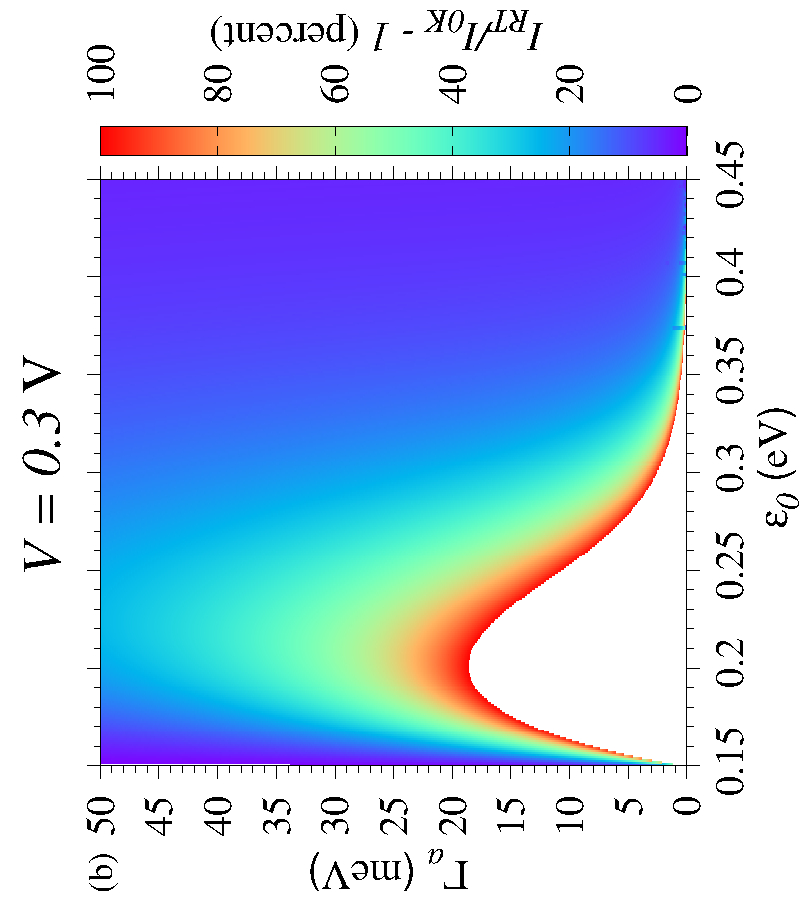}}
  \centerline{\includegraphics[width=0.22\textwidth,height=0.42\textwidth,angle=-90]{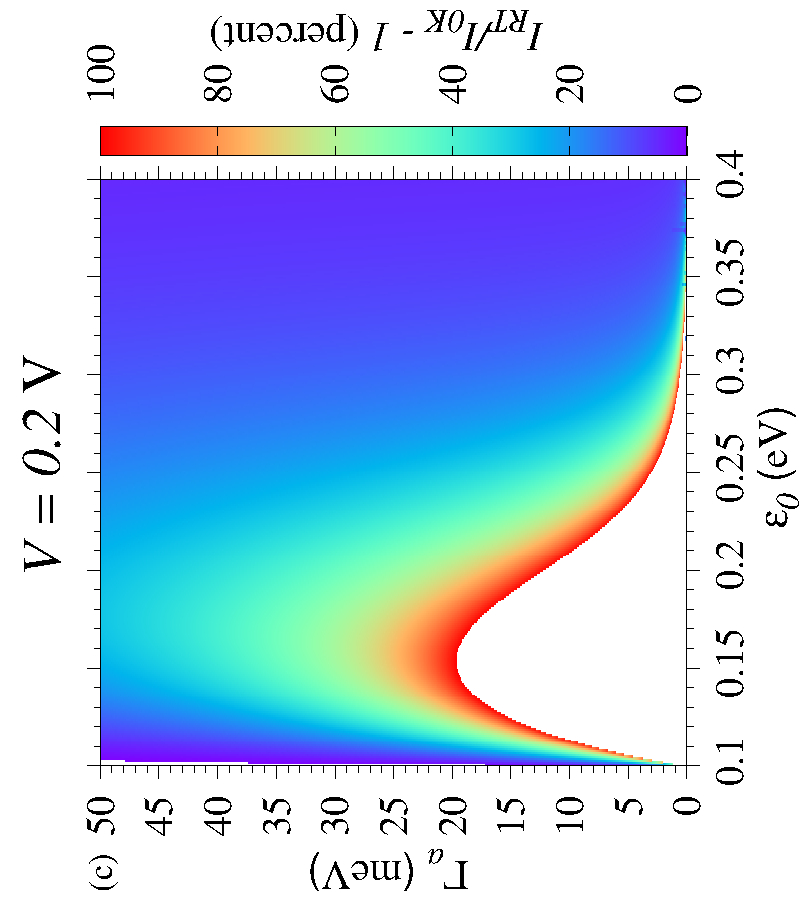}}
  \centerline{\includegraphics[width=0.22\textwidth,height=0.42\textwidth,angle=-90]{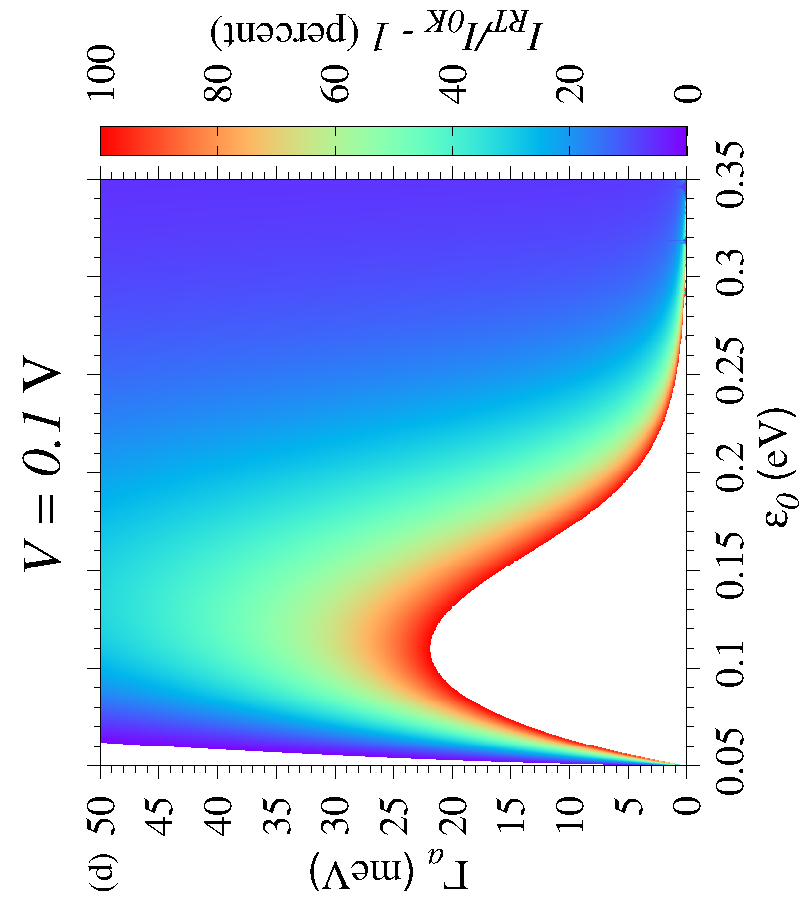}}
  \centerline{\includegraphics[width=0.22\textwidth,height=0.42\textwidth,angle=-90]{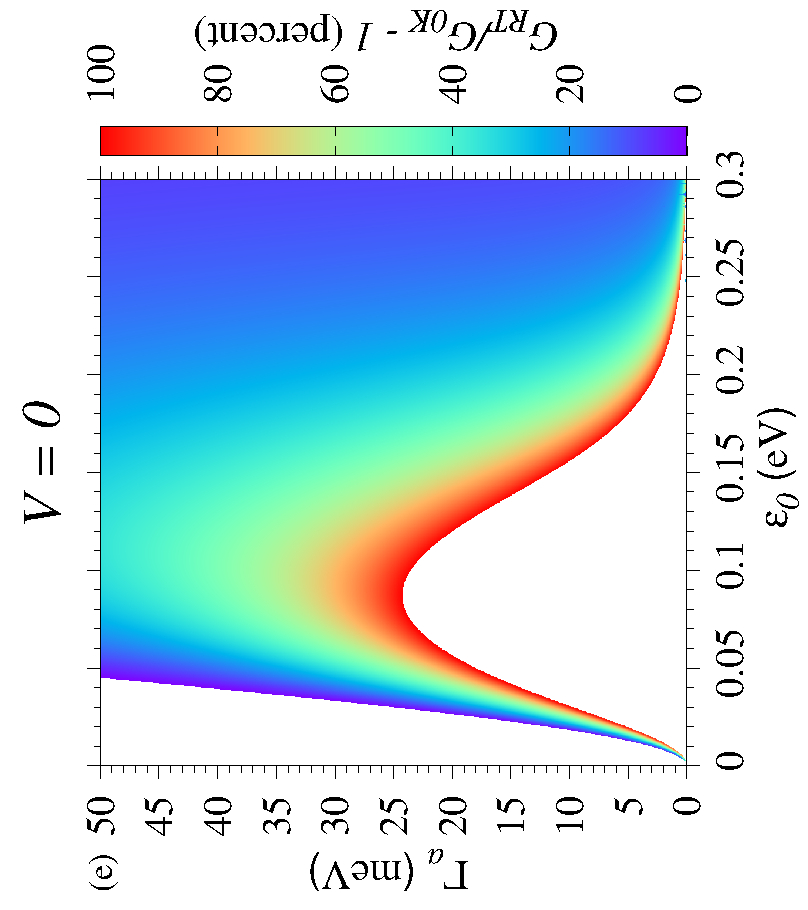}}
  \caption{Results illustrating the current enhancement below resonance ($\vert e V\vert < 2 \left\vert\varepsilon_0 \right\vert$) at lower biases.
    As the bias decreases (downwards), the white (empty) region (wherein the relative deviations exceed 100\%)
    extends upwards to larger $\Gamma_a$ and comprises a broader $\varepsilon_0$-range. 
    Notice that the leftmost positions of all panels are aligned to resonance.}
  \label{fig:errors-V-g-aligned-si}
\end{figure}
\begin{figure}[htb]
  \centerline{\includegraphics[width=0.22\textwidth,height=0.42\textwidth,angle=-90]{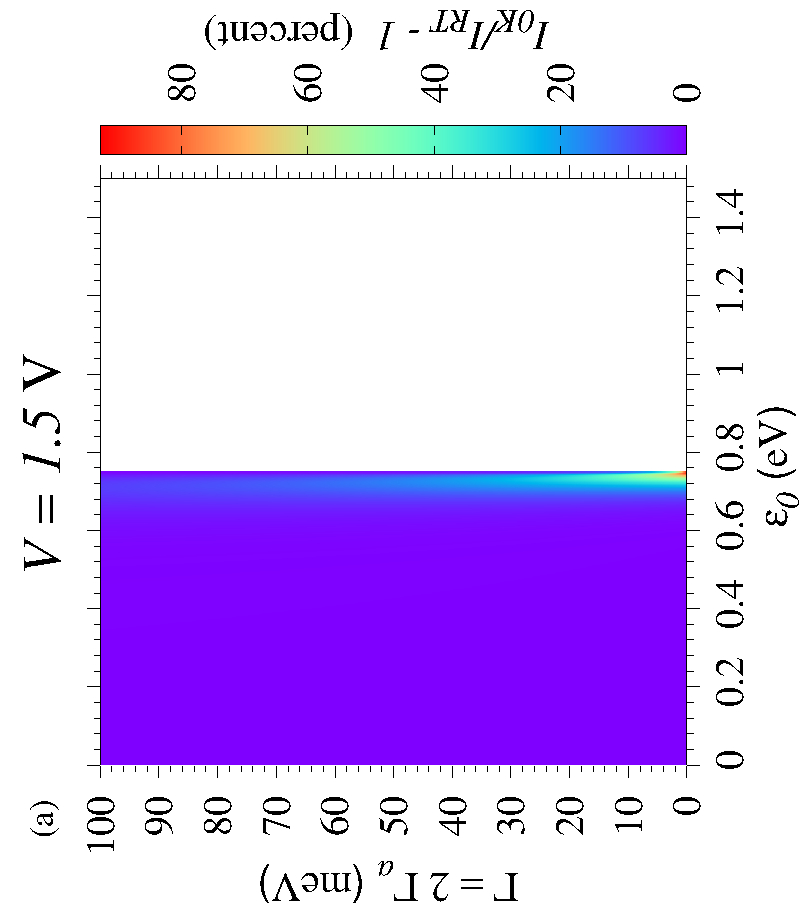}}
  \centerline{\includegraphics[width=0.22\textwidth,height=0.42\textwidth,angle=-90]{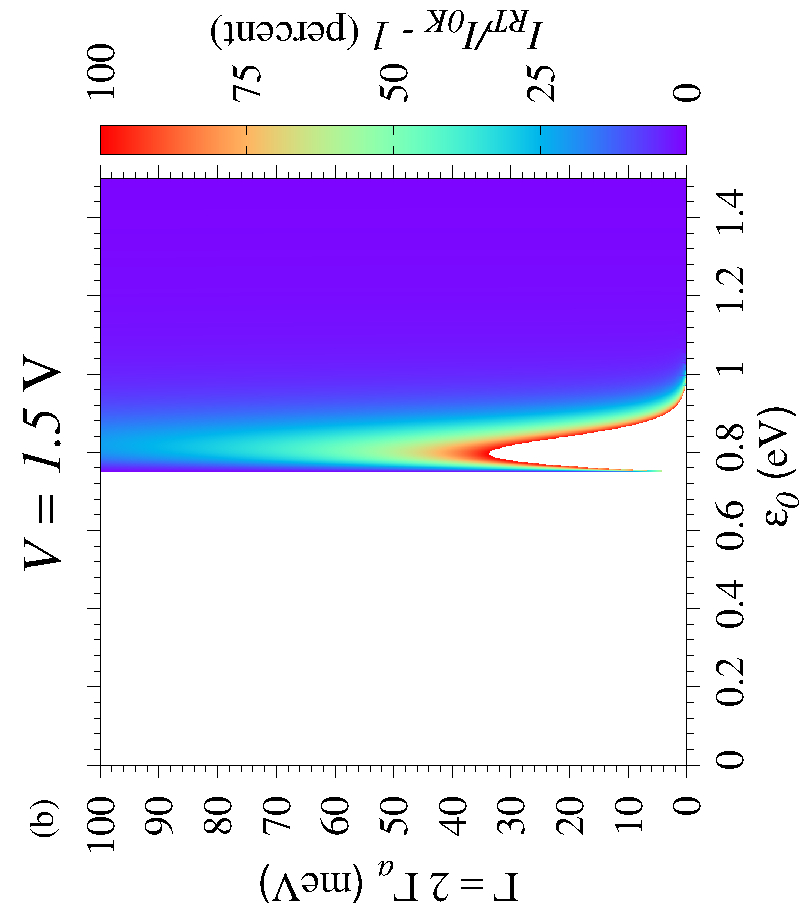}}
  \centerline{\includegraphics[width=0.22\textwidth,height=0.42\textwidth,angle=-90]{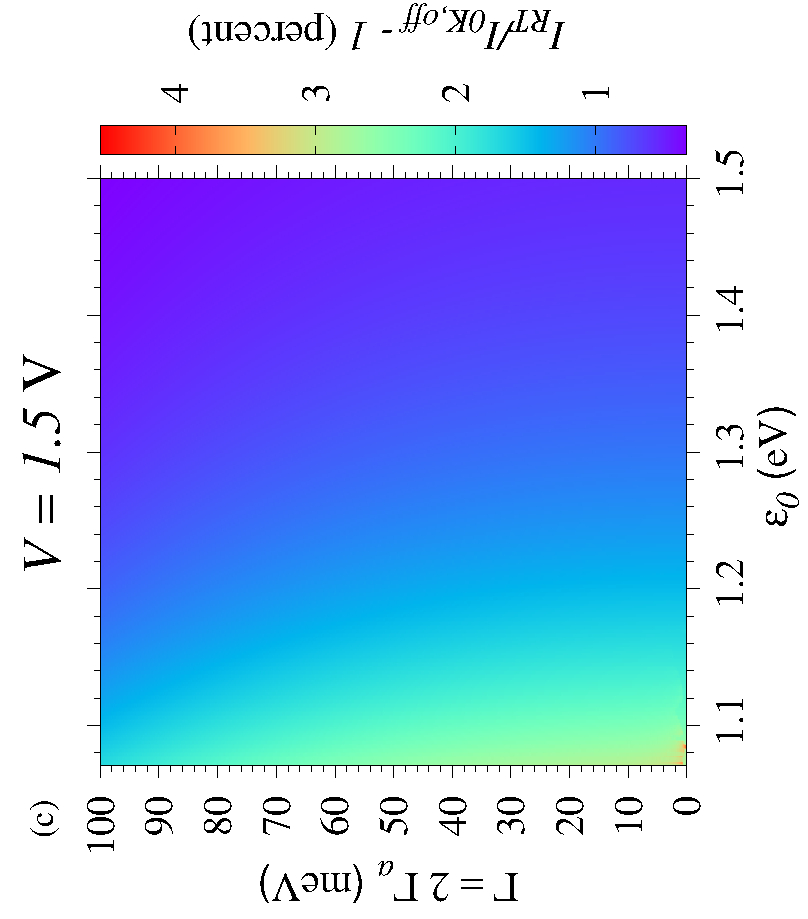}}
  \centerline{\includegraphics[width=0.355\textwidth,height=0.22\textwidth,angle=0]{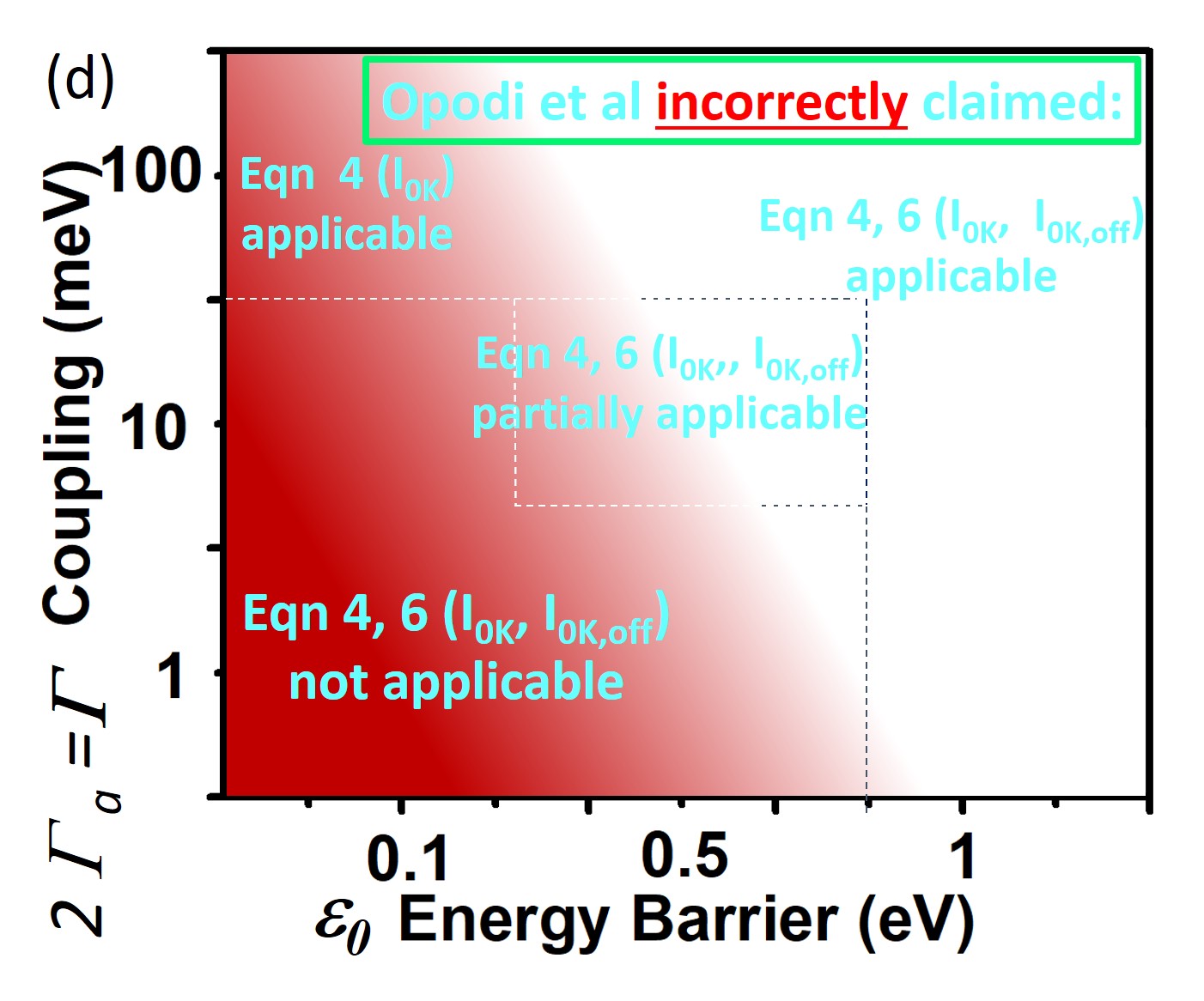}\hspace*{8ex}}
  \caption{Relative deviations in percent 
    of the currents $I_{0K}$ (panel a and b) and $I_{0K,off}$ (panel c) computed 
    via eqn~(\ref{eq-I0K}) and (\ref{eq-I0Koff})
    assuming zero temperature from the exact current $I_{RT}$ computed at room temperature.
    Comparison with panel d (adapted from Opodi et al, Phys.~Chem.~Chem.~Phys.~2022, 24, 11958 and ref.~\citenum{baldea:comment}) demonstrates that Fig.~5 of
    Opodi et al is a factual error.  Notice that, in order to facilitate comparison with the paper by Opodi et al,
    the electronic coupling $\Gamma = 2\Gamma_a$ in panel d is different from \figname\ref{fig:errors-1.5V}.}
  \label{fig:errors-1.5V-vs-opodi}
\end{figure}
\begin{figure*}[htb]
  \centerline{\includegraphics[width=0.22\textwidth,height=0.42\textwidth,angle=-90]{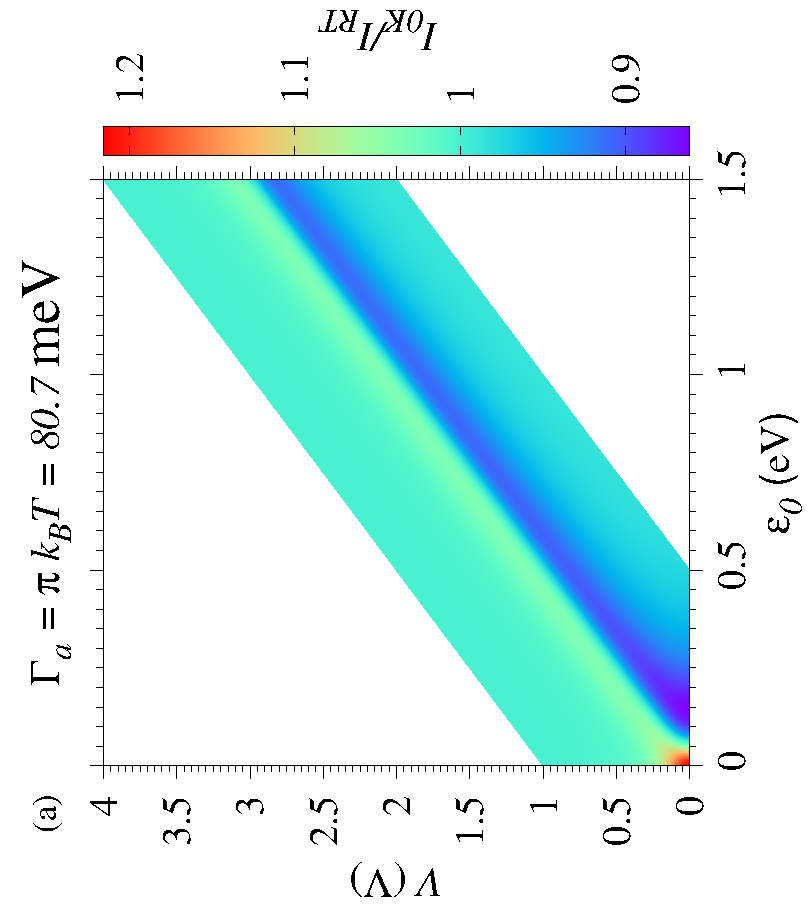}
    \includegraphics[width=0.22\textwidth,height=0.42\textwidth,angle=-90]{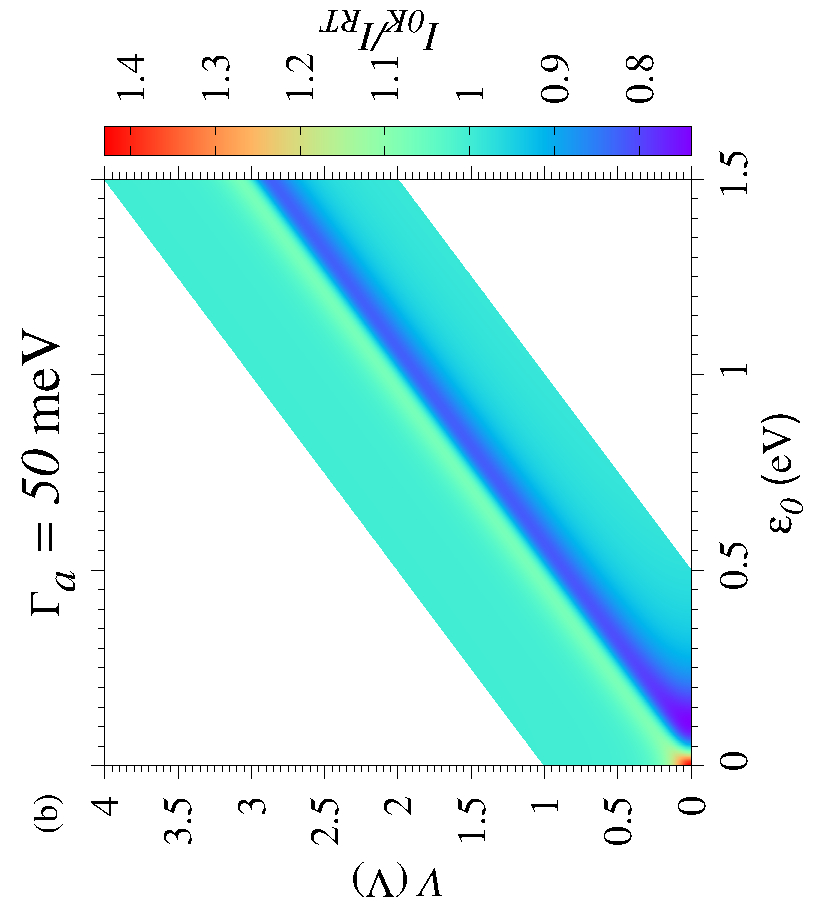}}
  \caption{Panel b visualizes the weak impact of the (room) temperature on
    the current at the highest value $\Gamma_a = 50$\,meV chosen in most diagrams, which is even smaller
    than the value $\Gamma_a = \pi k_B T_{RT} = 80.7$\,meV (panel a) where a weak temperature effect can be expected in view of eqn~(\ref{eq-low-T-V-gg}).}
  \label{fig:errors-fixed-Gamma}
\end{figure*}
\begin{figure*}[htb]
  \centerline{\includegraphics[width=0.22\textwidth,height=0.42\textwidth,angle=-90]{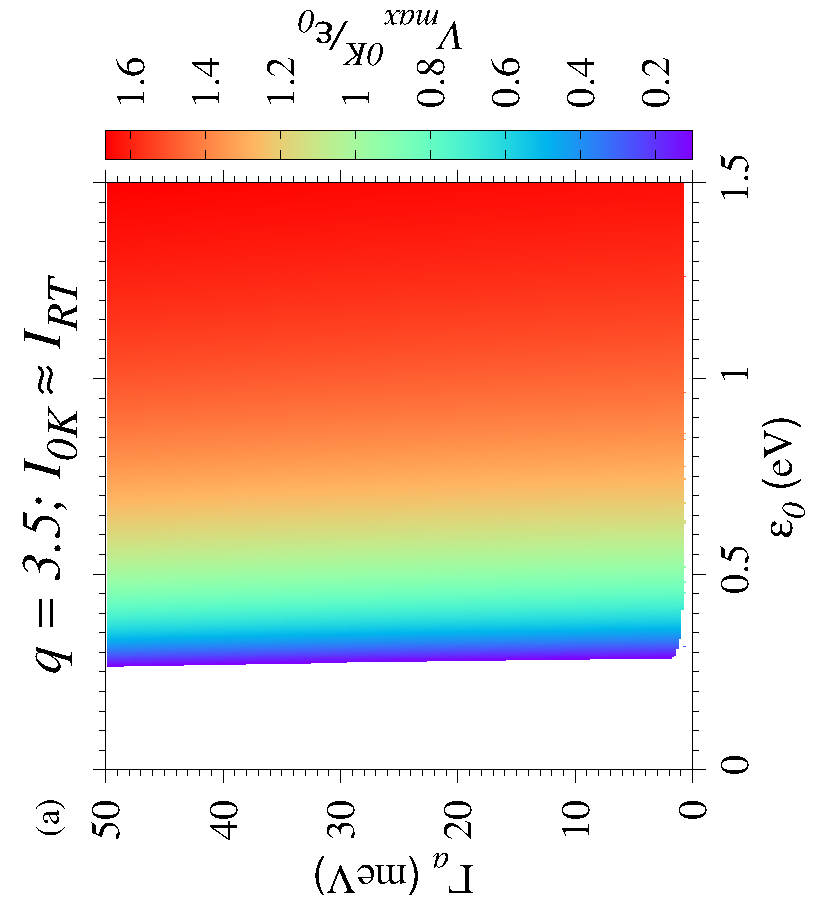}
    \includegraphics[width=0.22\textwidth,height=0.42\textwidth,angle=-90]{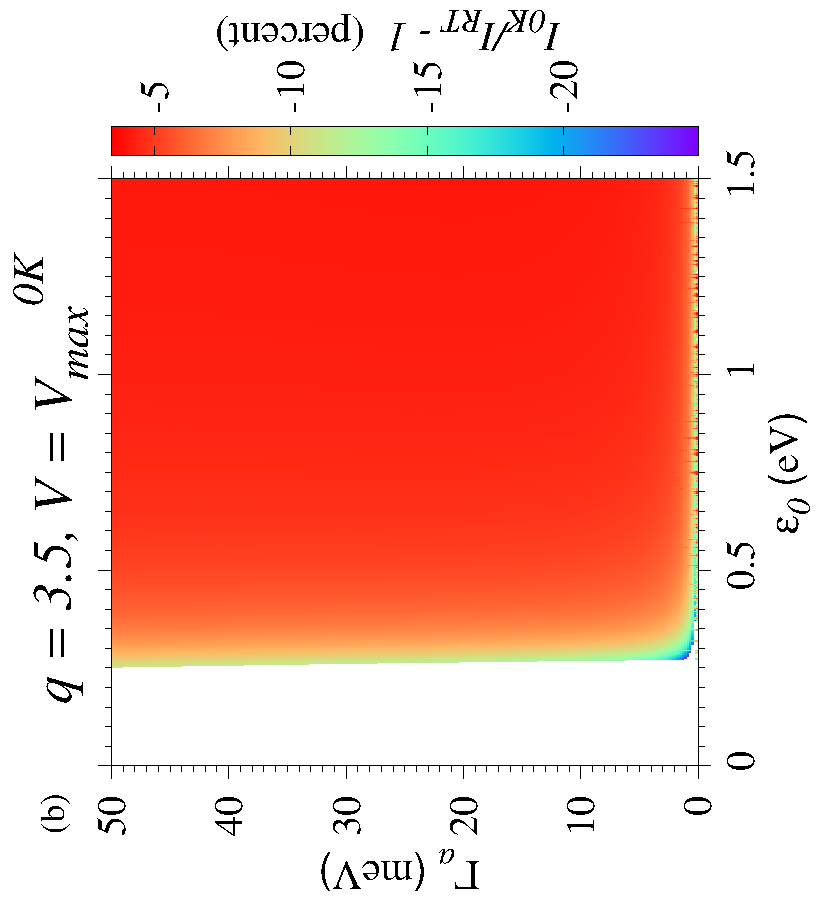}}
    \centerline{\includegraphics[width=0.22\textwidth,height=0.42\textwidth,angle=-90]{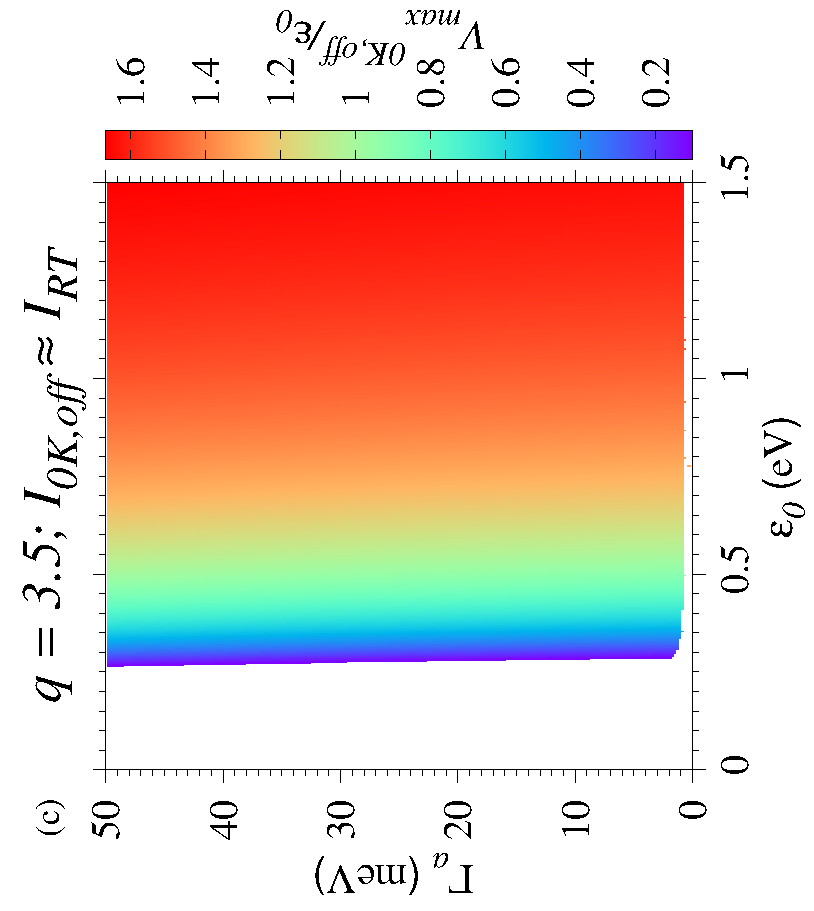}
    \includegraphics[width=0.22\textwidth,height=0.42\textwidth,angle=-90]{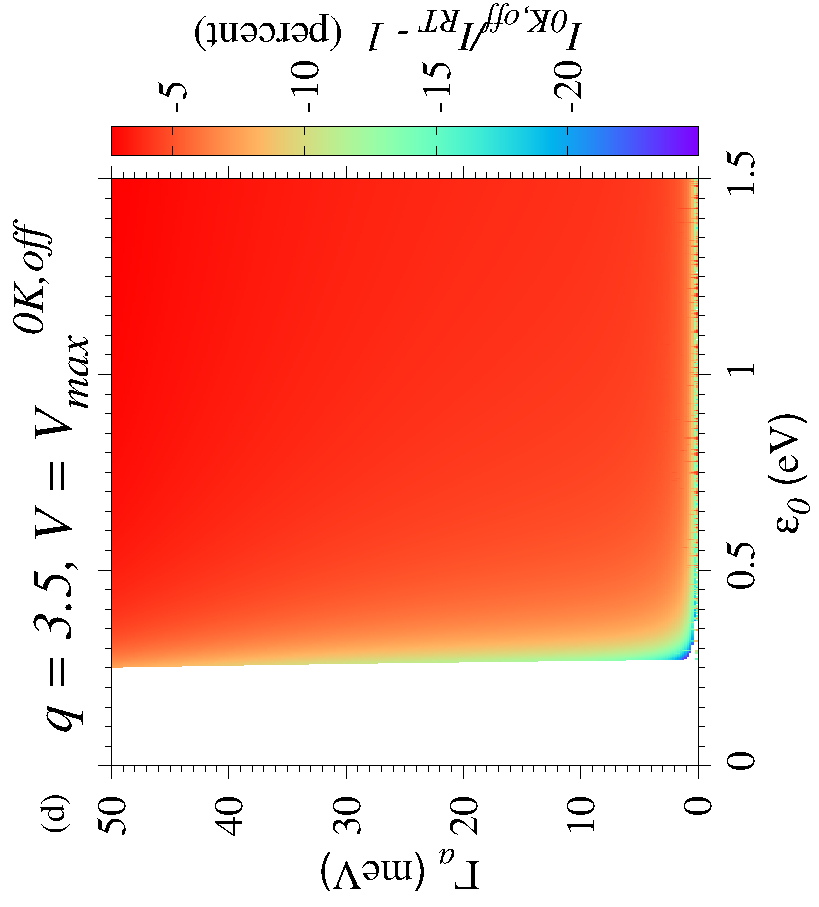}}
  \caption{Similar to \figname\ref{fig:q=3} but setting $q=3.5$ in eqn~(\ref{eq-low-T-V-q}).}
  \label{fig:q=3.5}
\end{figure*}
\begin{figure*}[htb]
  \centerline{\includegraphics[width=0.22\textwidth,height=0.42\textwidth,angle=-90]{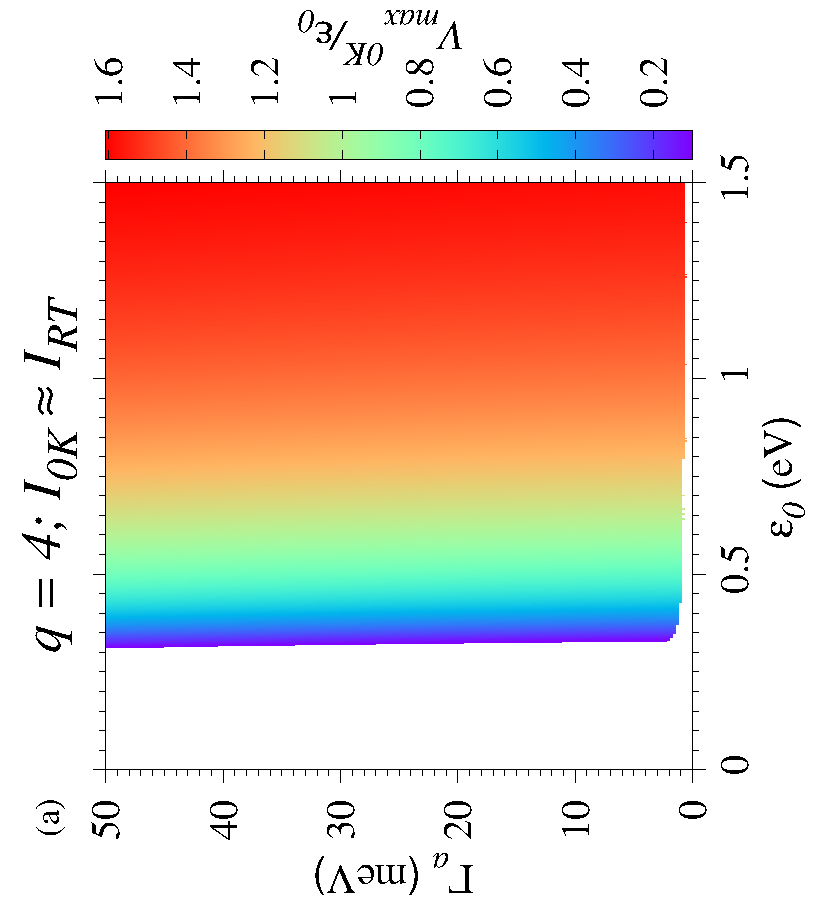}
    \includegraphics[width=0.22\textwidth,height=0.42\textwidth,angle=-90]{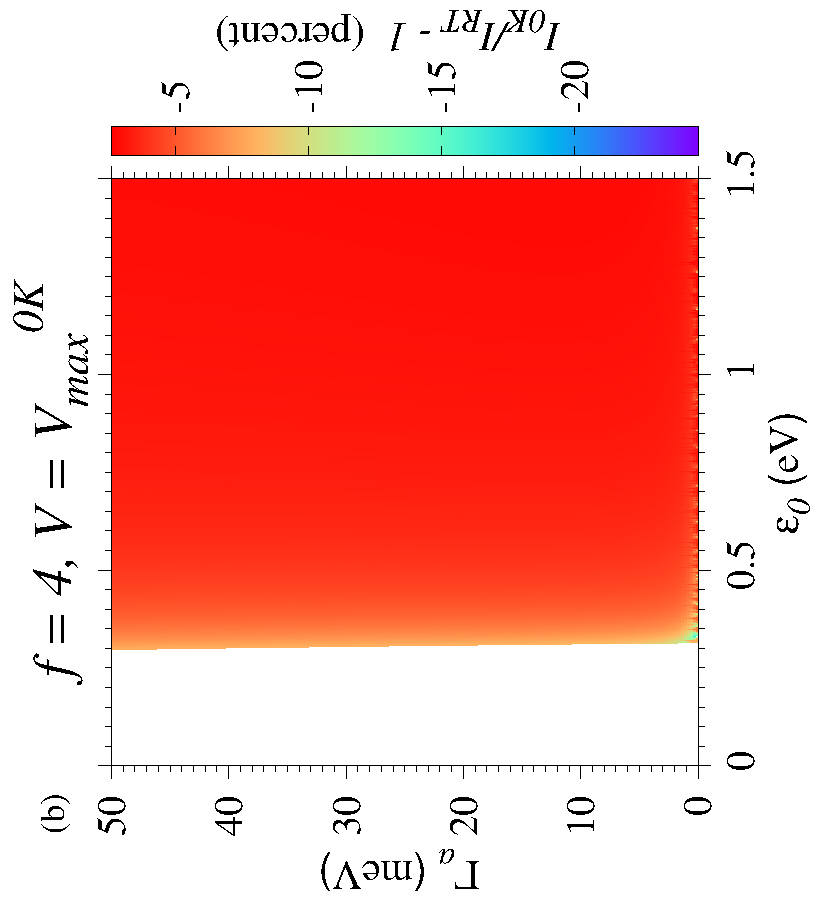}}
    \centerline{\includegraphics[width=0.22\textwidth,height=0.42\textwidth,angle=-90]{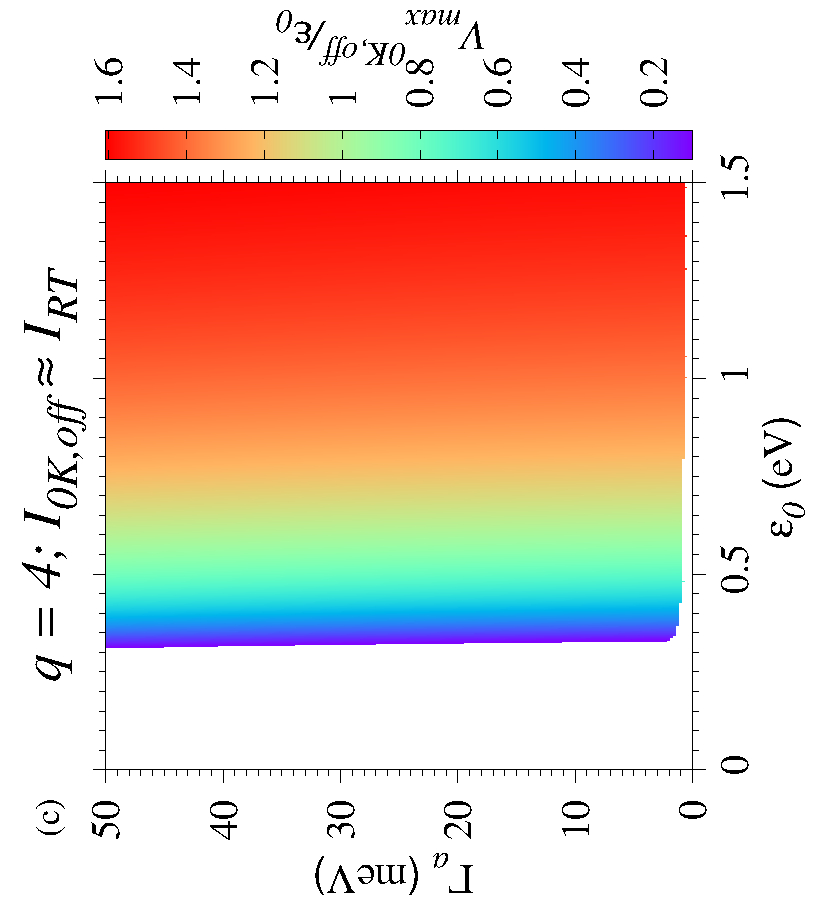}
    \includegraphics[width=0.22\textwidth,height=0.42\textwidth,angle=-90]{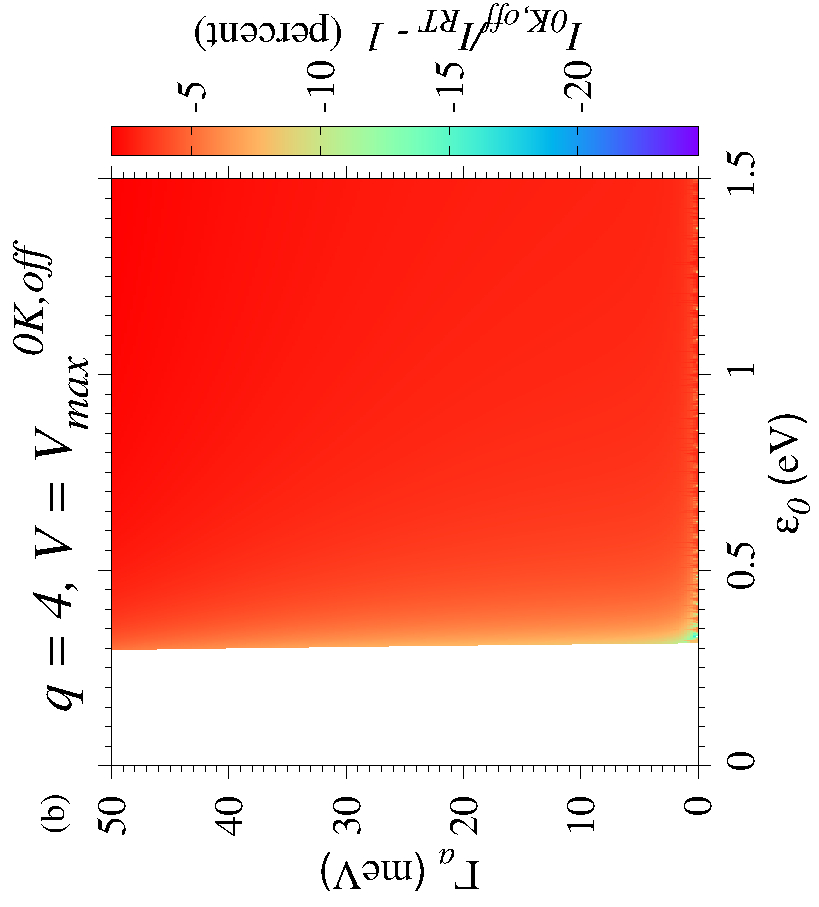}}
  \caption{Similar to \figname\ref{fig:q=3} but setting $q=4$ in eqn~(\ref{eq-low-T-V-q}).}
  \label{fig:q=4}
\end{figure*}
\begin{figure*}[htb]
  \centerline{\includegraphics[width=0.22\textwidth,height=0.42\textwidth,angle=-90]{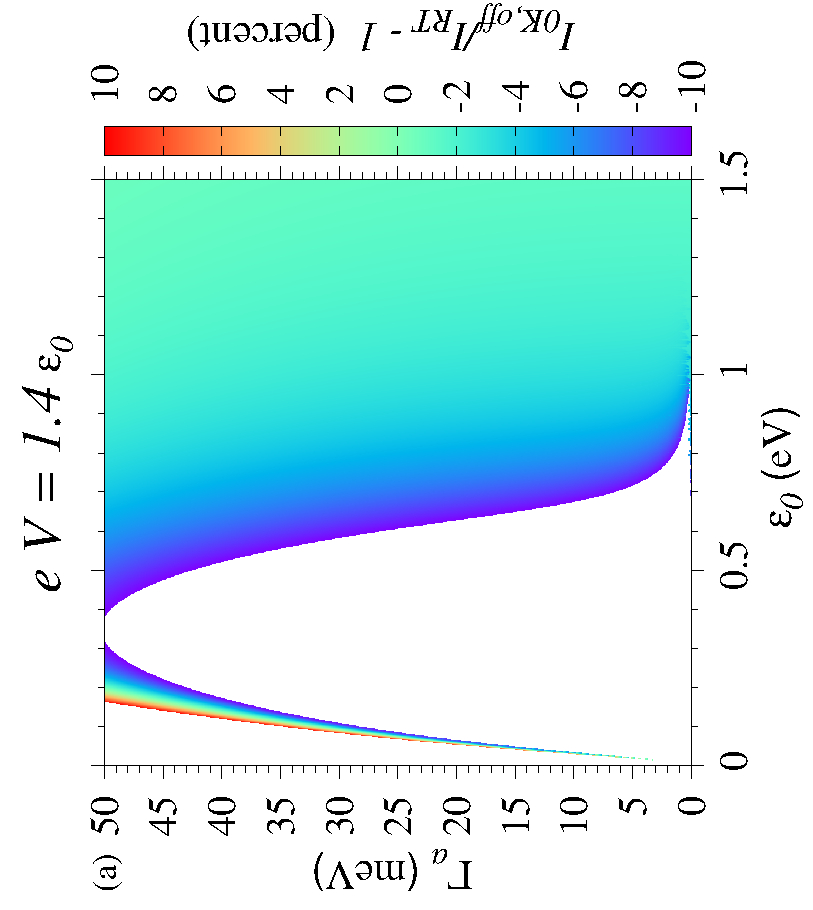}
    \includegraphics[width=0.22\textwidth,height=0.42\textwidth,angle=-90]{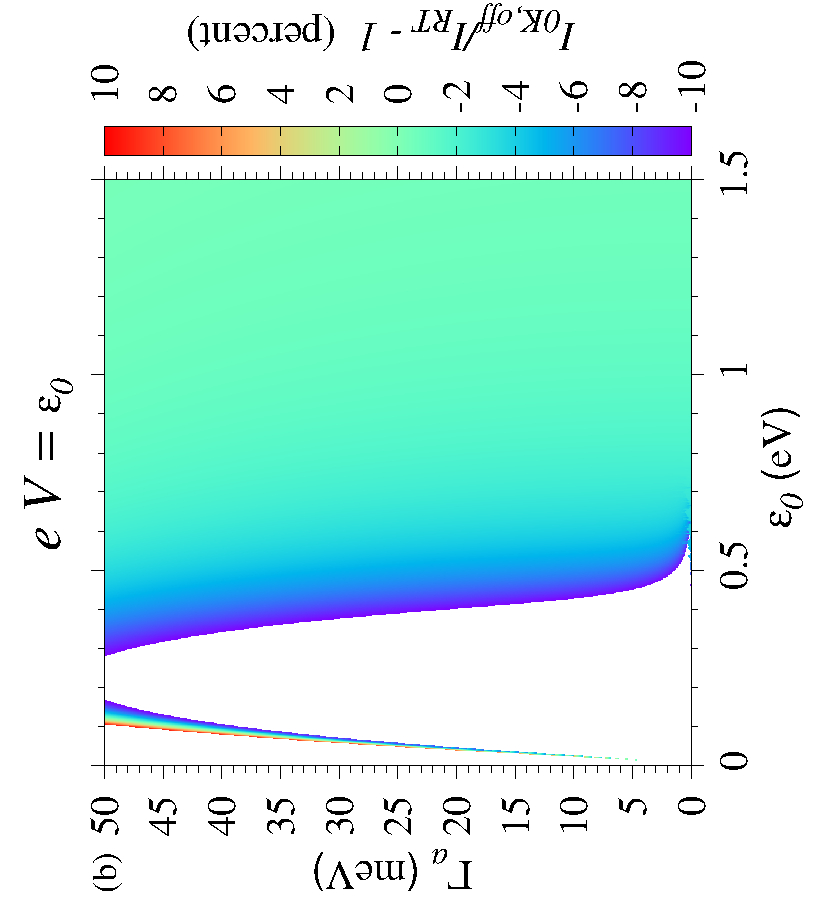}}
  \caption{This figure illustrates how restriction to the narrower bias range $\vert e V\vert < \left\vert \varepsilon_0\right\vert$ (panel b)
      can render data fitting using $I_{0K,off}$ (eqn~(\ref{eq-I0Koff})) applicable for junctions having, e.g., $\varepsilon_0 \simeq 0.5$\,eV,
      a fact impossible when using the broader bias range  $\vert e V\vert < 1.4\,\left\vert \varepsilon_0\right\vert$ (panel a).}
  \label{fig:I0Koff-1.4-1.0-10perc}
\end{figure*}
\end{document}